\documentclass[10pt,a4paper]{article}

\usepackage{jheppub}

\usepackage{braket}
\usepackage{comment}
\usepackage{subcaption}

\usepackage{slashed}

\newcommand{\al}{\alpha}
\newcommand{\be}{\beta}
\newcommand{\ga}{\gamma}

\newcommand{\bb}{\delta}

\newcommand{\nn}[1]{\left\lVert #1\right\rVert}

\title{Breaking of integrability in elliptic deformations of AdS3 strings with mixed flux}

\author{Ian Le Meur,}
\author{Fiona~K.~Seibold%
}

\affiliation{Theoretical Particle Physics Laboratory, EPFL, 1015 Lausanne, Switzerland}

\emailAdd{ian.lemeur@epfl.ch}
\emailAdd{fiona.seibold@epfl.ch}

\abstract{We construct elliptic deformations of the $AdS_3 \times S^3 \times T^4$ superstring supported by mixed NSNS and RR fluxes in type IIB supergravity and investigate their integrability. The NSNS sector includes a Wess-Zumino term, and imposing that the theory preserves 8 supersymmetries also requires the presence of a B-field coupling the deformed $AdS_3$ and $S^3$ factors. We explore integrability from two different perspectives: through the construction of a flat Lax connection and the computation of the tree-level worldsheet S-matrix. For generic elliptic parameters and NSNS flux, we find inelastic $2 \rightarrow 2$ amplitudes, violating the elasticity required by integrability. In the integrable cases as identified by the Lax analysis, these inelastic amplitudes vanish and the tree-level S-matrix satisfies the classical Yang-Baxter equation. In particular, the addition of a Wess-Zumino term, which preserves integrability in the rational and trigonometric models, breaks integrability for elliptic deformations.
}

\begin{document}
\maketitle
\allowdisplaybreaks

\section{Introduction}

Integrable string backgrounds provide unique laboratories to explore string theory  in curved space-times.
Particularly interesting are those backgrounds containing an Anti-de-Sitter geometry and playing a role in the AdS/CFT correspondence~\cite{Maldacena:1997re}, as do for instance $AdS_5 \times S^5$~\cite{Bena:2003wd} and $AdS_3 \times S^3 \times T^4$. 
The latter is particularly interesting due to the large moduli space of parameters~\cite{Larsen:1999uk,OhlssonSax:2018hgc}.
It is well-known that the $AdS_3 \times S^3 \times T^4$ superstring can be supported by a mixture of NSNS and RR fluxes~\cite{Duff:1998cr}, 
\begin{equation}
H_3 = q \, \mathsf{h}_V, \qquad F_3 = \sqrt{1-q^2} \, \mathsf{h}_V\,, \qquad \mathsf{h}_V =2 (\text{Vol}(AdS_3) + \text{Vol}(S^3))\,,
\end{equation}
with $0\leq q\leq 1$, while preserving sixteen supersymmetries corresponding to the isometry algebra $\mathfrak{psu}(1,1|2) \oplus \mathfrak{psu}(1,1|2)$, as well as the integrability of the model~\cite{Cagnazzo:2012se}. 
This can be traced back to the classical integrability of the Principal Chiral Model (PCM), which stays integrable when adding a Wess-Zumino (WZ) term~\cite{Abdalla:1984gm}, for any value of the (quantised) WZ level $k=2 \pi \text{T} q$, where $\text{T}$ is the string tension. 
The case $q=1$ is special as it corresponds to the pure NSNS point, for which the worldsheet theory becomes a WZW model which can be analysed using 2d CFT techniques~\cite{Maldacena:2000hw}. 
As soon as $q<1$, the theory is supported by a mixture of fluxes and its quantisation becomes trickier, but it can be achieved by studying the theory in uniform light-cone gauge~\cite{Arutyunov:2005hd}, and we refer the reader to~\cite{Demulder:2023bux} for a review of that method in the context of integrable $AdS_3$ strings.
The tree-level worldsheet S-matrix for the mixed-flux theory in the massive sector~\footnote{The theory whose S-matrix is computed contains 4 massive bosons, 4 massless bosons, 4 massive fermions and 4 massless fermions. Perturbative calculations in the massless sector are difficult due to the presence of IR divergences.} was obtained in~\cite{Hoare:2013pma}, and the all-loop S-matrix was bootstrapped using symmetries~\footnote{Up to overall dressing phases, which have more recently been fixed using additional physical input~\cite{Frolov:2024pkz,Frolov:2025tda}.} in~\cite{Lloyd:2014bsa}. As required by integrability, the two-body worldsheet S-matrix only features elastic processes and satisfies the Yang-Baxter equation~\cite{Parke:1980ki,Zamolodchikov:1978xm}.

\medskip
Beyond adding a WZ term, the $AdS_3 \times S^3 \times T^4$ superstring admits a large landscape of continuous integrable deformations, preserving different numbers of supersymmetries and falling into different classes of integrable models (rational, trigonometric and elliptic), see~\cite{Seibold:2024qkh, Hoare:2021dix} for reviews.
Most of these integrable deformations can be described by Yang-Baxter (YB) deformations.
Initially proposed for the PCM on a simple Lie group~\cite{Klimcik:2002zj,Klimcik:2008eq,Delduc:2013fga}, this family of integrable models was then generalised to symmetric and semi-symmetric space sigma-models~\cite{Delduc:2013qra,Delduc:2014kha}. It can therefore be applied to integrable (super)string theory on backgrounds featuring $AdS_n$ and $S^n$ factors, and includes TsT~\cite{Osten:2016dvf}, Jordanian~\cite{vanTongeren:2015soa,vanTongeren:2019dlq}, and quantum-group deformations~\cite{Delduc:2014kha,Hoare:2018ngg}. 
The latter are closely related to $\lambda$-deformations~\cite{Sfetsos:2013wia,Hollowood:2014qma}.
Also from the point of view of deformations, the $AdS_3 \times S^3 \times T^4$ background plays a special role due to the factorisation of the superisometry algebra into two $\mathfrak{psu}(1,1|2)$ copies, which allows for the construction of multi-parametric quantum-deformations~\cite{Hoare:2014oua,Seibold:2019dvf,Hoare:2022vnw}, as well as supersymmetry-preserving ones. 
The trigonometric deformation of the $AdS_3 \times S^3 \times T^4$ superstring is equivalent (up to total derivative terms) to the unilateral YB deformation, which only deforms one $\mathfrak{psu}(1,1|2)$ copy and therefore preserves 8 supersymmetries.

\medskip
Also the trigonometric deformation of the $AdS_3 \times S^3 \times T^4$ superstring can be supported by a mixture of fluxes, while preserving 8 supersymmetries and integrability. The NSNS flux can be generated by the presence of a WZ term in the action~\cite{Delduc:2017fib,Delduc:2018xug} or by a TsT transformation along two $\mathfrak{u}(1)$ directions inside one $\mathfrak{psu}(1,1|2)$ copy~\cite{Hoare:2022asa}, or both at the same time.
Note however that contrary to the rational+WZ theory, the trigonometric+WZ metric cannot be supported by pure NSNS flux, supergravity necessarily imposes the presence of RR fluxes as well.
The tree-level S-matrix of the bi-YB deformation of $AdS_3 \times S^3 \times T^4$ was computed in~\cite{Seibold:2021lju} (in the massive sector), while that of the bi-YB+WZ model was obtained in~\cite{Bocconcello:2020qkt} (in the massive bosonic sector), and that of the trigonometric+TsT was discussed in~\cite{Hoare:2022asa} (massive bosonic sector). In all these cases the S-matrix satisfies the classical Yang-Baxter equation.
The perturbative S-matrix of~\cite{Seibold:2021lju} also agrees with the exact S-matrix bootstrapped using the $q$-deformed symmetry of the model~\cite{Hoare:2014oua}. The exact S-matrix remains however unknown in the quantum-deformed mixed-flux case. 

\medskip
Recently, in~\cite{Hoare:2023zti,Hoare:2025rtl}, an elliptic deformation of the $AdS_3 \times S^3 \times T^4$ superstring has been constructed. This deformation falls outside of the YB family. Its supergravity background is characterised by a trivial dilaton, vanishing NSNS three-form flux but non-trivial RR three-form and five-form fluxes. More explicitly, in terms of (deformed) $\mathfrak{su}(1,1) \oplus \mathfrak{su}(2)$ vielbein $e^a$ with $a=0,\dots,5$, the fluxes are
\begin{equation}
\label{eq:ellipticfluxes}
F_3 = \sum_j x_4^{(j)} f_3^{(j)}\,, \qquad F_5 = \sum_{j=1}^4 \sum_{k=1}^3 x^{(j)}_k f_3^{(j)} \wedge J_{2,k}\,,
\end{equation}
with auxiliary forms
\begin{equation}
\label{eq:auxfluxes}
\begin{aligned}
f_3^{(1)} &= 2(e^0 \wedge e^2 \wedge e^4 - e^1 \wedge e^3 \wedge e^5)\,, \qquad J_{2,1} = e^6 \wedge e^7 - e^8 \wedge e^9\,, \\
f_3^{(2)} &= 2(e^0 \wedge e^4 \wedge e^5 + e^1 \wedge e^2 \wedge e^3)\,, \qquad J_{2,2} = e^6 \wedge e^8 + e^7 \wedge e^9\,, \\
f_3^{(3)} &= 2(e^0 \wedge e^1 \wedge e^5 - e^2 \wedge e^3 \wedge e^4)\,, \qquad J_{2,3} = e^6 \wedge e^9 - e^7 \wedge e^8\,, \\
f_3^{(4)} &= 2(e^0 \wedge e^1 \wedge e^2 + e^3 \wedge e^4 \wedge e^5)\,.
\end{aligned}
\end{equation}
The three-form $f_3^{(4)}$ is twice the sum of the two (deformed) three-dimensional volume forms, while the other three forms $f_3^{(j)}$ with $j=1,2,3$ couple the (deformed) $AdS_3$ and (deformed) $S^3$ sectors. The two-forms $J_{2,j}$ are the usual orthogonal two-forms on the torus $T^4$. 
There are three deformation parameters in the metric, and sixteen deformation parameters in the RR fluxes. The type IIB supergravity equations result in 6 equations involving the norm and scalar products of the vectors $\mathbf{x}^{(j)}$, reducing the number of free RR parameters to 10. Imposing three additional constraints, one can obtain a theory that preserves 8 supersymmetries, coming from one $\mathfrak{psu}(1,1|2)$ copy. The tree-level worldsheet S-matrix satisfies the classical Yang-Baxter equation on the supersymmetric locus, which is a strong indication that the elliptic deformation of the supersymmetric string theory is integrable~\cite{Hoare:2025rtl}. Its relation with the integrable elliptic S-matrix bootstrapped using the YB equation in~\cite{deLeeuw:2020ahe} remains an open question.

\medskip
In this paper we are interested in the integrability of the mixed-flux elliptic $AdS_3 \times S^3 \times T^4$ superstring, which, in addition to the fluxes~\eqref{eq:ellipticfluxes}, also contains an NSNS three-form flux that is a linear combination of the four auxiliary fluxes $f_3^{(j)}$ defined in~\eqref{eq:auxfluxes}, namely
\begin{equation}
H_3 = \sum_j x_5^{(j)} f_3^{(j)}\,, \qquad j=1,2,3,4\,.
\end{equation}
We analyse the integrability of the theory in two different ways: by searching for a flat Lax connection, and through the construction of the tree-level worldsheet S-matrix.

\medskip
The structure of this paper is as follows. We begin in Section~\ref{sec:coupledEB} by defining the action of the elliptic PCM on $SU(1,1) \times SU(2)$ with NSNS flux. In Section~\ref{sec:Lax} we identify the classically integrable cases by writing the equations of motion as the flatness of a Lax connection. Section~\ref{sec:sugra} then discusses the embedding into type IIB supergravity of the elliptic deformation, as well as its supersymmetries. Then, in Section~\ref{sec:Quantisation} we quantise the theory in uniform light-cone gauge and analyse the dispersion relation of excitations on the string worldsheet. We compute the tree-level worldsheet S-matrix in Section~\ref{sec:Smatrix}. We find that it satisfies the Yang-Baxter equation in the integrable cases as identified in Section~\ref{sec:Lax}, and gives rise to inelastic processes for the mixed-flux elliptic $AdS_3 \times S^3 \times T^4$ string. We conclude in Section~\ref{sec:conclusions}.

\section{Elliptic deformation of the bosonic theory}
\label{sec:coupledEB}

In this section we review the elliptic deformation of the PCM on $SU(2)$ and on $SU(1,1)$~\cite{Cherednik:1981df,Lacroix:2023qlz,Hoare:2023zti}. We then couple these two integrable models through an antisymmetric B-field and also add a WZ term. This action corresponds to the bosonic truncation of the elliptic deformation of the $AdS_3 \times S^3 \times T^4$ superstring with mixed flux. 

\subsection{Bosonic $AdS_3 \times S^3$ strings}

Among all coset geometries $AdS_n = \frac{SO(2,n-1)}{SO(1,n-1)}$ and $S^n = \frac{SO(n+1)}{SO(n)}$, the geometries $AdS_3$ and $S^3$ are special because they correspond to group manifolds with $AdS_3 = SU(1,1)$ and $S^3=SU(2)$. Bosonic strings on these backgrounds can then be described by the PCM on $F= SU(1,1) \times SU(2)$.

Let us denote by $T_0,T_1,T_2$ the three generators of $SU(1,1)$ and by $T_3, T_4, T_5$ the three generators of $SU(2)$. They obey the commutation relations and normalisation 
\begin{equation}
\label{eq:Trelations}
\begin{aligned}
\, [T_a,T_b] &=-2 \epsilon_{ab}{}^c T_c\,, &\qquad \text{Tr}[T_a T_b] &= +2 \eta_{ab}\,, &\qquad a,b,c &\in \mathcal A\,, &\qquad \mathcal A &= \{0,1,2\}\,, \\
[T_a, T_b] &= -2 \epsilon_{ab}{}^c T_c\,, &\qquad \text{Tr}[T_a T_b] &= -2 \eta_{ab}\,, &\qquad a,b,c &\in \mathcal S\,, &\qquad \mathcal S &= \{3,4,5\}\,,
\end{aligned}
\end{equation}
where indices are raised and lowered with the metric $\eta = \text{diag}(-1,+1,+1,+1,+1,+1)$, $\epsilon_{abc}$ denotes the antisymmetric Levi-Civita symbol with convention $\epsilon_{012} = \epsilon_{345}=+1$ and $\text{Tr}$ denotes the ad-invariant bilinear form on $\mathfrak{su}(1,1)$ or $\mathfrak{su}(2)$.
We take $\check{g} \in SU(1,1)$ and $\hat{g} \in SU(2)$, and define $g =(\check{g},\hat{g}) \in F$. The left-invariant Maurer-Cartan forms are $\check{J} = \check{g}^{-1} d \check{g} \in \mathfrak{su}(1,1)$ and $\hat{J} = \hat{g}^{-1} d \hat{g} \in \mathfrak{su}(2)$, which we gather into $J = (\check{J}, \hat{J}) \in \mathfrak{f}$. For an element $X =(\check{X}, \hat{X}) \in \mathfrak{f}$ we also define $\text{STr}[X] = \text{Tr}[\check{X}] - \text{Tr}[\hat{X}]$ so that $\text{STr}[T_a T_b] = 2\eta_{ab}$. 

The starting point is the PCM+WZ on $F$, with action
\begin{equation}
\label{eq:PCMWZ}
\begin{aligned}
\!S_\text{PCMWZ} &= \text{T} \left( \frac{1}{4} \int_{\Sigma} d\tau d\sigma \text{STr}[J_+  J_-]  + q S_{\text{WZ}} \right), \quad
S_{\text{WZ}} = \frac{1}{12} \int_{\mathcal B} d^3\sigma\epsilon^{\mu \nu \rho} \text{STr}[\tilde{J}_\mu [\tilde{J}_\nu,\tilde{J}_\rho]]\,,
\end{aligned}
\end{equation}
where T is the string tension, $k = 2 \pi \text{T} q$ is the quantised WZ level, $(\tau,\sigma)$ are the 2d coordinates on $\Sigma$ and we use light-cone coordinates $\sigma^\pm = \frac{1}{2}(\tau \pm \sigma)$ and $\partial_\pm = \partial_\tau \pm \partial_\sigma$. The currents $\tilde{J}$ are the 3d extension of $J=g^{-1} d g$, defined on $\mathcal B$ with boundary $\partial \mathcal B =\Sigma$, and $\epsilon^{\mu \nu \rho}$ is the antisymmetric tensor with $\epsilon^{012}=+1$. 
The PCM+WZ has a global $F_L \times F_R$ symmetry, where the $L$ (respectively $R$) subscript indicates that the symmetry acts from the left, $g \rightarrow g_0 g$ (respectively from the right, $g \rightarrow g g_0$) with global $g_0 \in F$. This model is well-known to be classically integrable, for any value of the WZ level.

In order to obtain the background of the model we choose a specific parametrisation for the group-valued field $g \in F$ and compare the action~\eqref{eq:PCMWZ} with the non-linear sigma-model action
\begin{equation}
\label{eq:SM}
S = \frac{\text{T}}{2} \int d\tau d\sigma(G_{\mu \nu} \partial_+ X^\mu \partial_- X^\nu+B_{\mu \nu} \partial_+ X^\mu \partial_- X^\nu)\,,
\end{equation}
where $X^\mu$ are the target-space fields, $G_{\mu \nu}$ is the metric and $B_{\mu \nu}$ is the antisymmetric B-field, from which one obtains the NSNS flux $H=dB$.

A possible $4 \times 4$ representation for the generators of $\mathfrak{su}(1,1)$ and $\mathfrak{su}(2)$, which obeys the relations~\eqref{eq:Trelations}, is
\begin{align}
T_0 &= i\sigma_3 \oplus 0\,, \qquad T_1 = \sigma_1 \oplus 0\,, \qquad T_2 = \sigma_2 \oplus 0 \,, \\
T_3 &=  0 \oplus i \sigma_3\,, \qquad T_4 = 0 \oplus i \sigma_1 \,, \qquad T_5 = 0 \oplus i \sigma_2\,,
\end{align}
where $\sigma_j$ are the Pauli matrices
\begin{equation}
\sigma_1 = \begin{pmatrix} 0 & 1 \\ 1 & 0 \end{pmatrix}\,, \qquad 
\sigma_2 = \begin{pmatrix} 0 & -i \\ i & 0 \end{pmatrix}\,, \qquad 
\sigma_3 = \begin{pmatrix} 1 & 0 \\ 0 & -1 \end{pmatrix}\,.
\end{equation}
The generators obey the reality condition
\begin{equation}
T_j^\dagger H + H T_j=0\,, \qquad H = \text{diag}(\sigma_3,1_2)\,.
\end{equation}
Defining the following basis of left-invariant one-forms on $SU(1,1) \times SU(2)$,
\begin{equation}
\label{eq:vielbein}
\mathsf{e}^a =  \frac{1}{2}\eta^{ab} \text{STr}[T_b g^{-1} d g]\,, \qquad d \mathsf{e}^a = \epsilon_{bc}{}^{a} \mathsf{e}^b \wedge \mathsf{e}^c\,,
\end{equation}
the metric and NSNS three-form are
\begin{equation}
ds^2 = \sum_{a=0}^5\sum_{b=0}^5  \eta_{ab} \mathsf{e}^a \mathsf{e}^b\,, \qquad H_3 = q \, \mathsf{h}_V\,, 
\end{equation}
where as in the introduction
\begin{equation}
\mathsf{h}_V = 2(\mathsf{e}^0 \wedge \mathsf{e}^1 \wedge \mathsf{e}^2 +  \mathsf{e}^3 \wedge \mathsf{e}^4 \wedge \mathsf{e}^5) = 2 (\text{Vol}(AdS_3) + \text{Vol}(S^3))\,.
\end{equation}
Note that the contribution from the WZ term cannot globally be written as $\mathsf{h}_V = d \mathsf{b}_V$, but this can be achieved locally when choosing a specific parametrisation for the group-valued field $g \in F$.
In particular, with the parametrisation
\begin{equation}
g = e^{T T_0} e^{Z_1 T_1} e^{Z_2 T_2} e^{\Phi T_3} e^{Y_1 T_4} e^{Y_2 T_5}\,,
\end{equation}
where $\{T,Z_1,Z_2\}$ are the $AdS_3$ fields and $\{\Phi,Y_1,Y_2\}$ are the $S^3$ fields, the vielbein~\eqref{eq:vielbein} are
\begin{equation}
\begin{aligned}
\mathsf{e}^0 &= \cosh(2 Z_1) \cosh(2 Z_2) dT + \sinh(2 Z_2) dZ_1\,, &\qquad \mathsf{e}^3 &= \cos 2 Y_1 \cos 2 Y_2 d\Phi - \sin 2 Y_2 dY_1\,, \\
\mathsf{e}^1 &= \cosh(2Z_1) \sinh(2 Z_2) dT + \cosh(2 Z_2) dZ_1\,, &\qquad \mathsf{e}^4 &= \cos 2 Y_1 \sin 2 Y_2 d\Phi + \cos 2 Y_2 dY_1\,,\\
\mathsf{e}^2 &= -\sinh(2 Z_1) dT  + dZ_2 \,, &\qquad \mathsf{e}^5 &= -\sin 2 Y_1 d\Phi + d Y_2\,,
\end{aligned}
\end{equation}
and the B-field is
\begin{equation}
\mathsf{h}_V = d \mathsf{b}_V\,, \qquad \mathsf{b}_V =-\sinh2Z_1 \, dT\wedge dZ_2-\sin2Y_1\,d\Phi\wedge dY_2\,.
\end{equation}
The metric then becomes
\begin{equation}
ds^2 = -dT^2 + dZ_1^2 +dZ_2^2 -2 \sinh(2 Z_1)  dZ_2 dT+ d\Phi^2 + dY_1^2 + dY_2^2 -2 \sin(2 Y_1) dY_2 d\Phi\,,
\end{equation}
which we recognise as the metric of $AdS_3 \times S^3$ written in Hopf coordinates. The metric and the B-field are invariant under shifts of $T$ and $\Phi$. They are also invariant under shifts of $Z_2$ and $Y_2$, even though this shift symmetry is not respected by the individual vielbein.

\subsection{Action of the elliptic deformation with B-field}
Let us now introduce the elliptic deformation. 
We deform the action to
\begin{equation}
\label{eq:coupledEWZ}
S = \text{T} \left(\frac{1}{4} \int_{\Sigma} d\tau d\sigma \text{STr}[J_+ \mathcal O J_-]  + q S_\text{WZ}[g] \right)\,,
\end{equation}
where the deforming operator $\mathcal O : \mathfrak{f} \rightarrow \mathfrak{f}$ is a linear operator which acts on the generators as
\begin{align}
\mathcal O(T_0) &= \alpha_0 T_0 + \beta_0 T_3 \,, \quad \mathcal O(T_1) = \alpha_1 T_1 - \beta_1 T_4\,, \quad \mathcal O(T_2) = \alpha_2 T_2 - \beta_2 T_5\,, \\
\mathcal O(T_3) &= \alpha_0 T_3 + \beta_0 T_0\,, \quad \mathcal O(T_4) = \alpha_1 T_4 + \beta_1 T_1\,, \quad \mathcal O(T_5) = \alpha_2 T_5 + \beta_2 T_2\,.
\end{align} 
Being constructed out of left-invariant Maurer-Cartan forms, the deformed action is still invariant under left-acting $SU(1,1) \times SU(2)$ symmetry, but the right-acting symmetry is broken by the insertion of the deforming operator $\mathcal O$. 
Depending on the values of the real deforming parameters $\{\alpha_a,\beta_a\}$ there might still be a residual right-acting symmetry. The latter is generated by the $X \in \mathfrak{f}$ that satisfy $[\text{ad}_X, \mathcal O]=0$.

When $\beta_{a}=q=0$ one has two decoupled elliptic models, and the action can be written 
\begin{equation}
\begin{aligned}
&S=S[\check{g}]+S[\hat{g}]\,, \qquad S[\check{g}] = \frac{\text{T}}{4} \int d\tau d\sigma\text{Tr}[\check{J}_+ \check{\mathcal{O}} \check{J}_-]\,, \qquad S[\hat{g}] = -\frac{\text{T}}{4} \int d\tau d\sigma\text{Tr}[\hat{J}_+ \hat{\mathcal O} \hat{J}_-]\,, \\
&\check{\mathcal O}(T_0) = \alpha_0 T_0\,, \qquad \check{\mathcal O}(T_1) = \alpha_1 T_1\,, \qquad \check{\mathcal O}(T_2) = \alpha_2 T_2\,, \\
&\hat{\mathcal O}(T_3) = \alpha_0 T_3\,, \qquad \hat{\mathcal O}(T_4) = \alpha_1 T_4\,, \qquad \hat{\mathcal O}(T_5) = \alpha_2 T_5\,.
\end{aligned}
\end{equation}
The action $S[\hat{g}]$ is the classically-integrable elliptic deformation of the PCM on $SU(2)$~\cite{Cherednik:1981df}. When two deformation parameters are equal to each other this reduces to the trigonometric deformation, which has a $SU(2) \times U(1)$ symmetry, while when all deformation parameters are equal to each other this is the rational PCM on $SU(2)$.
Similarly, the action $S[\check{g}]$ is the elliptic deformation of the PCM on $SU(1,1)$. Because this is a non-compact group, there can now be time-like trigonometric deformation (when $\alpha_1=\alpha_2$) and space-like trigonometric deformations (when $\alpha_0=\alpha_1$ or $\alpha_0=\alpha_2$). 

The metric and NSNS three-form of the deformed model~\eqref{eq:coupledEWZ} are
\begin{equation}
\label{eq:GBelliptic}
ds^2 = \sum_{a=0}^5\sum_{b=0}^5 \alpha_{a\bmod 3} \eta_{ab} \mathsf{e}^a \mathsf{e}^b\,, \qquad H = d(-\beta_0 \mathsf{b}_0 +\beta_1 \mathsf{b}_1 +\beta_2 \mathsf{b}_2) + q \mathsf{h}_V\,,
\end{equation}
where as before
\begin{equation}
\label{eq:hWZ}
 \mathsf{h}_V = 2(\mathsf{e}^0 \wedge \mathsf{e}^1 \wedge \mathsf{e}^2 +  \mathsf{e}^3 \wedge \mathsf{e}^4 \wedge \mathsf{e}^5)\,,
\end{equation}
and we introduce the following forms mixing the $AdS_3$ and the $S^3$ sectors,
\begin{equation}
\label{eq:Bfield}
\begin{aligned}
\mathsf{b}_0 &= \mathsf{e}^0 \wedge \mathsf{e}^3\,, \qquad \mathsf{h}_0 = d \mathsf{b}_0 = 2(- \mathsf{e}^0 \wedge \mathsf{e}^4 \wedge \mathsf{e}^5-\mathsf{e}^1 \wedge \mathsf{e}^2 \wedge \mathsf{e}^3 )\,, \\
\mathsf{b}_1 &= \mathsf{e}^1 \wedge \mathsf{e}^4\,, \qquad \mathsf{h}_1 = d \mathsf{b}_1 = 2(-\mathsf{e}^0 \wedge \mathsf{e}^2 \wedge \mathsf{e}^4 + \mathsf{e}^1 \wedge \mathsf{e}^3 \wedge \mathsf{e}^5)\,, \\
\mathsf{b}_2 &= \mathsf{e}^2 \wedge \mathsf{e}^5\,, \qquad \mathsf{h}_2 = d \mathsf{b}_2 = 2(+\mathsf{e}^0 \wedge \mathsf{e}^1 \wedge \mathsf{e}^5 - \mathsf{e}^2 \wedge \mathsf{e}^3 \wedge \mathsf{e}^4)\,.
\end{aligned}
\end{equation}

In terms of the deformed vielbein
\begin{equation}
\label{eq:defvielbein}
e^0 = \sqrt{\alpha_0} \mathsf{e}^0, \quad e^1 = \sqrt{\alpha_1} \mathsf{e}^1, \quad e^2 = \sqrt{\alpha_2} \mathsf{e}^2, \quad e^3 = \sqrt{\alpha_0} \mathsf{e}^3, \quad e^4 = \sqrt{\alpha_1} \mathsf{e}^4, \quad e^5 = \sqrt{\alpha_2} \mathsf{e}^5\,,
\end{equation}
the metric of the elliptic deformation of $AdS_3 \times S^3$ takes the canonical form
\begin{equation}
ds^2 = -(e^0)^2 + (e^1)^2 + (e^2)^2 + (e^3)^2 + (e^4)^2 + (e^5)^2\,,
\end{equation}
and the auxiliary B-fields~\eqref{eq:Bfield} are
\begin{equation}
\mathsf{b}_0 = \frac{1}{\alpha_0} e^0 \wedge e^3 \,, \qquad \mathsf{b}_1 = \frac{1}{\alpha_1} e^1 \wedge e^4\,, \qquad \mathsf{b}_2= \frac{1}{\alpha_2 } e^2 \wedge e^5\,,
\end{equation}
and
\begin{equation}
\begin{aligned}
\mathsf{h}_V &= \frac{2}{\sqrt{\alpha_0 \alpha_1 \alpha_2}}(e^0 \wedge e^1 \wedge e^2 + e^3 \wedge e^4 \wedge e^5) \,. 
\end{aligned}
\end{equation}
The special form of the auxiliary B-fields~\eqref{eq:Bfield}, given by the wedge product of vielbein, provides an a-posteriori motivation for the definition of the deformation as in the action~\eqref{eq:coupledEWZ}. Such a deformed action indeed produces NSNS three-form fluxes that are precisely of the type~\eqref{eq:auxfluxes} relevant for the construction of elliptic superstrings.

\section{Classical integrability}
\label{sec:Lax}

We would like to determine for which deformation parameters $\{\alpha_a,\beta_a,q\}$ the equations of motion resulting from the action~\eqref{eq:coupledEWZ} can be recast as the flatness condition of a Lax connection $\mathcal L(\tau,\sigma;z) \in \mathfrak{f}$, which, in addition to the worldsheet coordinates $(\tau,\sigma)$ also depends on a spectral parameter $z \in \mathbb{C}$. 

\subsection{Identification of the Lax integrable cases}

As already mentioned, the action~\eqref{eq:coupledEWZ} is invariant under a global left-acting $SU(1,1) \times SU(2)$ symmetry. The equations of motion are equivalent to the conservation of the associated currents, and read
\begin{equation}
\partial_+ K_- + \partial_- K_+ =0\,, \qquad K_+ = \text{Ad}_g (\mathcal O^T-q) J_+\,, \qquad  K_- = \text{Ad}_g (\mathcal O+q) J_-\,.
\end{equation}
The transpose of an operator is defined such that $\text{STr}[\mathcal O^T(X) Y] = \text{STr}[X \mathcal O(Y)]$. 
Exploiting that $d (\text{Ad}_g X) = d(g X g^{-1}) = \text{Ad}_g ([J,X]+ dX)$, this can be recast into
\begin{equation}
\label{eq:eom}
(\mathcal O+q) \partial_+ J_- + (\mathcal O^T-q) \partial_- J_+ + [J_+, (\mathcal O+q) J_-] - [(\mathcal O^T-q) J_+,J_-]  =0\,.
\end{equation}
One also has the flatness of the Maurer-Cartan form
\begin{equation}
\label{eq:flatness}
\partial_+ J_- - \partial_- J_+ + [J_+,J_-]=0\,.
\end{equation}
Together, the equations of motion~\eqref{eq:eom} and the flatness condition~\eqref{eq:flatness} imply that
\begin{equation}
\label{eq:derJ}
\begin{aligned}
\partial_+ J_- &= - (\mathcal O + \mathcal O^T)^{-1} \big( [J_+, \mathcal O J_-] - [\mathcal O^T J_+, J_-] + (\mathcal O^T +q) [J_+, J_-] \big) \,, \\
\partial_- J_+ &= - (\mathcal O + \mathcal O^T)^{-1} \big( [J_+, \mathcal O J_-] - [\mathcal O^T J_+, J_-] - (\mathcal O -q) [J_+, J_-] \big) \,.
\end{aligned}
\end{equation}
Note that $\mathcal O + \mathcal O^T$ is symmetric (in fact diagonal) and we take it to be full-rank so that it can indeed be inverted (we take $\alpha_a>0$). For the Lax we make the Ansatz
\begin{equation}
\label{eq:Lax}
\mathcal L_\pm(\tau, \sigma; z) = \mathcal F_{\pm}(z)(J_{\pm})\,, 
\end{equation}
where $\mathcal F_\pm(z) : \mathfrak{f} \rightarrow \mathfrak{f}$ are spectrally-dependent operators which act on the generators as
\begin{equation}
\begin{aligned}
\mathcal F_\pm(z) (T_a) &= \sum_{b=0}^5 f_{ab,\pm}(z) T_b\,.
\end{aligned}
\end{equation}
We recall that the Lax connection also depends on the coordinates $(\tau,\sigma)$ through the currents $J_\pm$. The flatness condition of the Lax is
\begin{equation}
\label{eq:flatLax}
\partial_+ \mathcal L_-(\tau, \sigma;z) - \partial_- \mathcal L_+(\tau, \sigma;z) + [\mathcal L_+(\tau, \sigma;z), \mathcal L_-(\tau, \sigma;z)]=0\,,
\end{equation}
which becomes
\begin{equation}
\mathcal F_-(z) \partial_+ J_- - \mathcal F_+(z) \partial_- J_+ + [\mathcal F_+(z) J_+, \mathcal F_-(z) J_-]=0\,.
\end{equation}
Plugging in the expressions for the derivatives of the currents~\eqref{eq:derJ} results in quadratic equations for the unknown functions $f_{ab,\pm}(z)$. Solving quadratic equations, as well as finding the compatibility conditions that ensure that there indeed exists a solution, are hard problems.
To proceed, we will assume the following expansion of operators and functions around $z=0$,
\begin{equation}
\label{eq:expansion}
\mathcal F_\pm(z) = 1 + z \mathcal F^{(1)}_\pm + z^2 \mathcal F^{(2)}_\pm + O(z^3)\,, \qquad f_{ab,\pm} (z)  = \delta_{ab} + z f_{ab,\pm}^{(1)} + z^2 f_{ab,\pm}^{(2)} + O(z^3)\,,
\end{equation}
and solve the equations perturbatively in the spectral parameter.
At zeroth-order, the flatness condition for the Lax is then equivalent to the flatness of the current $J_\pm$, which is indeed satisfied. To first order in the spectral parameter, one gets the equation
\begin{equation}
\label{eq:order1}
 \mathcal F^{(1)}_- \partial_+ J_- - \mathcal  F^{(1)}_+ \partial_- J_+ + [J_+, \mathcal F_-^{(1)} J_-] + [\mathcal F_+^{(1)} J_+, J_-]=0\,.
\end{equation}
This is equivalent to the equations of motion~\eqref{eq:eom} if one makes the choice
\begin{equation}
\label{eq:choice}
\mathcal F^{(1)}_+ = \mathcal O^T - q\,, \qquad \mathcal F^{(1)}_- = -(\mathcal O+q)\,.
\end{equation}
In fact, plugging in~\eqref{eq:derJ}, one can just solve the linear equations~\eqref{eq:order1} for the functions $f_{ab,\pm}^{(1)}$ and the most general solution is 
\begin{equation}
\begin{aligned}
\mathcal F_+^{(1)} &= c_{\mathcal A} P_{\mathcal A} (\mathcal O^T - q) + c_{\mathcal S} P_{\mathcal S} (\mathcal O^T - q) + \text{ad}_Z\,, \\
\mathcal F_-^{(1)} &= -c_{\mathcal A} P_{\mathcal A} (\mathcal O + q) - c_{\mathcal S} P_{\mathcal S} (\mathcal O + q) + \text{ad}_Z\,, 
\end{aligned}
\end{equation}
for an arbitrary $Z \in \mathfrak{f}$ and where $P_{\mathcal A}$ and $P_{\mathcal S}$ denote the projectors onto $\mathfrak{su}(1,1)$ and $\mathfrak{su}(2)$ respectively. The common $\text{ad}_Z$ can be removed by a worldsheet-independent but spectrally-dependent conjugation. One constant, for instance $c_{\mathcal A}$, can also be removed thanks to a rescaling of the spectral parameter. 
The solution~\eqref{eq:choice} corresponds to the special case $Z=0$ and $c_{\mathcal A} = c_{\mathcal S}=1$. In what follows we will assume that the first-order is non-trivial, and set $Z=0$, $c_{\mathcal A} \neq 0$ or $c_{\mathcal S} \neq 0$.
To second order, the flatness condition together with~\eqref{eq:derJ} leads to the following linear system of equations,
\begin{equation}
\begin{aligned}
 0 = \mathcal Z(J_+, J_-) & \equiv\mathcal F_+^{(2)} (\mathcal O + \mathcal O^T)^{-1} \big( [J_+, \mathcal O J_-] - [\mathcal O^T J_+, J_-] - (\mathcal O -q) [J_+, J_-] \big) \\
 &- \mathcal F_-^{(2)} (\mathcal O + \mathcal O^T)^{-1} \big( [J_+, \mathcal O J_-] - [\mathcal O^T J_+, J_-] + (\mathcal O^T +q) [J_+, J_-] \big)  \\
 & + [\mathcal F_+^{(2)} J_+,  J_-]+ [J_+, \mathcal F_-^{(2)} J_-] + [\mathcal F^{(1)}_+ J_+, \mathcal F^{(1)}_- J_-]\,.
 \end{aligned} 
\end{equation}
When projected onto the six generators of $\mathfrak{f}$, and pulling out the coefficients of the $J_+^a J_-^b$ terms, this leads to at most 216 linear equations for the 72 unknown $f_{ab,\pm}^{(2)}$. This is an over-constrained system of equations, and for it to admit a non-trivial solution, compatibility imposes additional constraints on the deformation parameters. 
To illustrate this, let us consider the term proportional to $J_+^a J_-^{b+3}$ and project onto $T_c$ and $T_{c+3}$, with $\{a,b,c\} =\{0,1,2\}$ all different.
Then one can check that 
\begin{equation}
\begin{aligned}
\mathcal Z(T_a,T_{b+3}) + \mathcal Z(T_{b+3},T_a) &= -[T_a, T_b] ((\mathcal F_+^{(2)}-\mathcal F^{(2)}_-)^{b}{}_{b+3} + 2 c_{\mathcal A}^2 \alpha_a \beta_b) \\
&\qquad+ [T_{a+3}, T_{b+3}] ((\mathcal F_+^{(2)}-\mathcal F^{(2)}_-)^{a+3}{}_{a} - 2 c_{\mathcal S}^2 \eta_{aa} \alpha_b \beta_a)\,,
\end{aligned}
\end{equation}
which, assuming $c_{\mathcal A} \neq 0$ or $c_{\mathcal S} \neq 0$, can only admit a solution if
\begin{equation}
\label{eq:condition1}
\beta_0(\alpha_1-\alpha_2) =0\,, \qquad \beta_1 (\alpha_0-\alpha_2) =0\,, \qquad \beta_2(\alpha_0-\alpha_1)=0\,.
\end{equation}
Similarly, the term proportional to $J_+^a J_-^b$ where both $a,b \in \mathcal S$ gives the equations
\begin{equation}
A \mathbf{f} = \mathbf{b}\,, \quad A = \begin{pmatrix} 1 & -1 & \frac{\alpha_1-\alpha_0}{\alpha_2} \\
\frac{\alpha_2-\alpha_1}{\alpha_0} & 1 & -1 \\
-1 & \frac{\alpha_0 - \alpha_2}{\alpha_1} & 1 
\end{pmatrix}\,, \quad \mathbf{b} = -2 c_{\mathcal S}^2 q \begin{pmatrix} \alpha_1-\alpha_0 \\ \alpha_2-\alpha_1 \\ \alpha_0-\alpha_2\end{pmatrix}\,, \quad
\mathbf{f} = \begin{pmatrix}
(\mathcal  F_+^{(2)}-\mathcal  F_-^{(2)})^3{}_3 \\
(\mathcal  F_+^{(2)}-\mathcal  F_-^{(2)})^4{}_4 \\
(\mathcal  F_+^{(2)}-\mathcal  F_-^{(2)})^5{}_5
\end{pmatrix}\,.
\end{equation}
Note that we have that
\begin{equation}
\mathbf{l}^T A=0\,, \qquad \mathbf{l} = \begin{pmatrix} \alpha_2 (\alpha_0+\alpha_1-\alpha_2) \\ \alpha_0(-\alpha_0+\alpha_1+\alpha_2) \\ \alpha_1 (\alpha_0-\alpha_1+\alpha_2) \end{pmatrix}\,.
\end{equation}
One must therefore have that $\mathbf{l}^T \mathbf{b} =0$, which leads to the condition (assuming $c_{\mathcal S} \neq 0$ or $c_{\mathcal A} \neq 0$, there is an analogous equation in the $SU(1,1)$ sector)
\begin{equation}
\label{eq:condition2}
\begin{gathered}
q (\alpha_0-\alpha_1)(\alpha_0-\alpha_2)(\alpha_1-\alpha_2)=0\,.
\end{gathered}
\end{equation}
These are necessary conditions that, within our Ansatz for the Lax, need to be satisfied to have classical integrability. The argument holds locally near the Maurer-Cartan point, we assume that $\mathcal F_\pm(z=0)= 1$ and that the function is analytic in a neighbourhood around $z=0$. We do not explore the possibility of having more exotic Lax connections. The consistency conditions of~\eqref{eq:condition1} and~\eqref{eq:condition2} are satisfied in the following cases (in each case one can choose the normalisation of one of the $\alpha_a$):
\begin{itemize}
\item The elliptic deformation without any B-field ($\beta_0 =\beta_1 =\beta_2=q=0$).
\item The time-like trigonometric+B-field deformation with
\begin{equation}
\label{eq:paramtlt}
\alpha_0=1\,, \qquad \alpha_1=\alpha_2=\alpha\,, \qquad \beta_0=\beta\,, \qquad \beta_1=\beta_2=0\,, \qquad q\,.
\end{equation}
\item The space-like trigonometric+B-field deformation with (and the equivalent theory obtained through exchanging indices $1 \leftrightarrow 2$)
\begin{equation}
\label{eq:paramslt}
\alpha_0=\alpha_1=\alpha\,, \qquad \alpha_2=1\,, \qquad \beta_0=\beta_1=0\,, \qquad \beta_2=\beta\,, \qquad q\,.
\end{equation}
\item The rational limit $\alpha_0=\alpha_1=\alpha_2$, for which the B-field parameters can be arbitrary.
\end{itemize}
In particular, in the elliptic deformation with all $\alpha_a$ different from each other, the compatibility condition forces $q=\beta_a=0$ and no non-trivial B-field is compatible with the Lax ansatz under consideration. Such an obstruction was already highlighted in~\cite{Klimcik:2024vss} for the WZ term.

Let us now constructively show that for the elliptic and trigonometric cases obeying the consistency conditions one can indeed find a spectrally-dependent solution to the flatness equation. It turns out that in all cases of interest the functions can be decomposed as
\begin{equation}
\begin{aligned}
\mathcal F_\pm(z) (T_a) &= f_{a,\pm}(z) T_a + g_{a+3,\pm}(z) T_{a+3}\,, \qquad a \in \mathcal A\,, \\ 
\mathcal F_\pm(z) (T_a) &= f_{a,\pm}(z) T_a + g_{a-3,\pm}(z) T_{a-3}\,, \qquad a \in \mathcal S\,.
\end{aligned}
\end{equation}
The equations to solve are
\begin{align}
0&=F_c -\frac{\alpha_b-\alpha_a+q}{\alpha_c} \Delta_c(f) - \frac{\beta_c}{\alpha_c} \eta_{cc} \Delta_c(g) -2 f_{a,+} f_{b,-}\,,   \\
0&= F_{c+3} -\frac{\alpha_{b}-\alpha_{a}+q}{\alpha_{c}} \Delta_{c+3}(f) + \frac{\beta_{c}}{\alpha_{c}} \Delta_{c+3}(g)  -2 f_{a+3,+} f_{b+3,-}\,,  \\
0&=G_c-\frac{\alpha_b-\alpha_a+q}{\alpha_c} \Delta_c(g)+ \frac{\beta_{c}}{\alpha_c} \Delta_c(f)- 2  \eta_{c  c} g_{a,+} g_{b,-}\,, \\
0&=G_{c+3} -\frac{\alpha_{b}-\alpha_{a}+q}{\alpha_{c}} \Delta_{c+3}(g)- \frac{\beta_{c}}{\alpha_{c}} \eta_{c c} \Delta_{c+3}(f)- 2 \eta_{c c} g_{a+3,+} g_{b+3,-} \,, \\
0&=-\frac{\beta_b}{\alpha_c} \eta_{aa} \Delta_c(g) - \frac{\beta_{a}}{\alpha_c} \Delta_c(f) - 2 g_{a,+} f_{b,-}\,, \\
0&=- \frac{\beta_{b}}{\alpha_{c}} \eta_{c  c} \Delta_{c+3}(g)+ \frac{\beta_{a}}{\alpha_{c}} \eta_{a \, a} \Delta_{c+3}(f) - 2 g_{a+3,+} f_{b+3,-}\,,\\
0&=- \frac{\beta_{b}}{\alpha_c} \Delta_c(f) - \frac{\beta_a}{\alpha_c} \eta_{b b} \Delta_c(g)- 2 f_{a,+} g_{b,-}\,,\\
0&=\frac{\beta_{b}}{\alpha_{c}} \eta_{b  b} \Delta_{c+3}(f) - \frac{\beta_{a}}{\alpha_{c}} \eta_{c c} \Delta_{c+3}(g) -2 f_{a+3,+} g_{b+3,-}\,,
\end{align}
where the indices $\{a,b,c\} = \{0,1,2\}$ so that each line corresponds to six equations, and we found it convenient to define
\begin{equation}
F_{c} \equiv f_{c,+} + f_{c,-}\,, \quad G_c \equiv g_{c,+} + g_{c,-} \,, \quad \Delta_c(f) \equiv f_{c,+}- f_{c,-}\,, \quad \Delta_c(g) \equiv g_{c,+}- g_{c,-}\,.
\end{equation}  
For the elliptic deformation without any B-field, the equations are solved by
$g_{a,\pm}=0$ and the combination of Jacobi elliptic functions
\begin{equation}
\label{eq:fgelliptic}
\begin{aligned}
    f_{0,\pm} &= \frac{\text{sc}(\nu,k^2)}{\text{sc}(\nu \mp z,k^2)}\,, &\qquad 
    f_{1, \pm} &= \frac{\text{sd}(\nu,k^2)}{\text{sd}(\nu \mp z,k^2)}\,, &\qquad
        f_{2,\pm}  &= \frac{\text{sn}(\nu,k^2)}{\text{sn}(\nu \mp z,k^2)}\,, \\
        f_{3, \pm} &= f_{0,\pm}\,, &\qquad f_{4,\pm} &= f_{1,\pm}\,, &\qquad f_{5,\pm} &= f_{2,\pm}\,.
    \end{aligned}
\end{equation}
The relation between the parameters $\alpha_a$, $\nu$ and the elliptic modulus $k$ (not to be confused with the WZ level) is
\begin{equation}
\label{eq:knuelliptic}
    k^2 = \frac{\alpha_0(\alpha_1-\alpha_2)}{\alpha_1(\alpha_0-\alpha_2)}\,, \qquad \text{cn}^2(\nu,k^2) = \frac{\alpha_2}{\alpha_0}\,, \qquad \text{dn}^2(\nu,k^2) = \frac{\alpha_2}{\alpha_1}\,, \qquad \text{cd}^2(\nu,k^2) = \frac{\alpha_1}{\alpha_0}\,.
\end{equation}
The expansion around $z=0$ is compatible with the expansion~\eqref{eq:expansion} and~\eqref{eq:choice} upon rescaling the spectral parameter. 

For the time-like trigonometric case with deformation parameters~\eqref{eq:paramtlt}, we find it convenient to define
\begin{equation}
\label{eq:abtlt}
r_{t} = \sqrt{\alpha (1-\alpha-\beta^2) - (1-\alpha) q^2}\,,
\end{equation}
and a solution to the flatness condition~\eqref{eq:flatLax} is
\begin{equation}
\label{eq:fgtlt}
\begin{aligned}
f_{0,\pm} &= f_{3,\pm} = \frac{r_t \cos z \pm (1-\alpha) \sin z}{r_t \cos z \mp (\alpha \mp q) \sin  z}\,, \qquad g_{0,\pm} = g_{3,\pm} = -\beta \frac{\sin z}{r_t \cos z \mp (\alpha \mp q) \sin z}\,, \\
f_{1,\pm} &= f_{2,\pm} = f_{4,\pm} = f_{5,\pm} =  \frac{r_t}{r_t \cos z \mp (\alpha \mp q) \sin z}\,, \qquad
g_{1,\pm}=g_{2,\pm}=g_{4,\pm}=g_{5,\pm}=0\,.
\end{aligned}
\end{equation}
In the zero B-field limit with $\beta=q=0$ this reproduces the results of~\eqref{eq:fgelliptic} with $k^2=0$, namely
\begin{equation}
f_{0,\pm} = \frac{\tan \nu}{\tan (\nu \mp z)}\,, \qquad f_{1,\pm} = f_{2,\pm} = \frac{\sin \nu}{\sin(\nu  \mp z)}\,, \qquad \alpha_1=\alpha_2=\alpha = \cos^2(\nu)\,, \qquad \alpha_0=1\,.
\end{equation}
Another interesting limit is when $r_t \rightarrow 0$. With our parametrisation, the spectral parameter then drops out of the functions~\eqref{eq:fgtlt}. It is however possible to perform the rescaling $z \rightarrow r_t z$, and then take $r_t \rightarrow 0$ while keeping the spectral parameter $z$ finite. This leads to rational functions
\begin{equation}
f_{0, \pm} = \frac{1\pm (1-\alpha) z}{1 \mp (\alpha\mp q) z}\,, \qquad f_{1,\pm} = \frac{1}{1\mp (\alpha \mp q) z}\,, \qquad g_{0,\pm} = -\beta \frac{z}{1\mp (\alpha \mp q) z}\,.
\end{equation}
In particular, for the isotropic limit $\alpha \rightarrow 1$ and $\beta \rightarrow 0$, which corresponds to the PCMWZ model, one gets
\begin{equation}
f_{a,\pm} = \frac{1}{1\mp(1\mp q) z}\,, \qquad g_{a,\pm}=0\,.
\end{equation}

Finally, in the space-like trigonometric case with B-field and deformation parameters given in~\eqref{eq:paramslt}, we find it convenient to define
\begin{equation}
\label{eq:abslt}
r_s = \sqrt{\alpha (\alpha-1-\beta^2)-(\alpha-1) q^2}\,,
\end{equation}
so that a solution to the flatness condition~\eqref{eq:flatLax} is
\begin{equation}
\label{eq:fgslt}
\begin{aligned}
f_{2,\pm} &= f_{5,\pm} = \frac{r_s \cosh z \pm (1-\alpha) \sinh z}{r_s \cosh z \mp (\alpha \mp q) \sinh z}\,, \qquad g_{5,\pm} = -g_{2,\pm} = \beta \frac{\sinh z}{r_s \cosh z \mp (\alpha \mp q) \sinh z}\,, \\
f_{0,\pm} &= f_{1,\pm} = f_{3,\pm} = f_{4,\pm} =  \frac{r_s}{r_s \cosh z \mp (\alpha \mp q) \sinh z}\,, \qquad
g_{0,\pm}=g_{1,\pm}=g_{3,\pm}=g_{4,\pm}=0\,.
\end{aligned}
\end{equation}
When $\beta=q=0$ this reproduces the results of~\eqref{eq:fgelliptic} with $k^2=1$, namely
\begin{equation}
f_{0,\pm} = f_{1,\pm} = \frac{\sinh \nu}{\sinh (\nu \mp z)}\,, \qquad f_{2,\pm} = \frac{\tanh \nu}{\tanh(\nu \mp z)}\,, \qquad \alpha_0=\alpha_1=\alpha = \cosh^2 \nu\,, \qquad \alpha_2=1\,.
\end{equation}

The existence of a flat Lax connection is a good indicator but not a sufficient condition for classical integrability. In principle, one would need to check that the conserved charges that follow from the Lax and the transfer matrix are in involution. This constrains the form of the Poisson bracket of $\mathcal L_\sigma(\tau,\sigma;z)$. 
As we will now see, the integrability in the trigonometric cases~\eqref{eq:paramtlt} and~\eqref{eq:paramslt} follows from the formulation of the theory as a Yang-Baxter+WZ deformation (up to total derivative terms, and up to reality for the space-like trigonometric deformation). 

\subsection{The trigonometric deformation with mixing B-field as a YB deformation}
In the integrable trigonometric+B-field cases with $q=0$, the deformation parameters $\alpha_a$ are related to quantum-group deformations, while $\beta_{a}$ are related to TsT transformations involving one $U(1) \subset SU(1,1)_R$ and one $U(1) \subset SU(2)_R$. Both TsT and quantum-group deformations can be written in the unifying framework of Yang-Baxter (YB) deformations~\cite{Klimcik:2002zj,Klimcik:2008eq}. The YB deformation of the PCM on $F$ reads
\begin{equation}
\label{eq:YB}
S_{\text{YB}} = \frac{1}{4} \int d\tau d\sigma \, \text{STr}[J_+ \frac{1}{1-R} J_-]\,, \qquad J_\pm = g^{-1} \partial_\pm g\,, \qquad g \in F\,,
\end{equation}
where the deforming operator $R: \mathfrak{f} \rightarrow \mathfrak{f}$ is antisymmetric with respect to the supertrace, $R^T=-R$, and satisfies the (modified) classical Yang-Baxter equation ((m)cYBE)
\begin{equation}
\label{eq:mcYBE}
[R(X),R(Y)] - R([R(X),Y]+[X,R(Y)])=-c^2 [X,Y]\,,  \qquad \forall X,Y \in \mathfrak{f}\,, \qquad c^2 \in \mathbb{R}\,. 
\end{equation}
The case $c=0$ corresponds to the cYBE, leading for instance to TsT transformations~\cite{Osten:2016dvf} or Jordanian deformations~\cite{Kawaguchi:2014qwa,vanTongeren:2015soa}, while $c^2 \neq 0$ is the mcYBE, leading for instance to quantum group deformations~\cite{Delduc:2013fga,Delduc:2013qra}. Note that in the inhomogeneous case it is possible to rescale $R$ so that the only relevant cases are $c=i$, which corresponds to the non-split mcYBE, or $c=1$, which corresponds to the split mcYBE. This rescaling introduces a deformation parameter into the action~\eqref{eq:YB}. The split and non-split cases need to be distinguished if $\mathfrak{f}$ has a specific real form. $R$ indeed has to preserve this real form. 

The equations of motion of~\eqref{eq:YB} and the flatness condition for $J_\pm$ can be recast as the flatness of the Lax connection
\begin{equation}
\label{eq:LaxYB}
\mathcal L^r_\pm(\tau,\sigma;z) = -\frac{1-c^2}{1\mp z} \text{Ad}_g \frac{1}{1\pm R}  J_\pm\,.
\end{equation}
The $r$ superscript indicates that this is the rational description of the model, and indeed the Lax $\mathcal L^r(\tau,\sigma;z)$ is a rational function of the spectral parameter $z$. 

In the integrable trigonometric cases~\eqref{eq:paramtlt} and~\eqref{eq:paramslt} with $q=0$, the action~\eqref{eq:coupledEWZ} can be written
\begin{equation}
S = \text{T} \left(\zeta S_{\text{YB}} + \int d\tau d\sigma \mathcal L_{td} \right) \,, 
\end{equation}
where $\zeta$ is a constant that allows for a rescaling of the YB action and the term $\mathcal L_{td}$ corresponds to a total derivative term which may be added to the action. In all the cases we consider, this total derivative term is of the form
\begin{equation}
\label{eq:Ltd}
\mathcal L_{td} = \sum_{a=0}^5 \rho_a \text{STr}[T_a[J_+,J_-]]= - \sum_{a=0}^5 \rho_a \text{STr}[T_a(\partial_+ J_- - \partial_- J_+)]\,,
\end{equation}
where we used the flatness of the current $J$ in the second equality to highlight that this is indeed a total derivative term. 
The explicit form of the deforming operator $R$, the rescaling $\zeta$ and the constants $\rho_a$ depend on the type of trigonometric deformation.
 
\paragraph{Time-like trigonometric+TsT.}
In the integrable time-like trigonometric case with deformation parameters~\eqref{eq:paramtlt} and $q=0$ we find that the trigonometric+B-field and YB actions agree up to the rescaling and the total derivative term
\begin{equation}
\rho_0 = \rho_3 = \frac{1}{8} \sqrt{\alpha} \sqrt{1-\alpha -\beta^2}\,, \qquad \rho_1=\rho_2=\rho_4=\rho_5=0\,, \qquad \zeta = 1-\beta^2\,,
\end{equation}
provided one chooses the R-matrix
\begin{equation}
\label{eq:Rtlt}
\begin{aligned}
R(T_0) &= \beta T_3\,, \qquad R(T_1) = -\frac{\sqrt{1-\alpha-\beta^2}}{\sqrt{\alpha}} T_2\,, \qquad R(T_2) = +\frac{\sqrt{1-\alpha-\beta^2}}{\sqrt{\alpha}} T_1\,, \\
R(T_3) &= \beta T_0\,, \qquad R(T_4) = -\frac{\sqrt{1-\alpha-\beta^2}}{\sqrt{\alpha}} T_5\,, \qquad R(T_5) = +\frac{\sqrt{1-\alpha-\beta^2}}{\sqrt{\alpha}} T_4\,.
\end{aligned}
\end{equation}
This R-matrix solves the mcYBE with
\begin{equation}
c^2 = -\frac{1-\alpha-\beta^2}{\alpha}\,.
\end{equation}
For the R-matrix to preserve the real form of the algebra, one needs to take (the lower bound comes from the range of $\alpha$ and the reality of $\beta$) 
\begin{equation}
\label{eq:range}
0<\alpha +\beta^2 \leq 1 \quad \Rightarrow \quad c^2 \leq 0\,.
\end{equation}
The  R-matrix is a solution of the non-split mcYBE. In the decoupling limit $\beta=0$ it reduces as expected to the Drinfel'd-Jimbo R-matrix on $SU(1,1)$ (annihilating the Cartan direction $T_0$) and the Drinfel'd-Jimbo R-matrix on $SU(2)$ (annihilating the Cartan direction $T_3$). One can then define the rescaled operator
\begin{equation}
\begin{aligned}
&\tilde{R} = \eta^{-1} R\,, \qquad \eta = \frac{\sqrt{1-\alpha}}{\sqrt{\alpha}}\,, \qquad 
\tilde{R}(T_0) = \tilde{R}(T_3)=0\,, \\
&\tilde{R}(T_1) = - T_2\,, \qquad \tilde{R}(T_2) = T_1\,, \qquad \tilde{R}(T_4)=-T_5\,, \qquad \tilde{R}(T_5) = T_4\,,
\end{aligned}
\end{equation}
which satisfies the mcYBE with $c^2=-1$ and is such that $\tilde{R}^3 = -\tilde{R}$. The action then takes the form
\begin{equation}
\label{eq:YBusual}
S_\text{YB} = \frac{1}{4} \int d\tau d\sigma \text{STr}[J_+ \frac{1}{1-\eta \tilde{R}} J_-]\,.
\end{equation}
This is the usual form of the YB deformation, with $\eta \rightarrow 0$ in the undeformed limit.  

When $\beta = \sqrt{1-\alpha}$ (assuming the range $0<\alpha\leq 1$ to have a real B-field) one has $c=0$, the R-matrix solves the cYBE and simply acts as $R(T_0) = \beta T_3$ and $R(T_3)=\beta T_0$ with $[T_0,T_3]=0$. This corresponds to a TsT transformation of the rational model in the $\{T_0,T_3\}$ directions. One can again rescale $\tilde{R} = \beta^{-1} R$ to bring the YB deformation in its usual form~\eqref{eq:YBusual} with $\eta=\beta$. The TsT transformation of the rational model gives a trigonometric metric (where both $AdS_3$ and $S^3$ are deformed in the same way, with parameter $\alpha$), and a non-trivial B-field coupling these two sectors. The B-field parameter is not free, but related to the deformation in the metric as $\beta = \sqrt{1-\alpha}$. For general $(\alpha,\beta)$ in the range~\eqref{eq:range}, the R-matrix encodes the combination of a TsT and a quantum group deformation. The metric takes the same form as in the TsT case, but the magnitude of the B-field no longer depends on $\alpha$. 

\paragraph{Space-like trigonometric+TsT.}
In the integrable space-like trigonometric case with deformation parameters~\eqref{eq:paramslt} and $q=0$, or the equivalent deformation where indices $1,2$ are exchanged, the trigonometric+B-field and YB actions are again equivalent up to the rescaling and the total derivative terms
\begin{equation}
\rho_2 = \frac{\sqrt{\alpha} \sqrt{\alpha-1-\beta^2}}{8}  \,, \quad \rho_5 = \frac{\sqrt{\alpha} \sqrt{1-\alpha+\beta^2}}{8} \,, \quad \rho_0=\rho_1=\rho_3 =\rho_4=0\,, \quad \zeta = 1+\beta^2\,.
\end{equation}
The R-matrix acts as
\begin{equation}
\begin{aligned}
R(T_2) &= -\beta T_5\,, \qquad R(T_0) = - \frac{\sqrt{\alpha-1-\beta^2}}{\sqrt{\alpha}} T_1\,, \qquad R(T_1) = - \frac{\sqrt{\alpha-1-\beta^2}}{\sqrt{\alpha}} T_0\,, \\
R(T_5) &= +\beta T_2\,, \qquad R(T_3) = -\frac{\sqrt{1-\alpha+\beta^2}}{\sqrt{\alpha}} T_4\,, \qquad R(T_4) = +\frac{\sqrt{1-\alpha+\beta^2}}{\sqrt{\alpha}} T_3\,,
\end{aligned}
\end{equation}
and solves the mcYBE with 
\begin{equation}
c^2 = -\frac{1-\alpha+\beta^2}{\alpha}\,.
\end{equation}
When $\alpha-\beta^2=1$ one finds that $c^2=0$ and the R-matrix solves the cYBE. The R-matrix preserves the real form and acts as $R(T_2) = -\beta T_5$, $R(T_5)=+\beta T_2$, with vanishing action on all other generators. This generates a TsT transformation in the $\{T_2,T_5\}$ directions. 

Away from that special line, there is no range of parameters for which the R-matrix preserves both the $\mathfrak{su}(1,1)$ and $\mathfrak{su}(2)$ real forms. On one hand, when $\alpha-\beta^2<1$ we have that $c^2<0$, the R-matrix is a solution of the non-split mcYBE but its action on the $\mathfrak{su}(1,1)$ generators breaks the reality of the algebra. On the other hand, when $\alpha-\beta^2>1$, we have that $c^2 >0$, the R-matrix is a solution of the split mcYBE but its action on $\mathfrak{su}(2)$ generators breaks reality. This is consistent with the fact that there are no solutions to the split mcYBE on compact Lie algebras. This is also consistent with the fact that we expect the Drinfel'd-Jimbo deformation of $SU(1,1) \times SU(2)$ to give the time-like trigonometric model, not the space-like one~\cite{Hoare:2022asa}. 

\subsection{Adding the WZ term}
Let us now consider the integrable trigonometric cases with both B-field (parameter $\beta$) and WZ term (parameter $q$) turned on.
The integrable YB+WZ deformation of the PCM on $F$ reads~\cite{Klimcik:2019kkf,Delduc:2014uaa,Hoare:2020mpv}
\begin{equation}
\label{eq:YBWZ}
S_{\text{YBWZ}} = \frac{q}{4} \int d\tau d\sigma \, \text{STr}[J_+\frac{e^\chi + e^{\chi R}}{e^\chi - e^{\chi R}} J_-] + q S_\text{WZ}[g]\,,
\end{equation}
where $\chi$ is a free parameter and $R:\mathfrak{f} \rightarrow \mathfrak{f}$ again satisfies $R^T=-R$, the (m)cYBE~\eqref{eq:mcYBE}, as well as the additional condition
\begin{equation}
\label{eq:solvability}
(R^2 - c^2)[X,Y]_R=0\,, \forall X, Y \in \mathfrak{f}\,, \qquad [X,Y]_R = [X, R(Y)] + [R(X), Y]\,.
\end{equation}
 Note that with respect to~\cite{Hoare:2020mpv} we rescaled $R \rightarrow \chi \rho^{-1} R$ so that $\rho$ does not appear explicitly in the action, and also sent $g \rightarrow g^{-1}$ and $q \rightarrow -q$. 
Assuming that $\chi \rightarrow 2 q \zeta^{-1}$ for some constant $\zeta$ as $q \rightarrow 0$, in the limit $q \rightarrow 0$ the action becomes
\begin{equation}
\lim_{q \rightarrow 0} S_{\text{YBWZ}} = \zeta S_{\text{YB}}\,.
\end{equation}
As was shown in~\cite{Hoare:2020mpv}, the YB+WZ action~\eqref{eq:YBWZ} is integrable and the equations of motion for $J=g^{-1} d g$ can be written as the flatness of the following rational Lax connection
\begin{equation}
\mathcal L^r_\pm(\tau,\sigma;z) = \pm \text{Ad}_g \frac{q}{\xi (1\mp z)} \frac{1}{1-e^{\pm \chi}e^{\chi R}}  J_\pm\,, \qquad \xi = \frac{q}{2} \frac{\sinh \chi}{\cosh \chi - \cosh c\chi}\,.
\end{equation}
We recall that $c^2$ is the real parameter appearing on the right-hand side of the mcYBE~\eqref{eq:mcYBE}. In the limit $q \rightarrow 0$, with the above scaling of $\chi$, one has that $\xi \rightarrow \frac{\zeta}{2(1-c^2)}$ and one recovers the Lax~\eqref{eq:LaxYB} of the YB deformation without WZ term. 

In the two integrable trigonometric cases with WZ term, we find that the action~\eqref{eq:coupledEWZ} can be written
\begin{equation}
\label{eq:SYBWZtd}
S = \text{T} \left( S_\text{YBWZ} + \int d\tau d\sigma \mathcal L_{td} \right)\,,
\end{equation}
with total derivative term again given as in~\eqref{eq:Ltd}.

In the time-like trigonometric deformation with parameters~\eqref{eq:paramtlt} and $q \neq 0$, one finds that the equality~\eqref{eq:SYBWZtd} is satisfied provided one adds to the action the total derivative term~\eqref{eq:Ltd} with parameters
\begin{equation}
\rho_0 = \rho_3 = -\frac{(\cosh(\chi)-\cosh(r_0 \chi)) \sin(r_1 \chi)}{8 \sinh(\chi)(\cos(r_1 \chi) - \cosh(\chi))}\,, \qquad \rho_1=\rho_2=\rho_4=\rho_5=0\,,
\end{equation}
and one chooses the R-matrix to act on the generators as
\begin{equation}
\begin{aligned}
R(T_0) = r_0 T_3\,, \qquad R(T_1) = -r_1 T_2\,, \qquad R(T_2) = + r_1 T_1\,, \\
R(T_3) = r_0 T_0\,, \qquad R(T_4) = -r_1 T_5\,, \qquad R(T_5) = + r_1 T_4\,,
\end{aligned}
\end{equation}
where the relation between the parameters $(\chi,r_0,r_1)$ and $(\alpha,\beta,q)$ is
\begin{equation}
\alpha = \frac{-\cosh(\chi) + \cosh(r_0 \chi)}{\cos(r_1 \chi) - \cosh(\chi)}\,, \qquad \beta = \frac{\sinh(r_0 \chi)}{\sinh(\chi)}\,, \qquad q = \frac{\cosh(\chi) - \cosh(r_0 \chi)}{\sinh(\chi)}\,.
\end{equation}
This R-matrix solves the mcYBE with $c^2=- r_1^2$. 
It also satisfies
\begin{equation}
R^2-c^2 = (r_0^2 + r_1^2) P_\mathfrak{h}\,, 
\end{equation}
where $P_\mathfrak{h}$ projects onto the Cartan algebra $\mathfrak{h}=\{T_0,T_3\}$. Since $P_\mathfrak{h}[X,Y]_R =0 \, \forall X,Y \in \mathfrak{f}$, the deforming operator indeed satisfies the additional constraint~\eqref{eq:solvability}. 
It preserves the reality condition in the region $r_0,r_1 \in \mathbb{R}$, which corresponds to
\begin{equation}
\frac{1}{2} (1+q^2 - \beta^2 - \sqrt{(1+q^2-\beta^2)^2-4 q^2}) \leq \alpha \leq \frac{1}{2} (1+q^2 - \beta^2 + \sqrt{(1+q^2-\beta^2)^2-4 q^2})\,,
\end{equation}
with the assumption that $(1+q^2 - \beta^2)^2 > 4 q^2$. 
Similarly as in the case without WZ term, when $r_1 \neq 0$ one can rescale the deforming operator $\tilde{R} =r_1^{-1} R$ so that $\tilde{R}$ satisfies the mcYBE with $c^2 =-1$. When $r_1=0$ one rescales $\tilde{R} = r_0^{-1} R$ instead, and $\tilde{R}$ then solves the cYBE with $c^2=0$. In both cases this re-introduces the additional deformation parameter $\rho$ into the action, that we scaled away in~\eqref{eq:YBWZ}.

Note that around $q=0$ we have the expansion
\begin{equation}
r_0 = \beta + O(q)\,, \qquad r_1 = \frac{\sqrt{1-\alpha-\beta^2}}{\sqrt{\alpha}} + O(q)\,, \qquad \chi = \frac{2 q}{1-\beta^2} + O(q^2)\,,
\end{equation}
and sending $q\rightarrow 0$ one recovers the action of the YB deformation~\eqref{eq:YB} with R-matrix~\eqref{eq:Rtlt}.

A similar formal identification holds in the space-like trigonometric limit, though the operator $R$ does not preserve reality.

\paragraph{Comments on other types of integrable deformations.}
At this point let us address a question that might naturally come to the reader. As we have seen above, in the trigonometric limits, characterised by a $SU(1,1)_L \times SU(2)_L \times U(1)_R \times U(1)_R$ symmetry, one can always perform an additional TsT transformation which preserves both the left-acting $SU(1,1)_L \times SU(2)_L$ symmetry and the integrability of the model.
The elliptic deformation of the $AdS_3 \times S^3$ string preserves a left-acting $SU(1,1)_L \times SU(2)_L$ symmetry, with Cartan subalgebra $U(1) \times U(1)$. 
It is then possible to perform a TsT transformation in those two Cartan directions. Why do we not find this integrable model? The answer is the following: while this deformation preserves integrability, it breaks the left-acting symmetry and thus falls outside the class of theories that we are considering.

Recently, a classically-integrable four-parameter deformation of the PCM on $SU(2)$ was put forward in~\cite{Kotousov:2024lzg}. The additional deformation (on top of the three elliptic deformation parameters) gives rise to a B-field and a further deformation of the three-sphere metric. Just as for our setup, the B-field is constructed from wedge products of vielbein one-forms. However, since it only involves vielbein of $\mathfrak{su}(2)$, it gives rise to a closed B-field $H=dB=0$.

\section{Elliptic deformation of the superstring}
\label{sec:sugra}

Ultimately we are interested in defining an elliptic deformation of the superstring theory. The (undeformed) $AdS_3 \times S^3 \times T^4$ can be embedded both into type IIA and type IIB supergravity with mixed NSNS and RR fluxes. Here we will consider the embedding of the deformed metric and B-field~\eqref{eq:GBelliptic} into type IIB supergravity. 

\subsection{Embedding into supergravity}
In order to obtain an elliptic deformation of the mixed-flux $AdS_3 \times S^3 \times T^4$ superstring, the first step is to add by hand the four flat directions corresponding to the $T^4$, with coordinates $\{W_1,W_2,W_3,W_4\}$ and vielbein 
\begin{equation}
e^6 = dW_1\,, \qquad e^7 = dW_2\,, \qquad e^8 = dW_3\,, \qquad e^9 = dW_4\,.
\end{equation}
Then, motivated by the form of the fluxes in the pure RR elliptic case, we make the following Ansatz for the dilaton and the fluxes,
\begin{equation}
\label{eq:Ansatzfluxes}
\begin{aligned}
&\phi=F_1 = 0\,, \quad H_3 = \sum_{j=1}^4 x_5^{(j)} f_3^{(j)}\,, \quad F_3 = \sum_{j=1}^4 x_4^{(j)} f_3^{(j)}\,, \quad
F_5 = \sum_{j=1}^4 \sum_{k=1}^3 x_k^{(j)} f_3^{(j)}  \wedge J_{2,k}\,,
\end{aligned}
\end{equation}
where the auxiliary three-forms are given by
\begin{equation}
\begin{aligned}
f_3^{(1)} &= 2(e^0 \wedge e^2 \wedge e^4 - e^1 \wedge e^3 \wedge e^5)=-\sqrt{\alpha_0 \alpha_1 \alpha_2} \, d\mathsf{b}_1\,, \qquad J_{2,1} = e^6 \wedge e^7 - e^8 \wedge e^9\,, \\
f_3^{(2)} &= 2(e^0 \wedge e^4 \wedge e^5 + e^1 \wedge e^2 \wedge e^3)=-\sqrt{\alpha_0 \alpha_1 \alpha_2} \,d\mathsf{b}_0\,, \qquad J_{2,2} = e^6 \wedge e^8 + e^7 \wedge e^9\,, \\
f_3^{(3)} &= 2(e^0 \wedge e^1 \wedge e^5 - e^2 \wedge e^3 \wedge e^4)=+\sqrt{\alpha_0 \alpha_1 \alpha_2} \, d\mathsf{b}_2\,, \qquad J_{2,3} = e^6 \wedge e^9 - e^7 \wedge e^8\,, \\
f_3^{(4)} &= 2(e^0 \wedge e^1 \wedge e^2 + e^3 \wedge e^4 \wedge e^5) =+ \sqrt{\alpha_0 \alpha_1 \alpha_2} \, \mathsf{h}_V\,,
\end{aligned}
\end{equation}
and we identify
\begin{equation}
\label{eq:relx5B}
(x_5^{(1)}, x_5^{(2)}, x_5^{(3)}, x_5^{(4)}) = \frac{1}{\sqrt{\alpha_0 \alpha_1 \alpha_2}}(-\beta_1, \beta_0, \beta_2,q)\,.
\end{equation}
These auxiliary fluxes satisfy
\begin{equation}
f_3^{(j)} \cdot f_3^{(k)}=0\,, \qquad f_3^{(j)} \wedge f_3^{(k)}=0\,, \qquad d f_3^{(j)} =0\,,
\end{equation}
so that the type IIB supergravity equations of motion then reduce to
\begin{equation}
\begin{aligned}
    0 = R_{ab} - \frac{1}{4} H_{a kl} H_{b}{}^{kl} -\frac{1}{4} F_{akl} F_b{}^{kl} - \frac{1}{4 \times 4!} F_{apqrs} F_b{}^{pqrs}\,.
    \end{aligned}
\end{equation}
The Ricci tensor is diagonal when written in the tangent-space indices (related to the curved-space indices as $R_{\mu \nu} = e_\mu^a e_\nu^b R_{ab}$ where $e^a$ are the deformed vielbein~\eqref{eq:defvielbein}), with only non-vanishing components
\begin{equation}
\begin{aligned}
&R_{33}  = 2 \frac{\alpha_0^2 - (\alpha_1 - \alpha_2)^2}{\alpha_0 \alpha_1 \alpha_2}\,, \qquad  R_{44} = 2 \frac{\alpha_1^2-(\alpha_2-\alpha_0)^2}{\alpha_0 \alpha_1 \alpha_2}\,, \qquad R_{55} = 2 \frac{\alpha_2^2-(\alpha_0-\alpha_1)^2}{\alpha_0 \alpha_1 \alpha_2}\,, \\
&R_{00} = R_{33}\,, \qquad R_{11}=-R_{44}\,, \qquad R_{22}=-R_{55}\,.
\end{aligned}
\end{equation}

It is then straightforward to check that the supergravity equations of motion are satisfied provided that the deformation parameters are related by~\cite{Hoare:2023zti,Seibold:2024qkh}
\begin{equation}\begin{aligned}\label{eq:solution}
||\mathbf{x}^{(1)}||^2 & = \frac{(\alpha_0 + \alpha_2 - \alpha_1)}{\alpha_0 \alpha_2} - ||\mathbf{x}^{(4)}||^2 \,,
& \qquad
\mathbf{x}^{(1)} \cdot \mathbf{x}^{(4)} & = \mathbf{x}^{(2)} \cdot \mathbf{x}^{(3)} \,,
\\
||\mathbf{x}^{(2)}||^2 & = \frac{(\alpha_0 - \alpha_1 - \alpha_2)}{\alpha_1 \alpha_2} + ||\mathbf{x}^{(4)}||^2 \,,
& \qquad
\mathbf{x}^{(2)} \cdot \mathbf{x}^{(4)} & = -\mathbf{x}^{(1)} \cdot \mathbf{x}^{(3)} \,,
\\
||\mathbf{x}^{(3)}||^2 & = \frac{(\alpha_0 + \alpha_1 - \alpha_2)}{ \alpha_0 \alpha_1} - ||\mathbf{x}^{(4)}||^2 \,,
& \qquad
\mathbf{x}^{(3)} \cdot \mathbf{x}^{(4)} & = \mathbf{x}^{(1)} \cdot \mathbf{x}^{(2)} \,.
\end{aligned}
\end{equation}
The notation is such that each vector of deformation parameters is five-dimensional, 
\begin{equation}
\mathbf{x}^{(j)} = (x_1^{(j)}, x_2^{(j)}, x_3^{(j)}, x_4^{(j)}, x_5^{(j)})\,.
\end{equation}
The first three entries enter in the five-form RR flux $F_5$, the fourth entry modulates the three-form RR flux $F_3$ and the fifth entry corresponds to the NSNS three-form flux $H_3$. 
An interesting question is then what is the space of solutions to~\eqref{eq:solution}. In particular, do they admit solutions with real NSNS and RR fluxes? Do they impose constraints on the range of deformation parameters?
For the rational parameters $\alpha_0=\alpha_1=\alpha_2=1$, the reality of the fluxes implies that $\mathbf{x}^{(1)}=\mathbf{x}^{(2)} = \mathbf{x}^{(3)}=0$ and $||\mathbf{x}^{(4)}|| =1$. In the mixed-flux theory this then imposes the range $0 \leq |q| \leq 1$.
In the elliptic case, being able to construct an embedding with real fluxes imposes constraints on the range of deformation parameters $\alpha_j$ and $\beta_j$. A straightforward way to see this is to take the sum of the first two and the last two equations involving the norms in~\eqref{eq:solution}. Positivity of the norms then imposes
\begin{equation}
\label{eq:rangealpha}
0<\alpha_1 \leq \alpha_0\,, \qquad 0<\alpha_2 \leq \alpha_0\,.
\end{equation} 
To analyse in more detail the consequences of having real fluxes, it is useful to introduce the sums and differences
of the deformation vectors,
\begin{equation}
\mathbf{y}_\pm=\mathbf{x}^{(2)}\pm\mathbf{x}^{(4)},
\qquad
\mathbf{z}_\pm=\mathbf{x}^{(1)}\pm\mathbf{x}^{(3)}.
\end{equation}
We also introduce the alternative deformation parameters $\ga_j$ and $\bb_j$ through the relations (assuming $\alpha_j > 0$)~\footnote{With respect to~\cite{Hoare:2023zti,Hoare:2025rtl} we have $\left. \alpha_0 \right|_\text{here} = \left. \alpha_2 \right|_\text{there}$, $\left. \alpha_1 \right|_\text{here}= \left. \alpha_3 \right|_\text{there}$ and $\left. \alpha_2 \right|_\text{here} = \left. \alpha_1 \right|_\text{there}$. The definitions of the parameters $\gamma_{1,2,3}$ are then the same. }
\begin{equation}
\ga_1 = \frac{\al_0}{\sqrt{\al_0\al_1\al_2}}\,,
\qquad
\ga_2 = \frac{\al_2-\al_1}{\sqrt{\al_0\al_1\al_2}}\,,
\qquad
\ga_3 = \frac{\al_1 + \al_2- \al_0 }{\sqrt{\al_0\al_1\al_2}}\,,
\end{equation} 
and
\begin{equation}
\delta_0=\frac{\beta_0}{\alpha_0}\,,
\qquad
\delta_1=-\frac{\beta_1}{\alpha_1}\,,
\qquad
\delta_2=-\frac{\beta_2}{\alpha_2}\,,
\qquad
\delta_q=-\frac{q}{\alpha_2}.
\end{equation}
Note that the condition $\alpha_0,\alpha_1,\alpha_2>0$ is translated to
\begin{equation}
\label{eq:rangegamma}
\gamma_1 >0\,, \qquad \gamma_1\pm \gamma_2 + \gamma_3>0\,,
\end{equation}
while the relation~\eqref{eq:rangealpha}, which is a necessary condition for embedding the elliptic deformation into supergravity with real fluxes, reads
\begin{equation}
\label{eq:rangegammasugra}
\begin{aligned}
&\gamma_1\pm \gamma_2 \geq \gamma_3\,, \qquad \gamma_3>0\,, \\
&\gamma_1 \pm \gamma_2 > -\gamma_3\,, \qquad \gamma_3 \leq 0\,.
\end{aligned}
\end{equation}
In what follows we find it convenient to restrict to the branch with $\gamma_3>0$, which contains the undeformed point $(\gamma_1,\gamma_2,\gamma_3)=(1,0,1)$. Note that this is a restriction of our choosing; there are other valid supergravity solutions that have negative $\gamma_3$.

The supergravity conditions (\ref{eq:solution}) become particularly simple when expressed in these new variables,
\begin{equation}\label{eq:trace}
\begin{aligned}
\nn{\mathbf{y}_+}^2+\nn{\mathbf{z}_+}^2
&=\gamma_1^2-\gamma_2^2\,, &\qquad \mathbf{y}_+\cdot\mathbf{y}_-&=-\gamma_1\gamma_3\,,
& \qquad \mathbf{z}_+\cdot\mathbf{z}_-&=\gamma_2\gamma_3\,,\\
\nn{\mathbf{y}_-}^2+\nn{\mathbf{z}_-}^2
&=\gamma_1^2-\gamma_2^2\,, &\qquad \mathbf{y}_+\cdot\mathbf{z}_-&=0\,,
&\qquad
\mathbf{z}_+\cdot\mathbf{y}_-&=0\,.
\end{aligned}
\end{equation}
The Gram matrix of the four real vectors
$(\mathbf{y}_+,\mathbf{z}_+,\mathbf{y}_-,\mathbf{z}_-)$
is positive semidefinite and has the block structure
\begin{equation}
G= \begin{pmatrix}
G_+&D\\
D&G_-
\end{pmatrix}\succeq0\,,
\qquad
D=
\begin{pmatrix}
-\gamma_1\gamma_3&0\\
0&\gamma_2\gamma_3
\end{pmatrix}\,,
\qquad
G_\pm=
\begin{pmatrix}
\nn{\mathbf{y}_\pm}^2&
\mathbf{y}_\pm\cdot\mathbf{z}_\pm\\
\mathbf{y}_\pm\cdot\mathbf{z}_\pm&
\nn{\mathbf{z}_\pm}^2
\end{pmatrix}\,,
\end{equation}
where $G_+$ is the Gram matrix of the real vectors $(\mathbf{y}_+,\mathbf{z}_+)$ and $G_-$ is the Gram matrix of the real vectors $(\mathbf{y}_-,\mathbf{z}_-)$. We work in the generic elliptic region with
$\gamma_1\gamma_2\gamma_3\neq0$. The two vectors $(\mathbf{y}_+, \mathbf{z}_+)$ are then linearly independent: taking the scalar
product of a linear relation between them with
$\mathbf{y}_-$ and $\mathbf{z}_-$ forces both coefficients
to vanish, by (\ref{eq:trace}). Similarly, the two vectors $(\mathbf{y}_-, \mathbf{z}_-)$ are also linearly independent.  
Both Gram matrices $G_+$ and $G_-$ are therefore positive definite and
invertible.
Positivity of $G$
is then equivalent to positivity of the Schur complements,
\begin{equation}
G_- -DG_+^{-1}D\succeq0\,, \qquad G_+ -DG_-^{-1}D\succeq0\,.
\end{equation}
Geometrically, the first Schur complement is the Gram matrix of
the components of $\mathbf{y}_-$ and $\mathbf{z}_-$ that are
orthogonal to the span of $\mathbf{y}_+$ and $\mathbf{z}_+$.
Taking the trace and using (\ref{eq:trace}) gives the two inequalities
\begin{equation}
\label{eq:key}
(\mathbf{y}_\pm \cdot\mathbf{z}_\pm )^2
 \leq
\nn{\mathbf{y}_\pm }^2
\left(
\gamma_1^2-\gamma_2^2+\gamma_3^2
-\nn{\mathbf{y}_\pm }^2
\right)
-\gamma_1^2\gamma_3^2\,.
\end{equation}
At this point it is useful to introduce 
\begin{equation}
m_1 = \sqrt{\gamma_1^2 - (\gamma_2+\gamma_3)^2}\,, \qquad  m_2 = \sqrt{\gamma_1^2 - (\gamma_2-\gamma_3)^2}\,,
\end{equation}
with $m_1 m_2 \neq 0$ in the generic elliptic case, so that the right-hand side
factorises as
\begin{equation}\label{eq:factor}
(\mathbf{y}_\pm \cdot\mathbf{z}_\pm )^2
\leq
\left(
\gamma_3^2+\frac{(m_1+m_2)^2}{4}
-\nn{\mathbf{y}_\pm }^2
\right)\left(
\nn{\mathbf{y}_\pm}^2-\gamma_1^2+\gamma_2^2
+\frac{(m_1+m_2)^2}{4}
\right)\,.
\end{equation}
Since the left-hand side is nonnegative, it follows that
\begin{equation}\label{eq:u roots}
\gamma_1^2-\gamma_2^2-\frac{(m_1+m_2)^2}{4}
\leq\nn{\mathbf{y}_\pm}^2
\leq
\gamma_3^2+\frac{(m_1+m_2)^2}{4}\,.
\end{equation}
Consequently,
\begin{equation}\label{eq:envelope}
\begin{pmatrix}
\gamma_3^2+\frac{(m_1+m_2)^2}{4}&0\\
0&\frac{(m_1+m_2)^2}{4}
\end{pmatrix}
-G_\pm \succeq0\,.
\end{equation}
Indeed, (\ref{eq:u roots}) makes both diagonal entries
nonnegative, while (\ref{eq:factor}) makes the determinant
nonnegative.

Subtracting the NSNS contribution from $G_\pm$ leaves the Gram
matrix of the four real RR components of $\mathbf{y}_\pm$
and $\mathbf{z}_\pm$, which is also positive semidefinite.
\begin{equation}\label{eq:RRgram}
G_\pm - 
\begin{pmatrix}
(y_{\pm,5})^2
&
y_{\pm,5} z_{\pm,5}\\
y_{\pm,5} z_{\pm,5}
&
(z_{\pm,5})^2
\end{pmatrix}
\succeq0\,.
\end{equation}
Adding (\ref{eq:envelope}) and (\ref{eq:RRgram}) therefore yields
\begin{equation}\label{eq:finalmatrix}
\begin{pmatrix}
4\gamma_3^2+(m_1+m_2)^2&0\\
0&(m_1+m_2)^2
\end{pmatrix}
-
4 \begin{pmatrix}
(y_{\pm,5})^2
&
y_{\pm,5} z_{\pm,5}\\
y_{\pm,5} z_{\pm,5}
&
(z_{\pm,5})^2
\end{pmatrix}
\succeq0\,.
\end{equation}
Imposing that the eigenvalues of this matrix are non-negative gives the bounds
\begin{equation}
\label{eq:non-strict stability}
\frac{4(y_{\pm,5})^2}
{4\gamma_3^2+(m_1+m_2)^2}
+
\frac{4(z_{\pm,5})^2}{(m_1+m_2)^2}
\leq1\,.
\end{equation}
These two bounds need to be satisfied in order to have a supergravity embedding with real fluxes.

\subsection{Supersymmetry}
Let us now discuss the supersymmetries of the elliptic mixed-flux theory. 
In the pure RR elliptic deformation, there are additional constraints that need to be imposed on the deformation parameters $\mathbf{x}^{(k)}$ in order to have a background with 8 supersymmetries coming from the undeformed left copy of $\mathfrak{psu}(1,1|2)$~\cite{Hoare:2025rtl}. Namely, the pure RR elliptic background preserves 8 supersymmetries if the RR parameters satisfy the supergravity constraints~\eqref{eq:solution} and also lie in a two-plane, so that the rank of the Gram matrix is $\leq 2$ (it is rank 2 in the generic elliptic case and rank 1 in the rational limit). Without going into the explicit construction of the Killing spinors in the mixed-flux elliptic deformation, let us argue, following~\cite{Llatas:1999rf,Bergshoeff:2005ac}, that also in the elliptic mixed-flux theory one can construct a subclass of supersymmetric elliptic backgrounds.

The starting point is an elliptic supersymmetric pure RR background of the form~\eqref{eq:Ansatzfluxes}, satisfying the constraints~\eqref{eq:solution}, and with Killing spinor $\epsilon_0$ satisfying the Killing spinor equations  
\begin{equation}
\begin{aligned}
0 &= \sigma_1 \slashed{F_3} \epsilon_0 \,, \qquad 
0 &= ( \partial_M - \frac{1}{4} \slashed{\omega}_M - \frac{1}{8 \cdot 3!} \sigma_1 \slashed{F_3} \Gamma_M - \frac{1}{8 \cdot 2 \cdot 5!} i \sigma_2\slashed{F}_5 \Gamma_M) \epsilon_0\,.
\end{aligned}
\end{equation}
Then, because the supergravity constraints of~\eqref{eq:solution} involve only $SO(5)$ invariants, also the mixed flux background
\begin{equation}
H_3^\theta = \sin \theta F_3\,, \qquad F_3^\theta = \cos \theta F_3\,,
\end{equation}
will be a solution to the supergravity equations of motion. This rotation, which leaves $F_5$ invariant, corresponds to the $SO(2) \subset SL(2,\mathbb{R})$ subgroup of classical type IIB S-duality. 
Let us define
\begin{equation}
\epsilon^\theta = U_\theta \epsilon_0\,, \qquad U_\theta = e^{-\frac{\theta}{2} i \sigma_2} = \begin{pmatrix}
\cos \frac{\theta}{2} & -\sin \frac{\theta}{2} \\
\sin \frac{\theta}{2} & \cos \frac{\theta}{2}
\end{pmatrix}\,.
\end{equation}
Using in particular the identities
\begin{equation}
U_\theta \sigma_1 U_\theta^{-1} = \cos \theta \sigma_1 - \sin \theta \sigma_3\,, \qquad U_\theta \sigma_3 U_\theta^{-1} = \cos \theta \sigma_3 + \sin \theta \sigma_1\,,
\end{equation}
one can show that $\epsilon^\theta$ satisfies the equations
\begin{equation}
\label{eq:KS1}
0 = (\sigma_3 \slashed{H}_3^\theta + \sigma_1 \slashed{F_3}^\theta) \epsilon^\theta\,,
\end{equation} 
and
\begin{equation}
\begin{aligned}
0 &= ( \partial_M - \frac{1}{4} \slashed{\omega}_M  + \frac{1}{8} \sigma_3 \slashed{H}_{3,M}^\theta   - \frac{1}{8 \cdot 3!} \sigma_1 \slashed{F}_3^\theta \Gamma_M - \frac{1}{8 \cdot 2 \cdot 5!} i \sigma_2\slashed{F}_5 \Gamma_M) \epsilon^\theta\,,
\end{aligned}
\end{equation}
where we used~\eqref{eq:KS1} and $\{\Gamma_M, \slashed{F_3} \} = 6 \slashed{F}_M = 6 F_{M ab} \Gamma^{ab}$. Therefore, $\epsilon^\theta$ solves the Killing spinor equations for the mixed flux background. For any pure RR supersymmetric elliptic supergravity solution, it is possible to construct a corresponding mixed-flux background that also preserves 8 supersymmetries. Note that both backgrounds are characterised by the same Gram matrix, so the supersymmetry constraint is again equivalent to the vectors of flux parameters spanning a two-plane. 

Supersymmetry imposes further constraints on the range of NSNS deformation parameters. The full Gram matrix $G$ has rank two (in the generic elliptic case), so its Schur
complements now vanish,
\begin{equation}\label{eq:susy-schur-zero}
G_\mp -D G_\pm^{-1}D=0\,.
\end{equation}
Taking the trace now gives the equalities
\begin{equation}\label{eq:susy-gram-equality}
(\mathbf{y}_\pm \cdot\mathbf{z}_\pm )^2
=\nn{\mathbf{y}_\pm }^2
\left(\gamma_1^2-\gamma_2^2+\gamma_3^2
-\nn{\mathbf{y}_\pm }^2\right)
-\gamma_1^2\gamma_3^2\,,
\end{equation}
which are indeed the ones derived in~\cite{Hoare:2025rtl} for the pure RR elliptic deformation.
Expressing $\mathbf{y}_-$ and $\mathbf{z}_-$ as linear combinations of $\mathbf{y}_+$ and $\mathbf{z}_+$, using~\eqref{eq:trace} and specialising to the fifth component, yields
\begin{equation}\label{eq:susy-fifth-linear}
\begin{pmatrix}y_{-,5}\\z_{-,5}\end{pmatrix}
=DG_+^{-1}\begin{pmatrix}y_{+,5}\\z_{+,5}\end{pmatrix}\,.
\end{equation}
 Rearranging and expressing this component-wise gives
\begin{equation}\label{eq:susy-fifth-components}
\begin{aligned}
y_{+,5}&=-\frac{\nn{\mathbf{y}_+}^2}{\gamma_1\gamma_3}y_{-,5}
+\frac{\mathbf{y}_+\cdot\mathbf{z}_+}{\gamma_2\gamma_3}z_{-,5}\,,\\
z_{+,5}&=-\frac{\mathbf{y}_+\cdot\mathbf{z}_+}{\gamma_1\gamma_3}y_{-,5}
+\frac{\gamma_1^2-\gamma_2^2-\nn{\mathbf{y}_+}^2}
{\gamma_2\gamma_3}z_{-,5}\,.
\end{aligned}
\end{equation}
Taking the square of the two equations and adding them,
and using~\eqref{eq:susy-gram-equality} to remove
$(\mathbf{y}_+\cdot\mathbf{z}_+)^2$, leads to the equality
\begin{equation}\label{eq:susy-fifth-quadratic}
0={}y_{+,5}^2+z_{+,5}^2+y_{-,5}^2+z_{-,5}^2
+\frac{\gamma_1^2-\gamma_2^2+\gamma_3^2}
{\gamma_1\gamma_3}\,y_{+,5}y_{-,5}
-\frac{\gamma_1^2-\gamma_2^2-\gamma_3^2}
{\gamma_2\gamma_3}\,z_{+,5}z_{-,5}\,.
\end{equation}
Inserting the NSNS coefficients yields
\begin{equation}\label{eq:susy-general-deltas}
\delta_0^2
-\frac{(\gamma_1-\gamma_2)^2-\gamma_3^2}
{4\gamma_1\gamma_2}\,\delta_1^2
+\frac{(\gamma_1+\gamma_2)^2-\gamma_3^2}
{4\gamma_1\gamma_2}\,\delta_2^2
-\frac{(\gamma_1+\gamma_2+\gamma_3)
\bigl((\gamma_1-\gamma_3)^2-\gamma_2^2\bigr)}
{4\gamma_1^2(\gamma_1-\gamma_2+\gamma_3)}\,\delta_q^2
=0\,.
\end{equation}
This in particular implies that it is not possible to construct a supersymmetric elliptic deformation with only one type of NSNS flux. For instance, adding just a WZ term to the action allows for a supergravity embedding, but supersymmetry requires the presence of an additional auxiliary NSNS three-form coupling the $AdS_3$ and $S^3$ sectors.

On top of the above equality, there is also a bound on the magnitude of the deformation parameters, which comes from the RR Gram matrix being positive semidefinite
\begin{equation}
    G_+-\begin{pmatrix}y_{+,5}\\z_{+,5}\end{pmatrix}\begin{pmatrix}y_{+,5}&z_{+,5}\end{pmatrix}\succeq0\,.
\end{equation}
Since $G_+$ is positive definite, subtracting
the NSNS outer product from $G_+$ is permissible precisely when
\begin{equation}\label{eq:susy-projection-bound}
0\leq
\begin{pmatrix}y_{+,5}&z_{+,5}\end{pmatrix}
G_+^{-1}\begin{pmatrix}y_{+,5}\\z_{+,5}\end{pmatrix}
\leq1\,.
\end{equation}
Using \eqref{eq:susy-fifth-linear} and substituting the NSNS coefficients gives
\begin{equation}\label{eq:susy-general-reality}
0\leq{}-\frac{\gamma_1}{\gamma_3}\,\delta_0^2
+\frac{(\gamma_1-\gamma_2+\gamma_3)^2}
{4\gamma_2\gamma_3}\,\delta_1^2
-\frac{(\gamma_1+\gamma_2+\gamma_3)^2}
{4\gamma_2\gamma_3}\,\delta_2^2
+\frac{(\gamma_1+\gamma_2+\gamma_3)^2}
{4\gamma_1\gamma_3}\,\delta_q^2
\leq1\,.
\end{equation}
The lower bound expresses positivity of $G_+$. It is automatically satisfied by the solutions of~\eqref{eq:susy-general-deltas}. The upper bound is an independent restriction that ensures that the NSNS components leave room for a supersymmetric supergravity completion with real RR fluxes.

\section{Quantisation of the theory}
\label{sec:Quantisation}

In Section~\ref{sec:Lax} we analysed the classical integrability of the elliptic+B-field model, by trying to rewrite the equations of motion as the flatness of a Lax connection.
Because this is a constructive approach, not finding a Lax does not necessarily mean that the model is not integrable. It may be that our Ansatz~\eqref{eq:Lax} was not general enough to capture the elliptic+B-field cases. 
In order to provide further evidence for the non-integrability of the model, we now perturbatively quantise the theory in uniform light-cone gauge~\cite{Arutyunov:2005hd} and compute the tree-level worldsheet S-matrix. This allows us to highlight the existence of processes that violate the axioms of an integrable model.

\subsection{Light-cone gauge-fixing}
We are interested in the bosonic string sigma-model action
\begin{equation}
\label{eq:SSM}
S = -\frac{\text{T}}{2} \int d\tau d \sigma \left(\sqrt{-h} h^{\alpha \beta} G_{\mu \nu} \partial_\alpha X^\mu \partial_\beta X^\nu + \epsilon^{\alpha \beta} B_{\mu \nu} \partial_\alpha X^\mu \partial_\beta X^\nu\right)\,,
\end{equation}
where we couple the sigma-model~\eqref{eq:SM} to a dynamical worldsheet metric $h_{\alpha \beta}$ with inverse $h^{\alpha \beta}$ and determinant $h$, and also add four flat directions corresponding to the $T^4$, and $\epsilon^{\tau\sigma}=+1$. The target-space fields are gathered into $X = (T,Z_1,Z_2,\Phi,Y_1,Y_2,W_1,W_2,W_3,W_4)$. 

The action~\eqref{eq:SSM} has a lot of redundancies, which come from reparametrisation and Weyl invariance on the worldsheet, as well as invariance under kappa-symmetry if we were to also include fermions. 
These unphysical degrees of freedom need to be gauged away. 
A standard way to achieve this for string theories in curved space-time is to impose uniform light-cone gauge, which requires the presence of two commuting isometries in the background.
The presence of the NSNS flux still preserves the same isometries in $T$ and $\varPhi$, and we can fix the same uniform light-cone gauge as in the (pure RR) elliptic case.

We start by rescaling the fields as~\footnote{This is for later convenience, so that the excitations are canonically normalised.}
\begin{equation}
T \rightarrow \frac{T}{\sqrt{\al_0}}\,, \quad Z_1 \rightarrow \frac{Z_1}{\sqrt{\al_1}}\,, \quad Z_2 \rightarrow \frac{Z_2}{\sqrt{\al_2}}\,, \quad \Phi \rightarrow \frac{\Phi}{\sqrt{\al_0}}\,, \quad Y_1 \rightarrow \frac{Y_1}{\sqrt{\al_1}}\,, \qquad Y_2 \rightarrow \frac{Y_2}{\sqrt{\al_2}}\,,
\end{equation}
and introduce the target-space light-cone coordinates
\begin{equation}
X^+ = (1-a) T + a \Phi\,, \qquad X^- = - T + \Phi\,, \qquad T = X^+ - a X^-\,, \qquad \Phi = X^+ + (1-a) X^-\,,
\end{equation}
with free parameter $a \in [0,1]$. 
In order to obtain the light-cone gauge-fixed theory we T-dualise the sigma-model in $X^-$, and call $\mathring{X}^-$ the T-dual coordinate. Integrating out the auxiliary worldsheet metric, the T-dual Lagrangian is~\cite{Arutyunov:2014jfa}~\footnote{The paper~\cite{Arutyunov:2014jfa} makes the assumption that $B_{T\Phi}=B_{+-}=0$, which is not the case for our B-field. Indeed, while this is verified for $\mathsf{b}_V$, it is not the case for $\mathsf{b}_{0,1,2}$. We however checked that our results for the light-cone gauge-fixed theory obtained by this T-duality approach match those obtained using the Hamiltonian formalism, as summarised in appendix~\ref{app:Hamiltonian}.}
\begin{equation}
\label{eq:Lsq}
    \mathring{\mathcal{L}}=-\sqrt{-\det \left[ \mathring{G}_{\mu \nu} \partial_\al X^\mu \partial_\be X^\nu \right]}+\frac{1}{2} \epsilon^{\al \be} \mathring{B}_{\mu \nu} \partial_\al X^\mu \partial_\be X^\nu\,,
\end{equation}
where $X^\mu = (X^+,\mathring{X}^-,Z_1,Z_2,Y_1,Y_2, W_1, W_2, W_3, W_4)^\mu$, and
\begin{equation}
\label{eq:Buscher}
\begin{aligned}
\mathring{G}_{--} &= \frac{1}{G_{--}}\,, &\qquad \mathring{G}_{- \bar{\mu}} &= \mathring{G}_{ \bar{\mu} -} = \frac{B_{-\bar{\mu}}}{G_{--}}\,, &\qquad \mathring{G}_{\bar{\mu}\bar{\nu}} &= G_{\bar{\mu} \bar{\nu}} - \frac{G_{-\bar{\mu}} G_{-\bar{\nu}}-B_{-\bar{\mu}} B_{-\bar{\nu}}}{G_{--}}\,,\\
& & \mathring{B}_{-\bar{\mu}} &= - \mathring{B}_{\bar{\mu}-} = - \frac{G_{- \bar{\mu}}}{G_{--}}\,, &\qquad \mathring{B}_{\bar{\mu}\bar{\nu}} &= - B_{\bar{\mu} \bar{\nu}} + \frac{G_{-\bar{\mu}} B_{-\bar{\nu}}-B_{-\bar{\mu}} G_{-\bar{\nu}}}{G_{--}}\,,
\end{aligned}
\end{equation}
where the index $\bar{\mu}$ runs over all coordinates except the one involved in the T-duality. The light-cone gauge-fixed theory is obtained by setting
\begin{equation}
X^+ = \tau\,, \qquad \mathring{X}^- = \sigma\,.
\end{equation}
On the vacuum, $G_{--}=1-2a$, so the intermediate
Buscher rules~\eqref{eq:Buscher} are singular at $a=1/2$. We therefore perform the T-duality and gauge-fixing at $a\neq1/2$, combine the
Nambu--Goto and Wess--Zumino contributions, and only then take
$a\to1/2$ from below. The singularities cancel in the gauge-fixed Lagrangian.

\subsection{Expansion in the string tension}

The light-cone gauge-fixed Lagrangian is highly non-linear, featuring square roots, and hence difficult to quantise. What we can do on the other hand, is to quantise the theory perturbatively in the string tension.
We expand the light-cone gauge-fixed theory around the classical solution
 $Z_1=Z_2=Y_1=Y_2=W_j=0$ . The zeroth and first order Lagrangians vanish (in fact, the requirement that this is the case fixes the choice of branch in~\eqref{eq:Lsq}), while the quadratic Lagrangian is, up to total derivatives,
 \begin{equation}
 \label{eq:L2}
 \mathcal L_2 = \mathcal L_{2,\bullet} + \mathcal L_{2,\circ}\,,
 \end{equation}
 with 
\begin{equation}
\label{eq:L2massive}
\begin{aligned}
\mathcal{L}_{2,\bullet}&=
\frac{1}{2}(\dot{Z_1}^2+\dot{Z_2}^2+\dot{Y_1}^2+\dot{Y_2}^2)-\frac{1}{2}(1-\bb_0^2)(\acute{Z_1}^2+\acute{Z_2}^2+\acute{Y_1}^2+\acute{Y_2}^2)\\
&-\frac{1}{2}(\ga_1^2-\ga_2^2-\ga_3^2)(Z_1^2+Z_2^2+Y_1^2+Y_2^2)+\ga_2\ga_3(Z_1^2-Z_2^2+Y_1^2-Y_2^2) \\
&+\ga_3(\dot{Z_1}Z_2-Z_1\dot{Z_2}+\dot{Y_1}Y_2-Y_1\dot{Y_2}) \\
&-\bb_0(\acute{Z_1}\dot{Z_1}+\acute{Z_2}\dot{Z_2}+\acute{Y_1}\dot{Y_1}+\acute{Y_2}\dot{Y_2})+\bb_1 (\dot{Z_1} \acute{Y_1} - \acute{Z_1} \dot{Y_1}) + \bb_2 (\dot{Z_2} \acute{Y_2} - \acute{Z_2} \dot{Y_2}) \\
&+(\bb_0-\bb_q)(\ga_1+\ga_2+\ga_3) (Z_1\acute{Z_2}+Y_1\acute{Y_2})-\bb_0(\ga_1-\ga_2+\ga_3)(\acute{Z_1}Z_2+\acute{Y_1}Y_2) \\
&+\bb_1(\ga_1-\ga_2+\ga_3)(Z_2\acute{Y_1}-\acute{Z_1}Y_2)-\bb_2(\ga_1+\ga_2+\ga_3)(Z_1\acute{Y_2}-\acute{Z_2}Y_1)\,,
\end{aligned}
\end{equation}
and
\begin{equation}
\label{eq:L2massless}
\mathcal L_{2,\circ} = \sum_{j=1}^4 \left( \frac{1}{2} \dot{W}_j^2 -\frac{1}{2}(1-\bb_0^2) \acute{W}_j^2 - \bb_0 \dot{W}_j \acute{W}_j\right)\,.
\end{equation}
The quadratic Lagrangian $\mathcal L_{2,\bullet}$ governs the propagation of the massive fields, namely the $AdS_3$ fields $(Z_1,Z_2)$ and the $S^3$ fields $(Y_1,Y_2)$. 
The parameters $\bb_1$ and $\bb_2$ couple these two sectors. The parameters $\bb_0$ and $\bb_q$ break the parity invariance $\sigma \rightarrow -\sigma$.  
Note that $\bb_0$ also introduces a mixing between the $\tau$- and $\sigma$-derivatives. This is true even for the massless fields $W_j$ corresponding to the torus directions, as can be seen in the massless Lagrangian $\mathcal L_{2,\circ}$. It is curious that the deformation turned on by $\bb_0$, which corresponds to a B-field with legs only in $AdS_3$ and $S^3$, affects the massless sector in the light-cone gauge-fixed theory. Mathematically, it comes from the non-vanishing $\mathring{G}_{+-} =- B_{+-}/G_{--}$ contribution. Setting the massive fields to zero (which corresponds to the first contribution in the expansion), we have 
\begin{equation}
\mathring{G}_{++} \rightarrow -\frac{1-\bb_0^2}{1-2a}\,, \qquad \mathring{G}_{--} \rightarrow \frac{1}{1-2a}\,, \qquad \mathring{G}_{+-} \rightarrow \frac{\bb_0}{1-2a}\,, \qquad \mathring{B}_{+-} \rightarrow \frac{1}{1-2a}\,,
\end{equation}
and the leading contribution to the massless light-cone gauge-fixed Lagrangian comes from
\begin{equation}
\begin{aligned}
&-\sqrt{- \begin{vmatrix} \mathring{G}_{++} + \dot{W}_j \dot{W}_j & \mathring{G}_{+-} + \dot{W}_j \acute{W}_j \\ \mathring{G}_{+-} + \dot{W}_j \acute{W}_j & 
\mathring{G}_{--} + \acute{W}_j \acute{W}_j \end{vmatrix}} + \mathring{B}_{+-} \\ 
&=-\sqrt{\mathring{G}_{+-}^2 -\mathring{G}_{++} \mathring{G}_{--} -\mathring{G}_{--} \dot{W}_j \dot{W}_j- \mathring{G}_{++} \acute{W}_j \acute{W}_j  +2 \mathring{G}_{+-} \dot{W}_j \acute{W}_j} + \mathring{B}_{+-} + \dots \\
&= \frac{1}{2} \frac{\mathring{G}_{--}}{\mathring{B}_{+-}} \dot{W}_j \dot{W}_j +\frac{1}{2} \frac{\mathring{G}_{++}}{\mathring{B}_{+-}} \acute{W}_j \acute{W}_j -\frac{\mathring{G}_{+-}}{\mathring{B}_{+-}} \dot{W}_j \acute{W}_j + \dots \,,
\end{aligned}
\end{equation}
where $\dots$ represent higher-order terms in the expansion.

\subsection{Oscillator expansion}
Focusing on the quadratic Lagrangian~\eqref{eq:L2}, the equations of motion for the fields are
\begin{equation}
\mathcal E_{Z_1} = \mathcal E_{Z_2} = \mathcal E_{Y_1} = \mathcal E_{Y_2} = \mathcal E_{W_j} = 0\,,
\end{equation}
with
\begin{align}
\mathcal{E}_{Z_1} &= \ddot{Z_1} - (1 - \bb_0^2)\, Z_1''
+ m_1^2 \, Z_1- 2 \bb_0\, \dot{Z_1}'
+ 2\ga_3\, \dot{Z_2}-  (\Delta+2 \bb_0 \ga_3)\, Z_2'
-\kappa\, Y_2'\,, \\
\mathcal{E}_{Z_2} &= \ddot{Z_2} - (1 - \bb_0^2)\, Z_2''
+ m_2^2 \, Z_2- 2 \bb_0\, \dot{Z_2}'
- 2\ga_3\, \dot{Z_1}+  (\Delta+2 \bb_0 \ga_3) \, Z_1'
-\kappa \, Y_1'\,, \\
\mathcal{E}_{Y_1} &= \ddot{Y_1} - (1 - \bb_0^2)\, Y_1''
+ m_1^2 \, Y_1- 2 \bb_0\, \dot{Y_1}'
+ 2\ga_3\, \dot{Y_2}-  (\Delta+2 \bb_0 \ga_3) \, Y_2'
+\kappa \, Z_2'\,, \\
\mathcal{E}_{Y_2} &= \ddot{Y_2} - (1 - \bb_0^2)\, Y_2''
+ m_2^2\, Y_2- 2 \bb_0\, \dot{Y_2}'
- 2\ga_3\, \dot{Y_1}+  (\Delta+2 \bb_0 \ga_3)\, Y_1'
+\kappa \, Z_1'\,,
\end{align}
and
\begin{equation}
\mathcal E_{W_j} = \ddot{W}_j - (1-\bb_0^2) W_j'' - 2 \bb_0 \dot{W}_j'\,.
\end{equation}
We found it convenient to define the auxiliary quantities 
\begin{equation}
\label{eq:Delta}
\begin{aligned}
    \kappa &= (\gamma_1 - \gamma_2 + \gamma_3)\bb_1-(\gamma_1 + \gamma_2 + \gamma_3)\bb_2 = 2 z_{+,5}\,, \\
    \Delta &=2 \bb_0 \ga_1-\bb_q(\ga_1+\ga_2+\ga_3) = 2 y_{+,5}\,.
  \end{aligned}
\end{equation}
The equations of motion are best solved in momentum space, replacing the fields with their Fourier transform.
In the massless sector we have 
\begin{equation}
W_j(\tau,\sigma) = \int {\frac{dp}{\sqrt{2\pi}}} \, (e^{-i \omega \tau + i p \sigma} w_j(p) a_j(p) + \text{h.c.})\,, 
\end{equation}
where $\text{h.c.}$ denotes the hermitian conjugate (to have real fields), and the dispersion relation is
\begin{equation}
\omega(p) = |p|-\bb_0 p\,.
\end{equation}
The two roots obey $(\omega+\bb_0p)^2=p^2$, the branch chosen above has positive norm for $p\neq0$. The particles remain massless, with characteristic velocities $1-\bb_0$ and $-1-\bb_0$.

In the massive sector, we gather all the massive fields into $\mathbf{X} = (Z_1,Z_2,Y_1,Y_2)$, and make the plane-wave Ansatz (we shall be concerned about reality and integration over momentum only later)
\begin{align} 
\label{eq:Fourier}
\mathbf{X} = e^{-i \omega \tau + i p \sigma} \mathbf{c}\,,
\end{align}
where $\omega$ is the energy of the excitation and $p$ is its momentum, while $\mathbf{c}$ are coefficients.
Plugging this Ansatz into the equations of motion, one obtains the following system of linear equations for the coefficients:
\begin{equation}
\label{eq:eomomega}
   M(p, \omega) \mathbf{c}=0, \qquad M(p, \omega)=\begin{pmatrix}
        m_{11} & m_{12} & 0 & m_{14} \\
   -m_{12} & m_{22} & m_{14} & 0 \\
   0 & -m_{14} & m_{11} & m_{12} \\
   -m_{14} & 0 & -m_{12} & m_{22}
    \end{pmatrix}\,,
\end{equation}
where
\begin{equation}
\begin{aligned}
m_{11} &= -(\omega+\bb_0 p)^2 + p^2 + m_1^2\,, \qquad 
m_{22} = -(\omega+\bb_0 p)^2 + p^2 + m_2^2\,, \\
m_{12} &=- 2 i \gamma_3 (\omega+\bb_0 p)-i \Delta p\,, \qquad 
m_{14} =- i \kappa p\,.
\end{aligned}
\end{equation}
Note that for real energy and momentum, $m_{11}$ and $m_{22}$ are real, whereas $m_{12}$ and $m_{14}$ are imaginary, making $M$ a hermitian matrix, $M^\dagger=(M^*)^T=M$.
The dispersion relation relating the excitation's energy $\omega$ and the excitation's momentum $p$ follows from imposing that the linear system of equations~\eqref{eq:eomomega} does admit non-trivial solutions, which requires the determinant of the matrix $M$ to vanish,
\begin{equation} 
\label{eq:detM}  
\text{det} M = (m_{11}m_{22}+m_{12}^2+m_{14}^2)^2=0\,.
\end{equation}
This leads to a quartic equation in $\omega$
\begin{equation}
\label{eq:quartic}
\omega^4 + \mathfrak{A} \, \omega^3
+ \mathfrak{B} \, \omega^2
+ \mathfrak{C} \, \omega
+ \mathfrak{D} = 0\,,
\end{equation}
with coefficients 
\begin{equation}
\begin{aligned}
    \mathfrak{A}&=4\bb_0 p\,, \qquad 
    \mathfrak{B}=-2((1-3\bb_0^2)p^2+\ga_1^2-\ga_2^2+\ga_3^2)\,, \\
    \mathfrak{C}&=-4 p( \bb_0 (1 - \bb_0^2)p^2 +
  (\gamma_1 + \gamma_2 + \gamma_3) ( 
    \bb_0 (\gamma_1 - \gamma_2 + \gamma_3)-\bb_q \gamma_3))\,, \\
    \mathfrak{D}&=  ((1-\bb_0^2)p^2 + \gamma_1^2 -\gamma_2^2 - \gamma_3^2)^2 - 4 \gamma_2^2 \gamma_3^2 -\kappa^2  p^2 -  (\Delta+2 \bb_0 \ga_3)^2 p^2 \,.
\end{aligned}
\end{equation}
The roots of the quartic polynomial equation~\eqref{eq:quartic} are given by Ferrari’s formula. The linear system of equations~\eqref{eq:eomomega} can be used to eliminate two coefficients, e.g.~$c_2$ and $c_4$, through (assuming that $m_{22}\neq0$)
\begin{equation}
\label{eq:solcoeffs24}
c_2 = \frac{m_{12}}{m_{22}} c_1 - \frac{m_{14}}{m_{22}} c_3\,, \qquad c_4 = \frac{m_{12}}{m_{22}} c_3 + \frac{m_{14}}{m_{22}} c_1\,,
\end{equation}
from which it follows that (here and after we assume that the energies $\omega$ and the momenta $p$ are real)
\begin{equation}
|c_2|^2 + |c_4|^2 = \frac{m_{11}}{m_{22}} (|c_1|^2 + |c_3|^2)\,. 
\end{equation}

\paragraph{Selecting the energy branches.}
The quartic equation~\eqref{eq:quartic} is invariant under $(\omega,p) \rightarrow (-\omega,-p)$, which ensures that if $\omega_s(p)$ is a solution then so is $\omega_{s'}(p) = -\omega_s(-p)$. This is crucial to be able to construct real linear combinations of solutions, but it also means that we should understand which roots $\omega_s(p)$ correspond to particles. Usually, this is done by requiring positive energy $\omega(p)>0$. For instance, in the case of the Klein-Gordon equation with dispersion relation $\omega^2 = p^2 + m^2$, it is natural to choose the positive branch $\omega(p) = + \sqrt{p^2 + m^2}$ as the one being associated with a particle, while the negative branch $-\omega(-p) = - \sqrt{p^2 + m^2}$ corresponds to a negative-norm state. In our case however, as we will see, there are ranges of $p$ for which both $\omega(p)$ and $-\omega(-p)$ are negative, and also ranges of $p$ for which both are positive. It may then be unclear which solution is associated to a particle, and whether this notion is momentum-dependent.
Let us therefore do a careful analysis. 

Upon quantisation, the oscillator expansion of the fields takes the form
\begin{equation}
\mathbf{X}(\tau, \sigma) = \int {\frac{dp}{\sqrt{2\pi}}} \sum_{s \in S} \left( e^{-i \omega_s(p) \tau + i p\sigma} \mathbf{c}^{(s)}(p) \, a_s(p) + \text{h.c.} \right)\,,
\end{equation}
where $\mathbf{c}^{(s)}(p) = \left. \mathbf{c} \right|_{\omega = \omega_s(p)}$, the sum is over solutions to the dispersion relation and $\text{h.c.}$ denotes the hermitian conjugate. The operator $a_s(p)$ and its conjugate $a_s^\dagger(p)$ respectively annihilate and create an excitation with energy $\omega_s(p)$.  They should satisfy the canonical commutation relations
\begin{equation}
{[a_r(p), a_{s}^\dagger(p')] = \delta(p-p') \delta_{rs}\,.}
\end{equation}
One can check that the canonically-conjugate momenta can be written
\begin{equation}
\mathbf{P} =  \int{\frac{dp}{\sqrt{2\pi}}} \sum_{s \in S} \left(e^{- i \omega_s(p) \tau + i p \sigma} \left(\frac{i}{2} \frac{\partial M}{\partial \omega}  + \tilde{M} \right) \mathbf{c}^{(s)}(p) a_s(p) + \text{h.c.} \right) \,,
\end{equation}
where
\begin{equation}
\tilde{M}(p)= \begin{pmatrix}
0 & 0 & i p \bb_1 & 0 \\
0 & 0 & 0 & i p \bb_2 \\
-i p \bb_1 & 0 & 0 & 0 \\
0 & -i p \bb_2 & 0 & 0
\end{pmatrix}\,, \qquad \tilde{M}^\dagger = \tilde{M}\,, \qquad \tilde{M}^t = \tilde{M}^\star = -\tilde{M}\,.
\end{equation}
We define the symplectic pairing
\begin{equation}
\begin{aligned}
\Omega_{s}(p,p') &= \int d\sigma \left( \frac{\delta \mathbf{X}(\tau,\sigma)}{\delta a_s(p)} \cdot \frac{\delta \mathbf{P}(\tau,\sigma)}{\delta a_s^\dagger(p')} - \frac{\delta \mathbf{P}(\tau,\sigma)}{\delta a_s(p)} \cdot  \frac{\delta \mathbf{X}(\tau,\sigma)}{\delta a_s^\dagger(p')}  \right) \,,
\end{aligned}
\end{equation}
which, by virtue of the equations of motion, does not depend on $\tau$. Moreover, canonical normalisation imposes that $\Omega_{s}(p,p') = {i} \delta(p-p')$.
Inserting the oscillator expansion and performing the integral over $\sigma$, one sees that the contribution of $\tilde{M}$ drops out and one is left with
\begin{equation}
\Omega_s(p,p')  = {i} \delta(p-p')  N_s(p)\,, \qquad N_s(p) = -(\mathbf{c}^{(s)} (p))^\dagger \left.  \frac{\partial M}{\partial \omega} \right|_{\omega=\omega_s(p)}  \mathbf{c}^{(s)}(p)\,.
\end{equation}
Note that $M^\dagger = M$ for real $\omega$ and $p$, so that $N_s(p) \in \mathbb{R}$. 
To have canonically-normalised fields we require that $N_s(p)=1$, and in particular $\text{sgn}[N_s(p)]=+1$. This then indicates which $\omega_s(p)$ correspond to positive-norm states, which ones correspond to negative-norm states, and it also (partially) fixes the normalisation of the coefficients $c_{j}^{(s)}(p)$. Concretely, for our theory we have that
\begin{equation}
\label{eq:normalisation}
N_s = f(\omega_s(p))(|c_1^{(s)}|^2+|c_3^{(s)}|^2), \qquad f(\omega) = \frac{2 }{m_{22}} ((\omega + \bb_0 p) (m_{11} + m_{22}) + 2 i \ga_3 m_{12})\,,
\end{equation}
where we have taken advantage of the relations~\eqref{eq:solcoeffs24} as well as the determinant condition~\eqref{eq:detM} to simplify expressions involving the coefficients of $M$. Valid solutions $\omega_s(p)$ need to satisfy $f(\omega_s(p))>0$.

\paragraph{Fixing the relative angle.}
Each frequency root $\omega_s(p)$ has multiplicity two and there are therefore two independent polarisations, which we label $z$ and $y$. We then parametrise their independent coefficients as
\begin{equation}
(c_1^{(r)},c_3^{(r)})
=\frac{1}{\sqrt{f(\omega_s)}}
(\cos\theta_r,\sin\theta_r)\,,
\qquad r=z,y\,,
\end{equation}
where the angle might also depend on the momentum.
Equation~\eqref{eq:normalisation} then ensures that each
polarisation has unit symplectic norm. To describe independent
canonical oscillators, however, the two polarisations must also
be symplectically orthogonal. This leads to the additional condition
\begin{equation}
\begin{aligned}
0 &=-\mathbf{c}^{(z)\dagger}
\left.\frac{\partial M}{\partial\omega}\right|_{\omega=\omega_s}
\mathbf{c}^{(y)}
\\
&=f(\omega_s(p))\left[
(c_1^{(z)})^*c_1^{(y)}
+(c_3^{(z)})^*c_3^{(y)}
\right]
\\
&=\cos\theta_z\cos\theta_y+\sin\theta_z\sin\theta_y
=\cos(\theta_z-\theta_y)\,.
\end{aligned}
\end{equation}
Consequently,
\begin{equation}
\theta_y-\theta_z=\frac{\pi}{2}\pmod\pi\,.
\end{equation}
Within this real parametrisation, the remaining freedom is a
common rotation of the two orthonormal polarisations, and we can write
\begin{equation}
\label{eq:orthonormal-oscillators}
\begin{pmatrix}
c_1^{(z,s)}&c_1^{(y,s)}\\
c_3^{(z,s)}&c_3^{(y,s)}
\end{pmatrix}
=
\frac{1}{\sqrt{f(\omega_s)}}
\begin{pmatrix}
\cos\theta_s&\sin\theta_s\\
-\sin\theta_s&\cos\theta_s
\end{pmatrix}\,.
\end{equation}
The remaining angle $\theta_s$ may depend on the frequency
branch and momentum.

\paragraph{Oscillator expansion.}
Taking into account the multiplicity of the roots, and denoting by $S$ the space of solutions with positive symplectic norm, one has dispersion relations
\begin{equation}
\{\omega_s | s \in S\} = \{\omega_{z,\pm}, \omega_{y,\pm} \}\,, \qquad \omega_{z,\pm} = \omega_{y,\pm} = \omega_\pm\,,  
\end{equation}
operators
\begin{equation}
\{a_s, s\in S\} = \{a_{z,\pm}, a_{y,\pm} \}\,, \qquad \{a_s^\dagger, s\in S\} = \{a_{z,\pm}^\dagger, a_{y,\pm}^\dagger \}\,,
\end{equation}
and coefficients
\begin{equation}
\{ c^{(s)}_j | s \in S\} = \{ c^{(z,\pm)}_j, c^{(y,\pm)}_j \}\,.
\end{equation}
Finally, the oscillator expansion of the fields in its full glory is
\begin{equation}
\begin{aligned} 
\label{eq:oscillators}
Z_1 = \int {\frac{dp}{\sqrt{2\pi}}}\, ( & e^{-i \omega_{+} \tau + i p \sigma}(c_{1}^{(z,+)} a_{z,+} +c_{1}^{(y,+)} a_{y,+})+ e^{-i \omega_- \tau + i p \sigma}(c_{1}^{(z,-)} a_{z,-} +c_{1}^{(y,-)} a_{y,-})+\text{h.c.})\,, \\
Z_2 = \int {\frac{dp}{\sqrt{2\pi}}}\, ( & e^{-i \omega_{+} \tau + i p \sigma}(c_{2}^{(z,+)} a_{z,+} +c_{2}^{(y,+)} a_{y,+})+ e^{-i \omega_- \tau + i p \sigma}(c_{2}^{(z,-)} a_{z,-} +c_{2}^{(y,-)} a_{y,-})+\text{h.c.})\,, \\
Y_1 = \int {\frac{dp}{\sqrt{2\pi}}}\, ( & e^{-i \omega_{+} \tau + i p \sigma}(c_{3}^{(z,+)} a_{z,+} +c_{3}^{(y,+)} a_{y,+})+ e^{-i \omega_- \tau + i p \sigma}(c_{3}^{(z,-)} a_{z,-} +c_{3}^{(y,-)} a_{y,-})+\text{h.c.})\,, \\
Y_2 = \int {\frac{dp}{\sqrt{2\pi}}}\, ( & e^{-i \omega_{+} \tau + i p \sigma}(c_{4}^{(z,+)} a_{z,+} +c_{4}^{(y,+)} a_{y,+})+ e^{-i \omega_- \tau + i p \sigma}(c_{4}^{(z,-)} a_{z,-} +c_{4}^{(y,-)} a_{y,-})+\text{h.c.})\,.
\end{aligned}
\end{equation}
The coefficients obey the relations~\eqref{eq:orthonormal-oscillators} and~\eqref{eq:solcoeffs24}. The operators $a_{z,\pm}^\dagger$ and $a_{y,\pm}^\dagger$ create excitations with the same dispersion relation $\omega_\pm(p)$. In the undeformed limit, in a basis with $\theta=0$, these become the $AdS_3$ and $S^3$ excitations, respectively.
The coupling of the $AdS_3$ and $S^3$ sectors induced by the $B$-field is represented by the $c_{1,2}^{(y,\pm)}$ coefficients in $Z_1$ and $Z_2$, and by the $c_{3,4}^{(z,\pm)}$ coefficients in $Y_1$ and $Y_2$. These vanish in the undeformed limit for $\theta=0$. 

The quadratic Hamiltonian takes the canonical form 
\begin{equation}
H_2 =  \int dp \sum_{s \in S} \frac{\omega_s(p)}{2} (a^\dagger_s(p) a_s(p) + a_s(p) a^\dagger_s(p)) = \int dp \sum_{s \in S} \omega_s(p) a^\dagger_s(p) a_s(p) + E_{\rm vac}\,.
\end{equation}
Here $E_{\rm vac}$ is the contribution to the vacuum energy, and one has the identity $[H_2,a^\dagger_s(p)]=\omega_s(p)a^\dagger_s(p)$. The worldsheet momentum is
\begin{equation}
P_{w.s} = \int d\sigma \acute{X}^- = \int d\sigma (-P_j \acute{X}_j) = \int dp \sum_{s \in S} p \, a_s^\dagger(p) a_s(p)\,.
\end{equation}
Closed string states obey periodic boundary conditions and physical states must obey the level-matching condition
\begin{equation}
P_{w.s.} =0\,.
\end{equation}

\subsection{Dispersion relations}
In the above analysis we have always assumed that $\omega_\pm(p)$ are real energies. 
This is only true in some range of the deformation parameters. 
To show this, let us analyse the dispersion relations in some special cases.

\paragraph{Rational pure RR.}
In the rational model with no B-field, the quartic equation~\eqref{eq:quartic} simplifies to
\begin{equation}
\omega^4 -2 (p^2 + 2) \omega^2+  p^4=0\,,
\end{equation}
and the roots with positive symplectic norm are given by
\begin{equation}
\begin{aligned}
\label{eq:omegarational}
\omega_\pm(p) = \sqrt{p^2  +1} \pm 1\,.
\end{aligned}
\end{equation}
This dispersion relation is not the one usually found in the literature for perturbative pure RR $AdS_3 \times S^3 \times T^4$ strings~\cite{Hoare:2013pma}, which instead would be the relativistic dispersion relation $\omega(p) = \sqrt{p^2+ 1}$ without the $\pm 1$ shift. This is due to the choice of a different light-cone gauge-fixing~\cite{Borsato:2023oru,Hoare:2023zti}. Recall that in order to fix the gauge in the deformed theory, due to the large amount of broken symmetries, we are forced to use light-cone coordinates constructed using $T,\Phi$, which are $\mathfrak{u}(1)$ isometries sitting entirely inside one copy (the undeformed one) of the original $\mathfrak{psu}(1,1|2) \oplus \mathfrak{psu}(1,1|2)$ superalgebra. 
On the other hand, when quantising the undeformed theory in the usual BMN light-cone gauge~\cite{Berenstein:2002jq}, one uses global time $t \in AdS_3$ and angle $\phi \in S^3$, which sit in diagonal $\mathfrak{u}(1)$s of $\mathfrak{psu}(1,1|2) \oplus \mathfrak{psu}(1,1|2)$. This different choice of geodesics for the light-cone gauge-fixing results in a shift $\pm1$ of the excitations' energies~\cite{Hoare:2023zti}. The advantage of expanding around the BMN vacuum is that the light-cone gauge-fixed theory preserves half (eight) of the original (sixteen) supersymmetries, and that the worldsheet Hamiltonian is positive semidefinite. This follows from the BPS bound of the $\mathfrak{psu}(1,1|2)$ superalgebra. 
Since we consider a different light-cone gauge-fixing, the positivity argument does not apply anymore. 
However, the dispersion relation~\eqref{eq:omegarational} still satisfies $\omega_\pm(p) \geq 0$. The negative branch of the dispersion relation becomes gapless at zero momentum, $\omega_-(p=0)=0$.
For the normalisation of the coefficients, one gets
\begin{equation}
f(\omega_\pm(p)) = 4 \sqrt{1+p^2} >0\,, 
\end{equation} 
so that $\omega_\pm(p)$ are indeed the correct solutions to consider to describe particles.

\begin{figure}[!htb]
    \centering
    \captionsetup{font={color=black},labelfont={color=black}}
    \begin{subfigure}[t]{0.49\textwidth}
        \centering
        \includegraphics[height=5 cm]{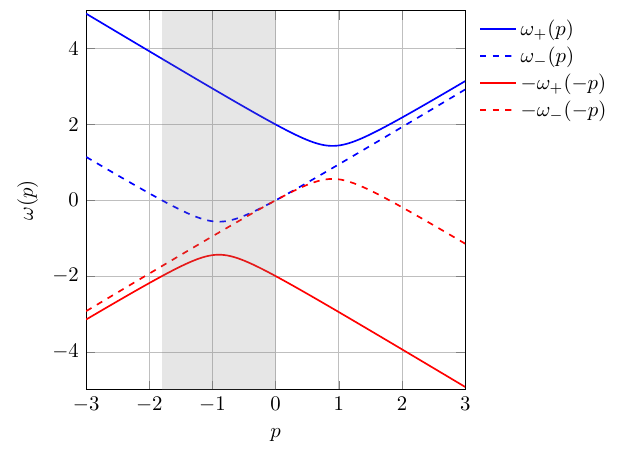}
        \caption{All four frequency branches.}
        \label{fig:rational-wz-all-four}
        \label{fig:omegaratWZ}
    \end{subfigure}\hfill
    \begin{subfigure}[t]{0.49\textwidth}
        \centering
        \includegraphics[height=5 cm]{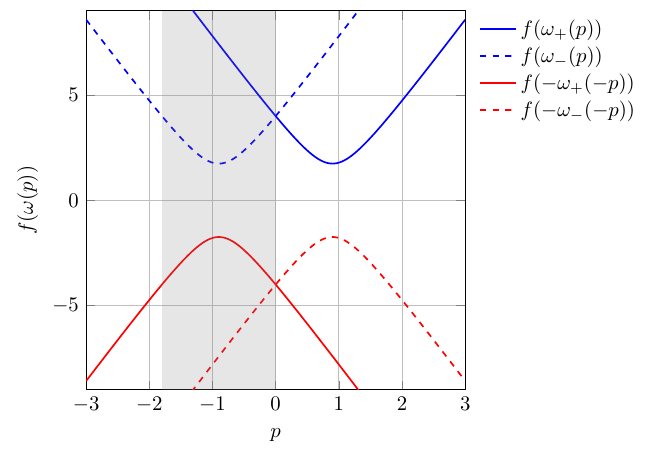}
        \caption{All four symplectic norms.}
        \label{fig:rational-wz-all-four-norms}
    \end{subfigure}
    \caption{Energies and symplectic norms for the rational+WZ model
    at $\delta_q=0.9$. Matching colours and line styles identify the
    same mode in the two panels: blue denotes positive-norm solutions,
    red negative-norm solutions, and solid/dashed lines denote the $+/-$ labels.
    The shaded interval $-1.8=-2 \bb_q<p<0$ has only one positive branch,
    although both particle branches retain positive norm. In particular,
    in that region the blue dashed branch has $\omega_-(p)<0$ but
    $f(\omega_-(p))>0$.}
    \label{fig:rational-wz-dispersion-comparison}
\end{figure}

\paragraph{Rational+WZ.} 
In the rational case, as discussed in Section~{\ref{sec:sugra}}, if one is interested in viewing this bosonic model as the bosonic truncation of a superstring theory, the supergravity equations and the reality of the fluxes tell us that $\mathbf{x}^{(1)}=\mathbf{x}^{(2)}=\mathbf{x}^{(3)}=0$ and therefore, among the four B-fields considered in this paper, only the one corresponding to adding a WZ term, given by $\mathsf{b}_V$ and tuned by $x^{(4)}_5 \sim \bb_q$, can be turned on. 
The quartic dispersion relation then simplifies to
\begin{equation}
\omega^4 - 2(p^2+2) \omega^2 + 8 p \bb_q \omega + p^2 (p^2-4 \bb_q^2)=0\,,
\end{equation}
and the positive-norm solutions are
\begin{equation}
\label{eq:omegarationalx4}
\begin{aligned}
 \omega_\pm(p) = \sqrt{p^2 \mp 2 p \bb_q +1} \pm 1 \,.
\end{aligned}
\end{equation}
The WZ term breaks parity invariance and the dispersion relation $\omega_\pm(-p) \neq \omega_\pm(p)$. 
The energies are real in the range $|\bb_q| \leq 1$. Let us continue by assuming that $0 \leq \bb_q \leq 1$. When $\bb_q=0$ one recovers the pure RR dispersion relation~\eqref{eq:omegarational}, while when $\bb_q=1$ one gets the piece-wise linear dispersion relation $\omega_\pm(p) = |p \mp 1| \pm 1$.
Interestingly, the energies become negative in some range of the momentum, see figure~\ref{fig:omegaratWZ}. Note that this is a direct consequence of analysing the mixed-flux theory in a different light-cone gauge, we have not introduced any deformation parameter yet.  
One has gapless excitations $\omega_-(p=0)=\omega_-(p=-2 \bb_q)=0$ and $\omega_-(p) < 0$ for $p \in (-2 \bb_q,0)$. 
The normalisation of the coefficients~\eqref{eq:normalisation} remains however positive, 
\begin{equation}
f(\omega_\pm(p)) = 4 \sqrt{1+p^2 \mp 2 p \bb_q}\,,
\end{equation}
with $f(\omega_\pm(p))>0$ $\forall p$ when $0 \leq \bb_q <1$.  However, because $f(\omega_\pm(p=\pm 1))=0$ when $\bb_q=1$, the pure NSNS case needs to be treated carefully, as the normalisation of the modes becomes singular for some value of the momentum.
In Figure~\ref{fig:rational-wz-dispersion-comparison} we plot the four
solutions to the quartic equation for the energy, as well as the value of $f(\omega)$.

The single-particle spectrum~\eqref{eq:omegarationalx4} is bounded from below, but in principle one can construct a state with arbitrarily many bosons occupying a negative-energy mode, making the quadratic Hamiltonian unbounded below on the unrestricted Fock space. Note however that it is not possible to construct such a state while also satisfying the level-matching condition. Indeed, both branches obey $\omega_\pm(p)\geq\delta_qp$, so level-matched multi-particle states have non-negative energy. In other words, the shifted operator $H_2 - \delta_q P_{w.s.}$ is positive semidefinite.

\begin{figure}[!htb]
    \centering
    \captionsetup{font={color=black},labelfont={color=black}}
    \includegraphics[width=0.60\linewidth]{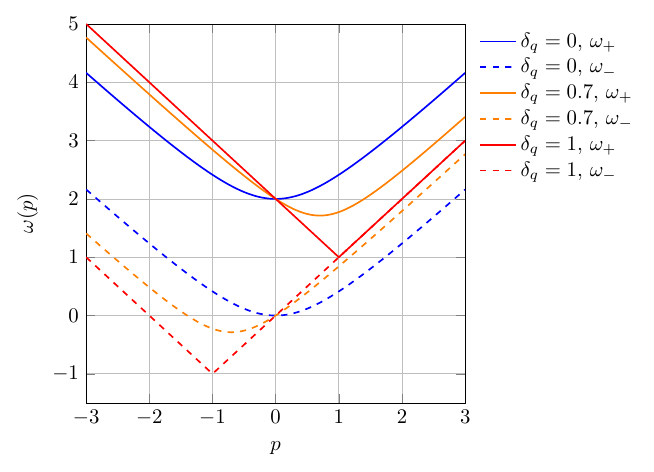}
    \caption{Positive-norm solutions to the dispersion relation in the rational+WZ model
    for $\delta_q=0,0.7,1$. Colour distinguishes $\delta_q$ while
    solid and dashed lines denote $\omega_+$ and $\omega_-$, respectively.
    At $\delta_q=1$, the dispersions become piecewise linear,
    $\omega_\pm(p)=|p\mp1|\pm1$. The corresponding norms are positive
    except at the corners $p=\pm1$, where they vanish.}
    \label{fig:rational-wz-particle-branches}
\end{figure}

\paragraph{Elliptic pure RR.}
In the pure RR elliptic case, the quartic equation reduces to
\begin{equation}
\omega^4-2 (p^2 + \gamma_1^2 -\gamma_2^2 + \gamma_3^2)\omega^2+ (p^2 + \gamma_1^2 - \gamma_2^2 - \gamma_3^2)^2 - 4 \gamma_2^2 \gamma_3^2 =0\,, 
\end{equation}
and the solutions with positive symplectic norm are
\begin{equation}
\omega_\pm(p) = \sqrt{p^2 + \gamma_1^2 + \gamma_3^2 - \gamma_2^2 \pm 2 \gamma_3 \bar{\omega}(p)}\,, \qquad \bar{\omega}(p) = \sqrt{p^2 + \gamma_1^2}\,,
\end{equation}
which can also be written
\begin{equation}
\omega_\pm(p) = \sqrt{(\bar{\omega}(p) \pm \gamma_3 +\gamma_2)(\bar{\omega}(p) \pm \gamma_3 - \gamma_2)}\,.
\end{equation}
One can check that the rational limit $\gamma_1=\gamma_3=1$, $\gamma_2=0$ reproduces the dispersion relation~\eqref{eq:omegarational}. 
The energies $\omega_\pm(p)$ are real and nonnegative for all $p \in \mathbb{R}$ in the range~\eqref{eq:rangegammasugra}, with a single minimum in $p=0$. Note that the dispersion relations $\omega_\pm(p)$ are exchanged under $\gamma_3 \rightarrow -\gamma_3$ so that one can assume $\gamma_3 \geq 0$ without loss of generality, and one then has that $\omega_-(p) \leq \omega_+(p)$.

\begin{figure}[t]
    \centering

    \begin{subfigure}[t]{0.49\linewidth}
        \centering
        \includegraphics[height=5.2 cm]{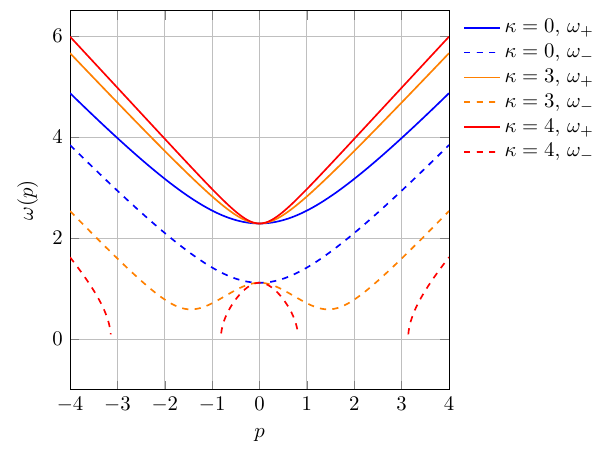}
        \caption{Different values of $\kappa$.}
        \label{fig:omegaplusminusellipticeta}
    \end{subfigure}
    \hfill
    \begin{subfigure}[t]{0.49\linewidth}
        \centering
        \includegraphics[height=5.2 cm]{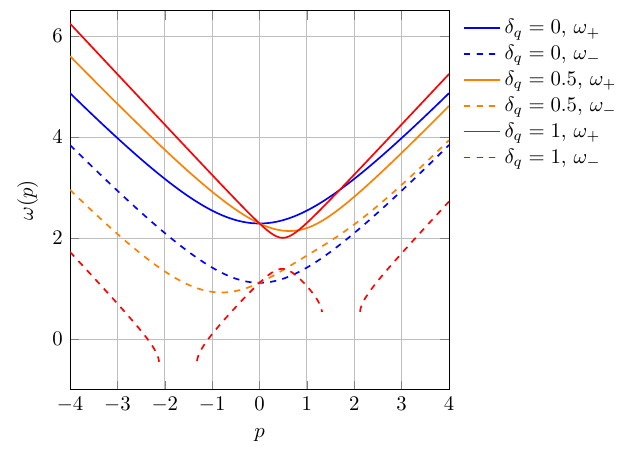}
        \caption{Different values of $\bb_q$.}
        \label{fig:omegapmWZ}
    \end{subfigure}

    \caption{Dispersion relations $\omega_{\pm}(p)$ for $\gamma_1=2$, $\gamma_2=1$, and $\gamma_3=\frac{1}{2}$ with~\ref{fig:omegaplusminusellipticeta} different values of $\kappa$ and~\ref{fig:omegapmWZ} different values of $\bb_q$.  For $\kappa=0$ the lower branch of the dispersion relation has a local minimum in $p=0$. This local minimum persists as long as $|\kappa|< \kappa_{\min}$ as given in~\eqref{eq:Deltamin}. For our choice of parameters this is $\kappa_\text{min} \approx 1.73$. Above that value, the point $p=0$ becomes a local maximum and there are two local minima developing at the two momenta $p^\star$ given in~\eqref{eq:pstar}. The bound~\eqref{eq:bound1} above which the dispersion relation becomes complex for some values of $p$ is $\kappa_0 \approx 3.26$. The plot on the right, corresponding to the dispersion relation in the elliptic+WZ model, has similar features but is no longer invariant under $p \rightarrow -p$, and energies may become negative for some range of momenta. }
    \label{fig:omegapm}
\end{figure}

\paragraph{Elliptic+$(\bb_1,\bb_2)$.}
Let us now consider the elliptic deformation supplemented with B-fields $\mathsf{b}_1$ and $\mathsf{b}_2$. 
Note that $\kappa$ as defined in~\eqref{eq:Delta} is the parameter toggling the B-field at this order, so that the case where $\bb_1$ is turned on and the case where $\bb_2$ is turned on can be discussed on an equal footing. The quartic equation~\eqref{eq:quartic} becomes
\begin{equation}
\omega^4 -2 (p^2 + \gamma_1^2 -\gamma_2^2 + \gamma_3^2)\omega^2 +  (p^2 + \gamma_1^2 - \gamma_2^2 - \gamma_3^2)^2 - (4 \gamma_3^2 \gamma_2^2 +\kappa^2p^2)=0\,.
\end{equation}
Solving for the dispersion relation gives
\begin{equation}
\label{eq:dispE12}
    \omega_\pm(p)=\sqrt{p^2+\ga_1^2-\ga_2^2+\ga_3^2\pm \sqrt{(4 \ga_3^2 + \kappa^2) p^2 + 4 \ga_1^2 \ga_3^2}}\,,
\end{equation}
The dispersion relation is parity-invariant, with $\omega_\pm(-p) = \omega_\pm(p)$. The energies $\omega_\pm(p)$ are real and nonnegative for all $p \in \mathbb{R}$ in the range
\begin{equation}
\begin{aligned}
&\ga_1^2\ge(1+\tilde{\kappa}^2)^2\ga_3^2\,, \qquad  (\ga_1\pm\ga_3)^2\ge\ga_2^2\,, \qquad \tilde{\kappa} = \frac{\kappa}{2 \gamma_3}\,, \\
&\ga_1^2<(1+\tilde{\kappa}^2)^2\ga_3^2\,, \qquad \tilde{\kappa}^2\gamma_1^2\ge(1+\tilde{\kappa}^2)\big(\gamma_2^2+\tilde{\kappa}^2\gamma_3^2\big)\,.
\end{aligned}
\end{equation}
Together with~\eqref{eq:rangegamma} and~\eqref{eq:rangegammasugra}, the allowed ranges become
\begin{equation}
\label{eq:boundkappa}
\begin{aligned}
&\gamma_3 >0\,, \qquad \gamma_1 \pm \gamma_2 - \gamma_3 \geq 0\,, \qquad -\kappa_0 \leq \kappa \leq \kappa_0\,, \\
&\gamma_3 < 0\,, \qquad \gamma_1 \pm \gamma_2 +\gamma_3>0\,, \qquad -\kappa_0 \leq \kappa \leq \kappa_0\,, \\
&\gamma_3=0\,, \qquad \gamma_1 \pm \gamma_2>0\,, \qquad -\kappa_0 \leq \kappa \leq \kappa_0\,,
\end{aligned}
\end{equation}
with
\begin{equation}
\label{eq:bound1}
\kappa_0 = m_1 + m_2 \,.
\end{equation}
This corresponds to the bound $\kappa^2 \leq (m_1+m_2)^2$ of~\eqref{eq:non-strict stability}, which is one of the necessary bounds for a supergravity completion with real fluxes.
In figure~\ref{fig:omegaplusminusellipticeta} we plot the dispersion relations $\omega_\pm(p)$ for fixed $\gamma_{1,2,3}$ and different values of $\kappa$. We clearly see the breakdown of reality when the B-field parameter $\kappa$ takes values greater than the above bound.  

For $\gamma_1,\gamma_3>0$ and a strictly positive gap
$\omega_-(0)>0$, the lower branch is smooth and even in $p$,
and therefore has a stationary point at $p=0$. Explicitly,
\begin{equation}
\label{eq:Deltamin}
\begin{aligned}
\omega_-(0)
&=\sqrt{(\gamma_1-\gamma_3)^2-\gamma_2^2}>0\,,
\qquad \omega_-'(0)=0\,,
\\
\omega_-''(0)
&=\frac{\kappa_\text{min}^2 - \kappa^2}{4 \ga_1 \ga_3 \omega_-(0)}\,,
\qquad
\kappa_{\mathrm{min}}
=2 \sqrt{\gamma_3(\gamma_1-\gamma_3)}\,.
\end{aligned}
\end{equation}
Here the parameter restrictions ensure
$\gamma_1>\gamma_3$ in the gapped regime.
The point $p=0$ is a local minimum for
$|\kappa|<\kappa_{\mathrm{min}}$ and a local maximum for
$|\kappa|>\kappa_{\mathrm{min}}$, provided the frequencies
remain real.
At $|\kappa|=\kappa_{\mathrm{min}}$, the quadratic curvature
vanishes, but
\begin{equation}
\omega_-(p)
=\omega_-(0)
+\frac{p^4}{8\gamma_1\gamma_3\,\omega_-(0)}
+O(p^6)\,,
\end{equation}
so $p=0$ remains a local minimum, with a leading quartic
variation. 

For $|\kappa|>\kappa_{\mathrm{min}}$, the two local minima
occur at
\begin{equation}
\label{eq:pstar}
p = \pm p^\star\,, \qquad p^\star = \sqrt{\frac{4 \ga_3^2 + \kappa^2 }{4}- \frac{4 \ga_1^2 \ga_3^2}{4 \ga_3^2 + \kappa^2}}\,.
\end{equation}
In that regime, the lowest-energy excitations no longer have zero momentum, but instead have $p=\pm p^\star$. 

The gapless case $\omega_-(0)=0$ must be treated separately. 
Let us consider the special case of the space-like trigonometric deformation with
\begin{equation}
\alpha_0=\alpha_1=\alpha\,, \qquad \alpha_2=1\,, \qquad \gamma_1 =1\,, \qquad \gamma_2 = \frac{1}{\alpha}-1\,, \qquad \gamma_3 = \frac{1}{\alpha}\,, \qquad \gamma_3 = \gamma_1+\gamma_2\,.
\end{equation}
We restrict to the regime~\eqref{eq:rangealpha} in which the bosonic theory can be embedded into supergravity, namely $\alpha \geq 1$.  
The dispersion relation $\omega_-(p)$ becomes gapless, with $\omega_-(p=0) = 0$, and it costs no energy to create such excitations at zero momentum.
The bounds in~\eqref{eq:Deltamin} and~\eqref{eq:bound1} degenerate to
\begin{equation}
\kappa_0 = \kappa_\text{min}  = \frac{2 \sqrt{\alpha-1}}{\alpha}\,,
\end{equation}
so that the excitations' energies remain real and nonnegative for all real momenta $p$, and with a single minimum at $p=0$, as long as $|\kappa| \leq 2\sqrt{\alpha-1}/\alpha$. 
Note that $|\kappa| =2\sqrt{\alpha-1}/\alpha $ saturates the bound and precisely corresponds to the TsT point for which the dispersion relation becomes
\begin{equation}
\omega_\pm(p) = \sqrt{p^2 + \frac{1}{\alpha}} \pm \frac{1}{\sqrt{\alpha}}\,.
\end{equation}
A similar observation holds for the space-like trigonometric deformation with
\begin{equation}
\alpha_0=\alpha_2=\alpha\,, \qquad \alpha_1=1\,, \qquad \gamma_1 =1 \,, \qquad \gamma_2 = 1-\frac{1}{\alpha}\,, \qquad \gamma_3 = \frac{1}{\alpha}\,, \qquad \gamma_3 = \gamma_1-\gamma_2\,.
\end{equation}

\paragraph{Elliptic+WZ.}
Finally we consider the elliptic model with WZ term. The solutions to the quartic equation cannot be written in a nice analytic form. Instead, to get some insights into their behaviour, we plot it for different values of the deformation parameter $\bb_q$ in figure~\ref{fig:omegapm}. 
The WZ term breaks parity invariance and the dispersion relation is no longer symmetric under $p \rightarrow -p$. There is again a value of $\bb_q$ beyond which reality is broken for some momenta $p$. 
As for the rational+WZ model, the energies may become negative for some values of $\bb_q$ and some momenta.

\section{The tree-level S-matrix}
\label{sec:Smatrix}

Now that we have diagonalised the quadratic Hamiltonian and obtained the spectrum of asymptotic one-particle states, we can study how these excitations interact.
The first interaction terms in the light-cone gauge-fixed theory appear at quartic order through $\mathcal L_4$, whose explicit form we present in appendix~\ref{app:L4}.
This quartic Lagrangian, as well as its counterpart, the quartic Hamiltonian density $\mathcal H_4$, can be rewritten in terms of oscillators using the mode expansion~\eqref{eq:oscillators}.
It takes the schematic form
\begin{equation}
\begin{aligned}
\mathcal H_4 = \int {\frac{d^4p}{(2\pi)^2}} & \Big(h_4^0(\vec{p})\,  a(p_4) a(p_3) a(p_2) a(p_1)\, e^{-i (\omega_1 + \omega_2 + \omega_3 + \omega_4) \tau } e^{i (p_1+p_2+p_3+p_4)\sigma} \\
&+ h_3^1(\vec{p})  \,  a^\dagger(p_4)  a(p_3) a(p_2) a(p_1) \, e^{-i (\omega_1 + \omega_2 + \omega_3 - \omega_4)\tau} e^{i (p_1+p_2+p_3-p_4)\sigma} \\
&+ h_2^2(\vec{p}) \,   a^\dagger(p_4) a^\dagger(p_3)  a(p_2) a(p_1)\, e^{-i (\omega_1 + \omega_2 - \omega_3 - \omega_4)\tau} e^{i (p_1+p_2-p_3-p_4)\sigma} + \text{h.c.} \Big)\,,
\end{aligned}
\end{equation}
where $d^4p=dp_1 dp_2 dp_3 dp_4$, $\omega_j=\omega(p_j)$, and we have suppressed all quantum numbers for readability (for instance, $a\in\{a_{z,\pm},a_{y,\pm}\}$). The coefficients $h_{n_i}^{n_f}(\vec{p})$ depend on the four momenta and correspond to scattering processes involving $n_i$ initial excitations and $n_f$ final excitations. 
The tree-level S-matrix defined through $\mathbb{S} = 1 + i \mathrm{T}^{-1} \mathbb{T}+ \dots$ is obtained through a double integration
\begin{equation}
\label{eq:Tmatrix}
\begin{aligned}
\mathbb{T} &= -\int d\tau H_4 = -\int d\tau d\sigma \, \mathcal H_4 & \\
&= -{\int} d^4p \Big( h_4^0(\vec{p})  a(p_4) a(p_3) a(p_2) a(p_1)  \, \delta(p_1+p_2+p_3+p_4) \delta(\omega_1 + \omega_2 + \omega_3 + \omega_4) \\
&\qquad +h_3^1(\vec{p}) \, a^\dagger(p_4) a(p_3) a(p_2) a(p_1) \, \delta(p_1+p_2+p_3-p_4) \delta(\omega_1 + \omega_2 + \omega_3 - \omega_4)\\
&\qquad +h_2^2(\vec{p})\, a^\dagger(p_4) a^\dagger(p_3) a(p_2) a(p_1) \, \delta(p_1+p_2-p_3-p_4) \delta(\omega_1 + \omega_2 - \omega_3 - \omega_4) +  \text{h.c.}
\Big)\,.
\end{aligned}
\end{equation}
Stability of the vacuum $\ket{0}$ and of physical one-particle states $a^\dagger(p) \ket{0}$ imposes constraints on the allowed scattering processes. Integrability constrains the tree-level S-matrix even further. In what follows we will often use the shorthand $\ket{z_\pm(p)} = a^\dagger_{z,\pm}(p) \ket{0}$ for the $AdS_3$ excitations and $\ket{y_\pm(p)} = a^\dagger_{y,\pm} \ket{0}$ for the $S^3$ excitations.

\subsection{Stability of the vacuum}
\label{subsec:vacuum-stability}
The coefficients $h_0^4(\vec{p})$ and $h_4^0(\vec{p})$ in the tree-level S-matrix~\eqref{eq:Tmatrix} correspond respectively to the production of four particles from the vacuum $0 \rightarrow 4$, or the annihilation of four particles into the vacuum $4 \rightarrow 0$. Note that in these two cases, the energy-momentum conservation implies the level-matching condition $p_1+p_2+p_3+p_4=0$ and therefore the scattered asymptotic states are physical. To have a stable vacuum, it is necessary for these two processes to vanish. For particles with massive relativistic dispersion relation $\omega(p) = \sqrt{p^2 + m^2}$, these processes are forbidden on kinematic grounds, as it is impossible to satisfy the energy conservation $\omega(p_1) + \omega(p_2) + \omega(p_3) + \omega(p_4)=0$ for real positive energies. In our elliptic case however, when $\delta_q \neq 0$ or $\delta_0 \neq 0$, the energies of single-particle states can become negative in some range of the momenta, and therefore the processes are no longer forbidden on kinematic grounds. 
In what follows, we specify a sufficient parameter region in which the perturbative vacuum cannot produce four particles.

To ensure the stability of the vacuum we would like to find a parameter $v \in \mathbb{R}$ such that $\omega > vp$ for all $p \in \mathbb{R}$ for positive-norm energies.
Energy conservation then excludes particle production, as it did for the massive relativistic case, because it leads to the contradiction
\begin{equation}
0 = \sum_j \omega_{s_j}(p_j) > v \sum_j p_j =0\,.
\end{equation}
This condition means that, while $H_2$ itself may not be positive-definite, it becomes so after an appropriate momentum shift $H_2 - v P_{w.s.}$. 
To find an appropriate $v$, let us consider a non-trivial polarisation vector $\mathbf{c}$ satisfying the on-shell constraint $M(p,\omega) \mathbf{c}=0$. Then, using that $\partial_\omega^2 M(p,\omega)= -2 \, 1_4$ one can check that
\begin{equation}
M(p,\omega) = M(p,vp) + (\omega-vp) \partial_\omega M(p,vp) - (\omega-vp)^2 1_4\,,
\end{equation}
and sandwiching with $\mathbf{c}$ gives the polynomial equation
\begin{equation}
\begin{aligned}
&0 = - (\omega- vp)^2 a + (\omega- vp) b  + c\,,\\
&a=\mathbf{c}^\dagger \mathbf{c}>0\,, \qquad b=\mathbf{c}^\dagger \partial_\omega M(p,vp) \mathbf{c} \in \mathbb{R}\,, \qquad  c=\mathbf{c}^\dagger M(p,vp) \mathbf{c} \in \mathbb{R}\,.
\end{aligned}
\end{equation}
The fact that $a>0$ follows from non-triviality of $\mathbf{c}$, while $b,c \in \mathbb{R}$ is a consequence of having a hermitian matrix $M$. The two solutions to the above quadratic equation are
\begin{equation}
\label{eq:solvp}
\omega-v p = \frac{b \pm \sqrt{b^2+4 ac}}{2 a}\,.
\end{equation}
On top, we also have the positive-norm constraint, which gives
\begin{equation}
N = -\mathbf{c}^\dagger \partial_\omega M(p,\omega) \mathbf{c} = -\mathbf{c}^\dagger \partial_\omega M(p,vp) \mathbf{c} +2 (\omega- vp) \mathbf{c}^\dagger  \mathbf{c} = -b+2 a(\omega-vp)>0\,, 
\end{equation}
so that
\begin{equation}
\mathbf{c}^\dagger M(p,vp)  \mathbf{c}  + (\omega- vp)^2 \mathbf{c}^\dagger \mathbf{c} = (\omega-v p) N\,.
\end{equation}
A sufficient condition to have that $\omega - v p >0$ is then to require that $M(p,v p)$ is positive-definite. This imposes $c >0$, selects the positive branch in~\eqref{eq:solvp}, and ensures that $\omega-vp >0$. Requiring positive-definiteness leads to the following two inequalities
\begin{equation}
\begin{aligned}
&(1-(v+\bb_0)^2) p^2 + m_1^2 >0\,, \\
&((1-(v+\bb_0)^2) p^2 + m_1^2)((1-(v+\bb_0)^2) p^2 + m_2^2 ) - (\Delta+2 (v+\bb_0) \ga_3)^2 p^2 - \kappa^2 p^2 >0\,,
\end{aligned}
\end{equation}
which need to be satisfied for all $p \in \mathbb{R}$. This is only possible if (assuming the generic elliptic case with $m_1>0$ and $m_2>0$)
\begin{equation}
\label{eq:vacuum-stability}
\frac{\Delta^2}{4 \ga_3^2+(m_1+m_2)^2 } + \frac{\kappa^2}{(m_1+m_2)^2} < 1\,,
\end{equation}
and a suitable choice is then
\begin{equation}
v = -\bb_0 - \frac{2 \ga_3 \Delta}{4 \ga_3^2 + (m_1 + m_2)^2}\,.
\end{equation}
Henceforth, to ensure a stable vacuum, we will always assume that the deformation parameters are such that~\eqref{eq:vacuum-stability} is satisfied. 
Note that, as derived in Section~\ref{sec:sugra}, the existence of a real supergravity completion implies the non-strict version of the vacuum-stability bound. This follows from the supergravity constraints and positivity of the Gram matrix, without requiring supersymmetry.  The converse does not hold: the stability bound can be satisfied even when no real supergravity completion exists.
Saturation of the bound corresponds to coincident frequency roots.

Note that this may not be the optimal bound, but it is a sufficient one to ensure the stability of the vacuum in all gapped cases, and also to ensure that the energies are real. In the $\bb_0=\Delta=0$ case it ensures that $\omega>0$ and the bound coincides with the reality condition derived in~\eqref{eq:boundkappa}. The cases where $m_1 m_2=0$, corresponding to the two space-like trigonometric deformations and their common rational limit, need to be treated separately because the dispersion relation might become gapless. 

\subsection{Stability of the one-particle states}
\label{subsec:production-level-matching}
Let us now consider the $1 \rightarrow 3$ (and also the conjugate $3 \rightarrow 1$ processes). The existence of such a process indicates that a single particle state is not stable, and has enough energy to produce three lighter particles. While this is not an issue for unphysical states, it must be absent for physical one-particle states, satisfying the level-matching condition. Let us check this in the elliptic deformation.

We label the incoming particle by $4$ and the outgoing particles by $1,2,3$. The incoming state contains just one particle, so level matching and energy-conservation impose
\begin{equation}
\label{eq:levmatch13}
p_4=0\,,\qquad p_1+p_2+p_3=0\,, \qquad \omega_{s_1}(p_1)+\omega_{s_2}(p_2)+\omega_{s_3}(p_3)=\omega_{s_4}(0)\,,
\end{equation}
with $s_j = \pm$ depending on the type of excitation.
The reverse processes have the same constraints with incoming and outgoing states interchanged.
There are two logical ways in which such a process can be absent. First, there may be no real momenta satisfying the conditions~\eqref{eq:levmatch13}. In that case, the energy-conservation delta function has no support, irrespective of the value of the interaction coefficient. Second, real solutions may exist but the interaction coefficient may vanish on those solutions. The first is a kinematic exclusion, while the second is a dynamical cancellation. We find that in the elliptic deformation, both cases can occur.

 We illustrate this in the elliptic+WZ theory, with deformation parameters
\begin{equation}
 (\gamma_1,\gamma_2,\gamma_3)=(1,\tfrac{1}{10},\tfrac35),\qquad
 \delta_q=\tfrac3{10},\qquad \delta_0=\delta_1=\delta_2=0\,,
 \label{eq:production-example-parameters}
\end{equation}
which satisfy the stability bound derived in~\eqref{eq:vacuum-stability}. Since there is no mixing
between the $AdS_3$ and the $S^3$ modes in the quadratic Lagrangian, it is natural to set $\theta=0$ in the oscillator expansion. Let us consider a scattering involving particles from the $AdS_3$ sector only, namely
\begin{equation}
 z_+(p_4)\rightarrow z_-(p_1)z_-(p_2)z_-(p_3)\,,
 \label{eq:production-example-conservation}
\end{equation}
where $+$ and $-$ denote the upper and lower positive-norm branches. 
For the momenta we choose $p_1=0.13$ and $p_4=0.3$, and solve the two conservation
conditions for $p_2,p_3$. The table below gives
one real solution. Its nonzero coefficient shows that energy and
momentum conservation alone do not forbid particle production.
\begin{center}
\captionsetup{font={color=black},labelfont={color=black}}
\begin{tabular}{clrr}
\hline
$i$ & Mode & $p_i$ & $\omega_i$\\
\hline
$1$ & outgoing $z_-$ & $0.1300000000$ & $0.4282972549$\\
$2$ & outgoing $z_-$ & $0.6905267670$ & $0.7438128136$\\
$3$ & outgoing $z_-$ & $-0.5205267670$ & $0.3928608946$\\
$4$ & incoming $z_+$ & $0.3000000000$ & $1.5649709630$\\
\hline
\multicolumn{4}{c}{$\braket{z_-(p_1) z_-(p_2) z_-(p_3)|\mathbb{T} |z_+(p_4)} = -0.00152997 $}\\
\hline
\end{tabular}
\label{tab:production-without-lm}
\end{center}
On the other hand, keeping the same $p_1=0.13$, we can solve the conservation
conditions again, now imposing the level-matching condition. The table below shows that a real
solution still exists, but the interaction coefficient now vanishes
within numerical precision.
\begin{center}
\captionsetup{font={color=black},labelfont={color=black}}
\begin{tabular}{clrr}
\hline
$i$ & Mode & $p_i$ & $\omega_i$\\
\hline
$1$ & outgoing $z_-$ & $0.1300000000$ & $0.4282972549$\\
$2$ & outgoing $z_-$ & $0.6159973191$ & $0.6922286802$\\
$3$ & outgoing $z_-$ & $-0.7459973191$ & $0.4763460072$\\
$4$ & incoming $z_+$ & $0$ & $1.5968719423$\\
\hline
\multicolumn{4}{c}{$\braket{z_-(p_1) z_-(p_2) z_-(p_3)|\mathbb{T} |z_+(p_4)}=0$}\\
\hline
\end{tabular}
\end{center}
Similar cancellations occur for other values of the deformation parameters, including cases with several NSNS parameters turned on. These results provide some numerical evidence for the tree-level stability of physical one-particle states; however, they do not constitute an analytic proof for every channel and parameter value.

\subsection{\texorpdfstring{Two-to-two scattering and integrability}{Two-to-two scattering and integrability}}
For integrable theories, we expect that upon solving the energy- and momentum-conservation $\delta$-functions (but not the level-matching condition), the only contribution to $\mathbb{T}$ comes from the terms with an equal number of creation and annihilation operators, and with elastic scattering $p_1=p_3$ and $p_2=p_4$ (up to relabelling).
Solving the energy conservation $\delta$-function yields the Jacobian $J = \frac{1}{|\omega_1' - \omega_2'|}$ and one is left with
\begin{equation}
\mathbb{T} = -{\int} \frac{dp_1 dp_2}{|\omega_1'-\omega_2'|}  h_2^2(p_1,p_2,p_1,p_2)a^\dagger(p_2) a^\dagger(p_1) a(p_2) a(p_1)\,.
\end{equation}
This S-matrix acts on a two-particle state prepared in the distant past with $p_1>p_2$ as
\begin{equation}
\mathbb{T} \ket{a_i^\dagger(p_1) a_j^\dagger (p_2)} = \mathcal T_{ij}^{kl}(p_1,p_2)\ket{a_k^\dagger(p_1) a_l^\dagger (p_2)}\,,
\end{equation}
where we restored the flavour indices $i,j,k,l$ to highlight that $\mathbb{T}$ is a matrix in flavour space. 
In an integrable theory, the tree-level S-matrix should satisfy the classical Yang-Baxter equation
\begin{equation}
\label{eq:cYBE}
[\mathbb T_{23}, \mathbb T_{13}] + [\mathbb T_{23}, \mathbb T_{12}] + [\mathbb T_{13}, \mathbb T_{12}] = 0\,,
\end{equation}
where the subscript indicates on which vector spaces the two-body S-matrix acts. 

The distinction between unphysical and physical states plays an important role in integrability. The Yang-Baxter equation is a condition on the worldsheet S-matrix at unrestricted momenta; the incoming and outgoing states need not satisfy the closed-string level-matching condition. Indeed, it arises from the decomposition of a three-body scattering into two-body scatterings, which can occur in two different ways, and even if the external three particles satisfy the level-matching condition (for a scattering of a physical state), each individual pair of momenta might not do so. In an integrable theory, the Yang-Baxter equation should therefore be checked without imposing level-matching.

There are then three necessary scattering conditions for integrability: the first is that particle-production amplitudes vanish on shell, the second is that only elastic processes contribute, and the third one is the classical YB equation. For a theory where all particles have relativistic dispersion relation $\omega(p)=\sqrt{p^2+1}$, elastic scattering automatically follows from the energy and momentum conservation. This is however not the case for the elliptic deformation, due to the complicated structure of the dispersion relations. Therefore, in our case, already the condition to have elastic $2 \rightarrow 2$ scattering becomes a stringent test of the integrability of the model.

\paragraph{Presence of inelastic scattering.}
What we find is that for a generic choice of deformation parameters, there exist solutions to the energy- and momentum-conservation $\delta$-functions which are inelastic. 
For instance, when considering the $2 \rightarrow 2$ scattering with energy-momentum conservation $\delta(\omega_1+\omega_2-\omega_3-\omega_4)\delta(p_1+p_2-p_3-p_4)$ there are solutions of the form $p_3=p_1' \neq p_1$ and $p_4 = p_2' \neq p_2$. This is a behaviour that has already been observed for the pure RR elliptic deformation~\cite{Hoare:2025rtl}, as well as other integrable theories, e.g.~\cite{Costantino:2026xjm}. It is not yet a sign of non-integrability. Indeed, for some portion of our parameter space, the corresponding quantities $h_{2}^{2}(p_1,p_2,p_1',p_2')=0$ when evaluated on these inelastic solutions, and similarly also $h_{4}^0=h_{3}^1=0$ on the support of the energy-momentum $\delta$-functions, so that there are no production/annihilation processes. We also check that in those cases the classical YB equation is satisfied. 
We identify this as the integrable portion of the parameter space.
In contrast, there are portions of the parameter space where $h_{2}^{2}(p_1,p_2,p_1',p_2') \neq 0$ when evaluated on the inelastic solutions. We take the presence of these inelastic processes to be a sign that the theory is not integrable.

Note that although we restrict our attention to bosonic external states, fermionic interactions cannot cancel these purely bosonic tree-level amplitudes. In the light-cone gauge-fixed theory, every interaction vertex must contain an even number of fermionic fields. With no external fermions, any internal fermion lines therefore form a loop. In particular, a cubic interaction with one boson and two fermions cannot generate the required four-boson tree-level exchange contribution. Therefore, even if we restrict to the massive bosonic sector of the theory, our results about its non-integrability hold also for the superstring. 

\begin{figure}[htbp]
    \centering
    \begin{subfigure}[t]{0.49\linewidth}
        \centering
        \includegraphics[height=6 cm]{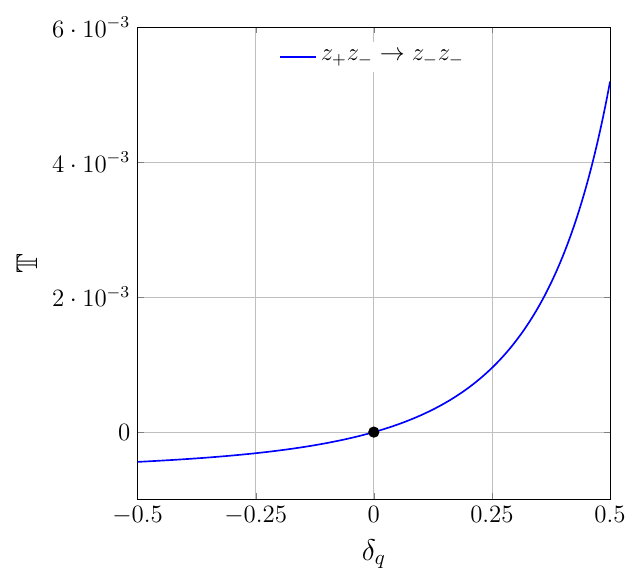}
        \caption{Varying $\bb_q$ at fixed elliptic parameters.}
        \label{fig:Smatrix-deltaq-varying}
    \end{subfigure}
    \hfill
    \begin{subfigure}[t]{0.49\linewidth}
        \centering
        \includegraphics[height=6 cm]{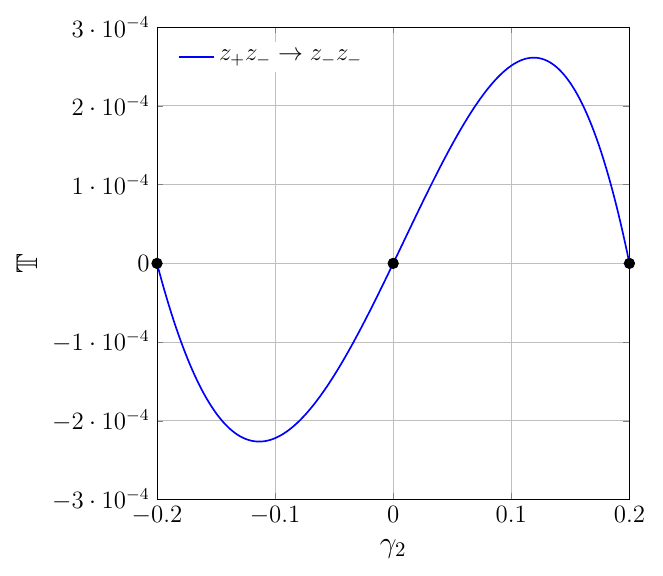}
        \caption{Varying $\gamma_2$ with $\bb_q=0.1$.}
        \label{fig:Smatrix-deltaq-fixed}
    \end{subfigure}
    \caption{Coefficient $\braket{z_-(p_3) z_-(p_4) | \mathbb{T}| z_+(p_1) z_-(p_2)}$, where $p_3, p_4$ are the (non-trivial) solutions to the energy-momentum conservation equations, as a function of $\bb_q$ with $\ga_2=0.1$ on the left and $\ga_2$ with $\bb_q=0.1$ on the right. The other parameters are $p_1=2,p_2=1$, $\bb_0=\bb_1=\bb_2=0$ and $\gamma_1=1.1$, $\gamma_3=0.9$. }
    \label{fig:Smatrix-deltaq}
\end{figure}

For a specific example of an inelastic process, let us focus on the elliptic+WZ theory and consider the coefficient of 
\begin{equation}
\label{eq:process}
a_{z,+}(p_1)a_{z,-}(p_2)a_{z,-}^\dagger(p_3)a_{z,-}^\dagger(p_4)\,,
\end{equation}
which corresponds to a $2 \rightarrow 2$ scattering involving four $AdS_3$ excitations.
We can solve the energy-momentum conservation equations 
\begin{equation}
    \omega_+(p_1)+\omega_-(p_2)-\omega_-(p_3)-\omega_-(p_4)=0\,, \qquad p_1+p_2-p_3-p_4=0\,,
\end{equation}
for $p_3$ and $p_4$ given $p_1$ and $p_2$. Because the energies $\omega_+$ and $\omega_-$ are different, if there is a solution to the energy-momentum conservation equations, it will generically be non-trivial, $\{p_1,p_2\} \neq \{p_3,p_4\}$. Figure~\ref{fig:Smatrix-deltaq} plots the coefficient $\braket{z_-(p_3) z_-(p_4) | \mathbb{T}| z_+(p_1) z_-(p_2)}$ as a function of the deformation parameter $\bb_q$ (for fixed elliptic deformation parameters), and as a function of $\gamma_2$ (for fixed deformation $\bb_q$). 
For generic values of the elliptic and flux parameters the S-matrix coefficient does not vanish, resulting in inelastic processes. 
The S-matrix coefficient only vanishes in four cases: $\gamma_2=0$ which corresponds to the time-like trigonometric deformation, $\gamma_2=\pm (\gamma_1-\gamma_3)$, which are the two space-like trigonometric deformations and $\bb_q=0$ which is the pure RR flux limit.

\begin{figure}[htbp]
    \centering
    \begin{subfigure}[t]{0.49\linewidth}
        \centering
        \includegraphics[height=6 cm]{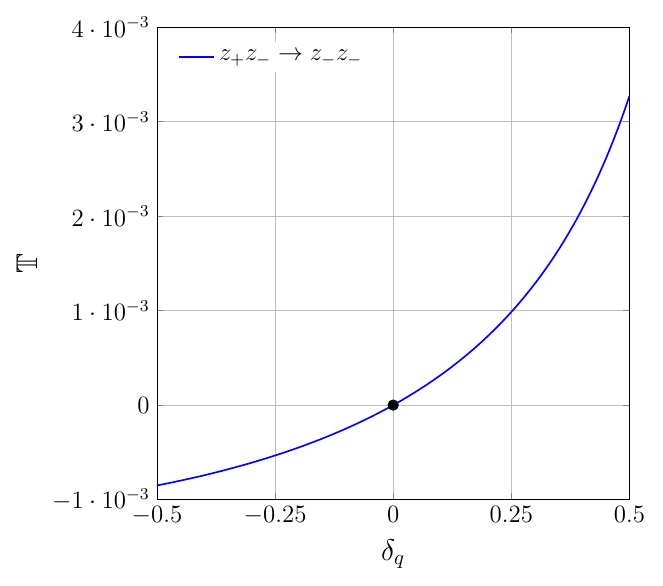}
        \caption{Varying $\bb_q$ and $\bb_0$ at fixed elliptic parameters.}
        \label{fig:Smatrix-deltaq and delta0-varying}
    \end{subfigure}
    \hfill
    \begin{subfigure}[t]{0.49\linewidth}
        \centering
        \includegraphics[height=6 cm]{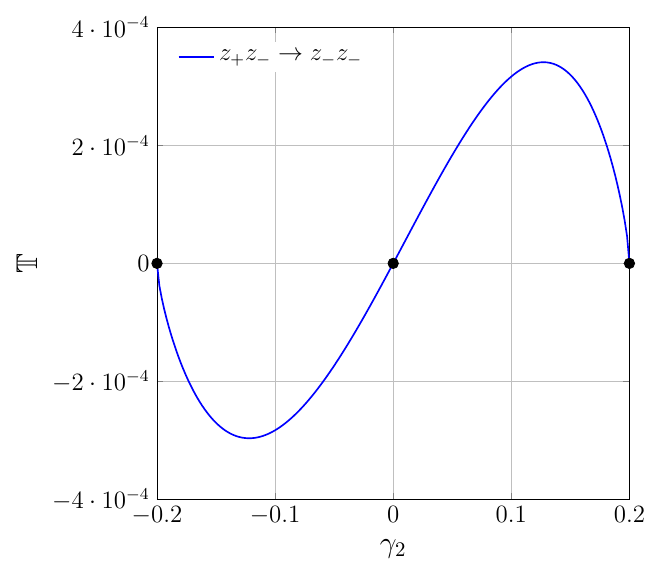}
        \caption{Varying $\gamma_2$ with $\bb_q=0.1$ and $\bb_0$ given by (\ref{eq:relation between delta0 and deltaq}).}
        \label{fig:Smatrix-deltaq and delta0-fixed}
    \end{subfigure}
    \caption{The same plots as in figure \ref{fig:Smatrix-deltaq} but now with $\bb_0\neq0$ given as in (\ref{eq:relation between delta0 and deltaq}). While $\bb_q$ only varies in the plot on the left, $\bb_0$ varies in both, because of the change in $\ga_2$ in the plot on the right. }
    \label{fig:Smatrix-deltaq and delta0}
\end{figure}

One could question if integrability is tied to the supersymmetry of the superstring theory. As discussed in Section~\ref{sec:sugra}, having a supersymmetric completion indeed imposes constraints on the NSNS parameters, and in particular a bosonic background with only $\bb_q$ turned on (only a WZ term) cannot be completed to a supersymmetric supergravity background with real fluxes. To highlight that the inelastic coefficients remain non-vanishing even if we choose a background with a supersymmetric completion, we impose $\bb_1=\bb_2=0$ and solve the constraints~\eqref{eq:susy-general-deltas} to find that
\begin{equation}
\delta_0
=\pm\frac{\gamma_1+\gamma_2+\gamma_3}{2\gamma_1}
\sqrt{
\frac{(\gamma_1-\gamma_3)^2-\gamma_2^2}
     {(\gamma_1+\gamma_3)^2-\gamma_2^2}
}\,\delta_q\,.
\label{eq:relation between delta0 and deltaq}
\end{equation}
We choose the positive branch and reproduce the plots of figure \ref{fig:Smatrix-deltaq}, now with both $\bb_0$ and $\bb_q$ switched on, with their relation given by (\ref{eq:relation between delta0 and deltaq}). The curves are only slightly modified, and the qualitative results remain the same, as shown in figure \ref{fig:Smatrix-deltaq and delta0}.

We performed a similar analysis for the other three B-field parameters $\bb_0, \bb_1, \bb_2$. The amplitude of selected inelastic processes as a function of the deformation parameters can be found in appendix~\ref{app:plots}.

The following table then summarises our results:
\begin{center}
\begin{tabular}{|c|c|c|c|c|c|} \hline
 & no B-field & $\bb_0$ & $\bb_1$ & $\bb_2$ & $\bb_q$  \\ \hline
Elliptic & $\checkmark$ & $\times$ & $\times$ & $\times$ & $\times$\\ \hline
Time-like trigonometric $\ga_2=0$ & $\checkmark$ & $\checkmark$ & $\times$ & $\times$ & $\checkmark$ \\ \hline
Space-like trigonometric $\ga_3=\ga_1-\ga_2$ & $\checkmark$& $\times$ & $\checkmark$  & $\times$& $\checkmark$\\ \hline
Space-like trigonometric $\ga_3=\ga_1+\ga_2$ & $\checkmark$ & $\times$ & $\times$ & $\checkmark$ & $\checkmark$ \\ \hline
Rational & $\checkmark$ & - & - & - & $\checkmark$ \\ \hline
\end{tabular}
\end{center}
A $\times$ indicates that the theory presents inelastic scattering processes and is not integrable. A $\checkmark$ indicates that the theory is integrable at tree-level: there are no inelastic processes, no particle production, and the classical Yang-Baxter equation is satisfied. The dashes $-$ in the last line (rational model) indicate that we do not consider these cases because they do not admit an embedding into supergravity with real fluxes.
While the rational and trigonometric $AdS_3 \times S^3 \times T^4$ string can be supported by a WZ term while preserving integrability, this is not the case of the elliptic deformation. The presence of the WZ term in the elliptic deformation leads to inelastic processes in the worldsheet S-matrix. 
For the trigonometric cases, in addition to the WZ term, the theory can also be supplemented by a B-field coupling the $AdS_3$ and $S^3$ sectors. 
Note that these results are in perfect agreement with the Lax integrability discussed in section~\ref{sec:Lax}. 
This is in contrast to e.g.~the results of~\cite{Borsato:2024sru}, where a Lax integrable model has a tree-level S-matrix with inelastic processes~\footnote{The origin of the inelastic processes is a bit different in the two setups. In the Jordanian deformation of~\cite{Borsato:2024sru}, the light-cone gauge-fixed theory presents cubic vertices that vanish when imposing the level-matching conditions, but contribute to the two-to-two scattering in an inelastic way. In our setup there are no cubic vertices and the inelastic processes originate from the quartic vertices.}.

\subsection{Analytic S-matrix coefficients in the integrable cases}

Finally, we provide the analytic expressions for the tree-level S-matrix in the integrable cases.
Let us briefly recall the bosonic tree-level S-matrix coefficients in the pure RR elliptic case, as derived and analysed in~\citep{Hoare:2023zti}. The choice $\theta_\pm(p)=0$ in the oscillator expansion leads to a diagonal S-matrix 
\begin{equation}
\label{eq:TmatrixElliptic}
\begin{aligned}
\mathbb{T} \ket{a_{z,\mu_1}^\dagger(p_1) a_{z,\mu_2}^\dagger(p_2)} &= (- 2 \mathcal A_{\mu_1 \mu_2} -  \mathcal B_{\mu_1 \mu_2}-  \mathcal D_{\mu_1 \mu_2}) \ket{a_{z,\mu_1}^\dagger(p_1) a_{z,\mu_2}^\dagger(p_2)}\,, \\
\mathbb{T} \ket{a_{y,\mu_1}^\dagger(p_1) a_{y,\mu_2}^\dagger(p_2)} &= (+ 2 \mathcal A_{\mu_1 \mu_2} + \mathcal B_{\mu_1 \mu_2}- \mathcal D_{\mu_1 \mu_2}) \ket{a_{y,\mu_1}^\dagger(p_1) a_{y,\mu_2}^\dagger(p_2)}\,, \\
\mathbb{T} \ket{a_{z,\mu_1}^\dagger(p_1) a_{y,\mu_2}^\dagger(p_2)} &= (+2 \mathcal G_{\mu_1 \mu_2} -\mathcal D_{\mu_1 \mu_2} ) \ket{a_{z,\mu_1}^\dagger(p_1) a_{y,\mu_2}^\dagger(p_2)}\,, \\
\mathbb{T} \ket{a_{y,\mu_1}^\dagger(p_1) a_{z,\mu_2}^\dagger(p_2)} &= (-2 \mathcal G_{\mu_1 \mu_2} -\mathcal D_{\mu_1 \mu_2}) \ket{a_{y,\mu_1}^\dagger(p_1) a_{z,\mu_2}^\dagger(p_2)}\,,
\end{aligned}
\end{equation}
with coefficients
\begin{equation}
\label{eq:Tmatrixcoefficients}
\begin{aligned}
\mathcal A_{\mu_1 \mu_2} &= \frac{1}{4} \frac{p_1^2 \omega_{\mu_2}(p_2)^2 + p_2^2 \omega_{\mu_1}(p_1)^2-2 p_1^2 p_2^2}{p_1 \omega_{\mu_2}(p_2) - p_2 \omega_{\mu_1}(p_1)}\,, \\
\mathcal G_{\mu_1 \mu_2} &= \frac{1}{4} \left(p_1 \omega_{\mu_2}(p_2) + p_2 \omega_{\mu_1}(p_1)\right)\,, \\
\mathcal D_{\mu_1 \mu_2} &= \left(a-\frac{1}{2} \right)(p_1 \omega_{\mu_2}(p_2) - p_2 \omega_{\mu_1}(p_1))\,,
\end{aligned}
\end{equation}
and
\begin{equation}
\mathcal B_{\mu_1 \mu_2} =  \frac{(\gamma_1-\gamma_3)^2 - \gamma_2^2}{4 \ga_3^2} \frac{(\omega_{\mu_1}(p_1)^2 - p_1^2 - (\ga_1+\ga_3)^2+\ga_2^2)(\omega_{\mu_2}(p_2)^2 - p_2^2 - (\ga_1+\ga_3)^2+\ga_2^2)}{p_1 \omega_{\mu_2}(p_2) - p_2 \omega_{\mu_1}(p_1)}\,.
\end{equation}
The $\mathcal{B}$ coefficient is the only one that depends explicitly on the deformation parameters and vanishes in the rational and in the space-like trigonometric limits.

\paragraph{Space-like trigonometric+$\bb_2$+WZ.} This is the integrable trigonometric branch with $\gamma_3=\gamma_1+\gamma_2$. 
In this case $\mathcal B=0$ and the S-matrix elements take the same form as in~\eqref{eq:TmatrixElliptic} and~\eqref{eq:Tmatrixcoefficients} when choosing $\theta_\pm(p)=0$. The B-field parameters $\bb_2$ and $\bb_q$ appear only implicitly through the modified dispersion relation. This is quite remarkable given that the explicit dispersion relation in the $\delta_q \neq 0$ case takes a very complicated form (while in the $\bb_q=0$ case it is given by~\eqref{eq:dispE12} with $\ga_3 = \ga_1+\ga_2$). The tree-level S-matrix is still diagonal and the classical YB equation is satisfied.

\paragraph{Space-like trigonometric+$\bb_1$+WZ.}
This is the integrable trigonometric branch with $\gamma_3 =\gamma_1-\gamma_2$. 
When setting $\theta_\pm(p)=0$, the tree-level S-matrix is no longer diagonal. It however recovers its diagonal structure with exactly the same coefficients as in~\eqref{eq:TmatrixElliptic} and~\eqref{eq:Tmatrixcoefficients} (and $\mathcal B=0$) if one makes the following rotation of the oscillators,
\begin{equation}
\label{eq:angle}
\sin (2\theta_\mu(p)) = \frac{2 p \bb_1 \left( \omega_{\mu}(p) - p \frac{\ga_1 \bb_q}{\ga_1-\ga_2}\right) }{\left(\omega_\mu(p)-p \frac{\ga_1 \bb_q}{\ga_1-\ga_2}\right)^2 + p^2 \bb_1^2}\,, \qquad \mu=\pm\,.
\end{equation}
The B-field parameters $\delta_1$ and $\delta_q$ again only appear implicitly through the modified dispersion relation and the one-particle change of basis.

\paragraph{Time-like trigonometric+$\bb_0$+WZ.}
Let us finish with the time-like trigonometric limit where $\ga_2=0$. When both $\bb_0=\bb_q=0$ this corresponds to the trigonometric deformation of $AdS_3 \times S^3 \times T^4$, which is equivalent to the unilateral YB/quantum group deformation up to total derivative terms. In the massive sector it preserves the Cartan $U(1)^4$ symmetry, and therefore allows for an expansion around the usual BMN vacuum. The light-cone gauge-fixed theory also has a quantum group symmetry, and the exact massive S-matrix is known, fixed by requiring compatibility with the quantum group symmetries~\cite{Hoare:2014oua}. The effect of picking a different light-cone gauge results in a twist of the S-matrix. Let us now analyse how this S-matrix changes when introducing a B-field.

The dispersion relation simplifies nicely to
\begin{equation}
    \omega_\pm(p)= -p \bb_0 + \bar{\omega}_\pm(p) \pm\ga_3\,, \qquad \bar{\omega}_\pm(p) = \sqrt{p^2\pm 2 \ga_1 p \bb_0  \mp (\ga_1+\ga_3)p \bb_q +\ga_1^2}\,.
\end{equation}
This dispersion relation is real for all $p$ if 
\begin{equation}
    (2\ga_1 \bb_0-(\ga_1+\ga_3)\bb_q)^2\le4\ga_1^2.
\end{equation}
The S-matrix is diagonal for the choice $\theta_\pm(p)=0$, of the form~\eqref{eq:TmatrixElliptic} but with coefficients $\mathcal B$ and $\mathcal G$ modified to
\begin{equation}
\begin{aligned}
   \mathcal{B}_{\mu_1\mu_2}&= \frac{1}{p_1 \omega_{\mu_2}(p_2) - p_2 \omega_{\mu_1}(p_1)}\Big( \bb_0 p_1 p_2 (p_2 \Omega_{\mu_1}(p_1) + p_1 \Omega_{\mu_2}(p_2)) + \bb_0(\bb_0 - 2\nu) p_1^2 p_2^2  \\
   &\qquad+
   \frac{\mu_1\mu_2 (\ga_1-\ga_3)}{1-(\bb_0 - \nu)^2} (\ga_1-\ga_3+\bb_0 (\mu_1 p_1 + \mu_2 p_2)) \Omega_{\mu_1}(p_1) \Omega_{\mu_2}(p_2)\Big) \,,
   \end{aligned}
\end{equation}
and
\begin{equation}
\mathcal G_{\mu_1 \mu_2} = \frac{1}{4} (p_1 \omega_{\mu_2}(p_2) + p_2 \omega_{\mu_1}(p_1)) -\frac{1}{2} \frac{\bb_0}{1-(\bb_0-\nu)^2} \Omega_{\mu_1}(p_1) \Omega_{\mu_2}(p_2)\,,
\end{equation}
where
\begin{equation}
\Omega_{\mu}(p) = \omega_{\mu}(p) - \mu (\ga_1+\ga_3) + \nu p\,, \qquad \nu = \frac{(\ga_1+\ga_3) \bb_q}{2 \ga_1}\,.
\end{equation}
It satisfies the classical Yang-Baxter equation.

As discussed in section~\ref{sec:Lax}, the deformation parameter $\bb_0$ corresponds to a TsT transformation, which is usually associated with an integrability-preserving Drinfel'd twist of the symmetry algebra and of the S-matrix~\cite{vanTongeren:2015uha}. However, because the TsT transformation is in directions subsequently used for light-cone gauge-fixing, not just in transverse directions, the twist involves redefinition of the notion of worldsheet energy and momentum. 
To see this, let us observe that we can write the dispersion relation in the form
\begin{equation}
W_\mu = \sqrt{P_\mu^2 -\mu (\Gamma_1+\Gamma_3) \bb_q P_\mu + \Gamma_1^2} + \mu \Gamma_3\,,
\end{equation}
with the new deformation parameters
\begin{equation}
\Gamma_1= C \ga_1\,, \qquad \Gamma_3 = (1-C)\ga_1+\ga_3\,, \qquad C=\frac{1+\nu^2 - \bb_0^2 + \sqrt{(1+\nu^2 - \bb_0^2)^2 - 4\nu^2}}{2}\,,
\end{equation} 
and the new energy and momentum variables
\begin{equation}
W_\mu(p) = \omega_\mu(p) - \nu \vartheta \Omega_\mu(p)\,, \qquad P_\mu(p) = C (p - \vartheta\Omega_\mu(p))\,, \qquad \vartheta = \frac{\bb_0-\nu +\nu/C}{1-(\bb_0-\nu)^2}\,.
\end{equation}
One can then check that all tree-level S-matrix elements can be brought into the form (we take $a=1/2$ for simplicity and $A,B$ denote all species of excitations)
\begin{equation}
\mathcal T_{AB}^{(\bb_0,\bb_q)}(p_1,p_2) = \frac{1}{C} \mathcal T^{(0,\bb_q)}_{A B}(P_1,P_2;\Gamma_1,\Gamma_3) + \frac{\vartheta (\ga_1+\ga_3)}{2} (J_{A,1} \Omega_2 - \Omega_1 J_{B,2})\,.
\end{equation}
In the above, $\mathcal T_{AB}^{(0,0)}$ denotes the tree-level trigonometric S-matrix without any B-field,  $\mathcal T_{AB}^{(0,\bb_q)}$ is the one with only WZ term and $\mathcal T_{AB}^{(\bb_0,\bb_q)}$ has both B-fields turned on. This transformation of the tree-level S-matrix is compatible with an abelian twist generated by the commuting charges $J$ and $Q=H-(\ga_1+\ga_3) M+\nu P_{w.s.}$, with eigenvalues
\begin{equation}
M_{z,\mu} = M_{y,\mu} = \mu\,, \qquad J_{z,\mu} = \mu\,, \qquad J_{y,\mu} = -\mu  \,, \qquad Q_{z,\mu} =Q_{y,\mu} = \Omega_\mu\,.
\end{equation}
The overall rescaling by $C^{-1}$ corresponds to a redefinition of the string tension.

Turning on $\bb_q$ corresponds to introducing a WZ term to the action. In the undeformed, mixed-flux $AdS_3 \times S^3 \times T^4$ theory, this leaves the symmetry algebra, including the one after fixing the BMN light-cone gauge, unchanged, but modifies the representation in which the excitations transform. In other words, all the modification is through the dispersion relation of the particles. In the quantum-group case however, this does not seem to be the case: it does not seem to be possible to redefine the deformation parameters and redefine energy and/or momentum to relate $\mathcal T_{AB}^{(0,\bb_q)}(P_1,P_2;\Gamma_1,\Gamma_3) = \mathcal T_{AB}^{(0,0)}(\tilde{P}_1,\tilde{P}_2;\tilde{\Gamma}_1,\tilde{\Gamma}_3)$. This obstruction was already observed in the tree-level S-matrix in the usual BMN light-cone gauge~\cite{Bocconcello:2020qkt}, and hints at a modification of the q-deformed light-cone gauge-fixed algebra when a WZ term is present.  

\section{Conclusions}
\label{sec:conclusions}
In this paper we analysed the integrability of the elliptic deformation of the PCM on $F = SU(1,1) \times SU(2)$, supplemented by a linear combination of four types of B-fields constructed out of left-invariant vielbeins $\mathsf{e}^a$: one of them ($\mathsf{b}_V$ with coupling $\bb_q$) corresponding to adding a WZ term to the action, while the other three ($\mathsf{b}_a$ with coupling $\bb_a$, $a=0,1,2$) couple the $SU(1,1)$ and  $SU(2)$ sectors and reduce to TsT transformations in the trigonometric limits. 

\medskip 
Starting with a quite general Ansatz for the Lax we found that the flatness equation does not admit non-trivial solutions for generic elliptic deformation parameters $\alpha_0,\alpha_1,\alpha_2$ and non-vanishing B-field, not even when tuning the B-field parameters. 
The only integrable cases with non-vanishing B-field are trigonometric models (and their rational limit), which can be recast as YB deformations of the PCM+WZ model.  
While not integrable, the mixed-flux elliptic deformations of the $AdS_3 \times S^3 \times T^4$ can be embedded into type IIB supergravity while preserving 8 supersymmetries. Therefore, while at the level of the supergravity theory the $SO(2)$ rotation of the NSNS and RR three-form flux preserves supergravity and supersymmetry, this is not the case of integrability. 

\medskip
To explore the link between Lax integrability and worldsheet scattering, we  quantised the theory in uniform light-cone gauge and computed the tree-level S-matrix for the massive worldsheet excitations. 
For generic elliptic deformation parameters the S-matrix features inelastic scattering, with $2 \rightarrow 2$ processes with different incoming and outgoing momenta $\{p_1,p_2\} \neq \{p_3,p_4\}$, violating one of the axioms of an integrable quantum field theory.
In the three classically integrable cases identified by the Lax analysis, while energy-momentum conservation still allows for such inelastic solutions, the S-matrix coefficients vanish on these solutions. There is no particle production, only elastic processes contribute and the S-matrix satisfies the classical YB equation.
There is then a perfect agreement between Lax integrability and worldsheet integrability.

\medskip
In the quantisation of the theory we observed an interesting consequence of expanding around a geodesic that is different from the usual BMN one. While analysing the mixed-flux $AdS_3 \times S^3 \times T^4$ superstring in the usual BMN light-cone gauge results in a positive semidefinite Hamiltonian, this may not be the case in a different gauge. In particular, when fixing the gauge using two isometries belonging to the same $\mathfrak{psu}(1,1|2)$ copy, as becomes necessary in the elliptic deformation due to the lack of other symmetries, the dispersion relation of the excitations propagating on the string worldsheet becomes negative in some range of the momentum. In the rational and trigonometric cases, it is always possible to perform a time-dependent redefinition of the excitations to land back on a positive semidefinite Hamiltonian, and the presence of these negative modes is just an artefact of the different choice of quantisation. But in the elliptic case, because the symmetry is broken, such a rotation introduces explicit time-dependence in the Hamiltonian. We have however seen that such negative-energy single-particle states are not necessarily problematic, as the level-matching condition forbids building multi-particle physical states with total negative energy.

\medskip
A conclusion of this study is that while rational and trigonometric $AdS_3 \times S^3 \times T^4$ strings can be supplemented by a WZ term while preserving the integrability of the theory, this is not the case of the elliptic $AdS_3 \times S^3 \times T^4$ string. It would be interesting to understand where this obstruction is coming from. 
An answer could come from analysing the (hidden) symmetries of the elliptic+WZ model. 
In the rational case, adding a WZ term does not affect the symmetry algebra before and after light-cone gauge-fixing ($\mathfrak{psu}(1,1|2)_L \oplus \mathfrak{psu}(1,1|2)_R$ and $[\mathfrak{psu}(1|1)_L \oplus \mathfrak{psu}(1|1)_R]^{\oplus 2}_{c.e}$ respectively), but changes the representation in which the excitations transform~\cite{Hoare:2013ida}. The matrix structure of the S-matrix is then the same, but the identification of the physical quantities (dispersion relation, central elements,...) depends on the coupling to the WZ term. 
For the trigonometric deformation, one copy of $\mathfrak{psu}(1,1|2)$ is preserved while the other gets promoted to a quantum group $\mathcal U_q(\mathfrak{psu}(1,1|2))$~\cite{Hoare:2014oua}. 
In the pure RR case the symmetry algebra after light-cone gauge-fixing becomes $[\mathfrak{psu}(1|1) \oplus \mathcal U_q(\mathfrak{psu}(1|1))]^{\oplus 2}_{c.e.}$, in agreement with perturbative S-matrix calculations~\cite{Seibold:2021lju}. In the trigonometric+WZ case the fate of symmetries after light-cone gauge-fixing is less understood. 
The bosonic tree-level S-matrix of the trigonometric+WZ model was computed in~\cite{Bocconcello:2020qkt}, but does not match an exact S-matrix constructed by assuming that the WZ term leaves the symmetry algebra intact and only enters through the representations. 
It may then be that the symmetry algebra after light-cone gauge-fixing explicitly depends on $\bb_q$ (in addition to the quantum group parameter)~\cite{Kawaguchi:2013gma,Demulder:2017zhz}, and it cannot be further enlarged to accommodate the elliptic deformation parameter.

\medskip
Finally, an important open problem concerns the formulation of the integrable pure RR elliptic deformation of $AdS_3 \times S^3 \times T^4$ to all orders in the fermions, as an integrable deformation of the supercoset model on $\mathfrak{psu}(1,1|2)^{\oplus 2}$. Such a formulation would make it possible to write the Lax in terms of group-valued fields and facilitate the analysis of its (hidden) symmetries and construct an exact S-matrix.
Constructing an elliptic deformation of the $AdS_3 \times S^3 \times S^3 \times S^1$ superstring, or of the supercoset model based on $\mathfrak{d}(2,1;\alpha)^{\oplus 2}$~\cite{Babichenko:2009dk,Borsato:2015mma}, would also be useful to understand if the elliptic deformation can be extended to non-simply-laced superalgebras. Elliptic deformations were also recently generalised to higher rank groups~\cite{Lacroix:2023qlz}, but their application to other AdS/CFT setups is not established yet. We will address some of these open problems in the near future.

\section*{Acknowledgments}
We thank Ben Hoare for comments on the draft. Parts of this work were presented at the \textit{Higher-dimensional Integrability and Holography} workshop in Buggerru (June 2026) and at the \textit{Workshop on Integrable String Sigma-Models} in Castasegna (August 2026). We are grateful to the participants of both events for fruitful discussions.
This work was supported by the Swiss National Science Foundation (SNSF) through the Ambizione grant ``Effective strings, membranes, and their solvability'' (project number 223544).

\appendix

\section{Light-cone gauge-fixing in the Hamiltonian formalism}
\label{app:Hamiltonian}
In this appendix we recall how the light-cone gauge-fixed Hamiltonian is obtained in the Hamiltonian formalism, as opposed to the Lagrangian formalism discussed in the main text. 
Starting from the string sigma-model~\eqref{eq:SSM}, the momenta conjugate to the bosonic fields are
\begin{equation}
P_\mu = -\text{T} (\sqrt{-h} h^{\tau \beta} G_{\mu \nu} + \epsilon^{\tau \beta} B_{\mu \nu}) \partial_\beta X^\nu\,.
\end{equation}
The gauge-fixed theory corresponds to setting $X^+=\tau$ and $P_-=1$. The light-cone gauge-fixed Hamiltonian is 
\begin{equation}
    \mathcal{H}^\text{g.f}=-P_+(X^j,\partial_\sigma X^j,P_j)\,,
\end{equation}
obtained by solving the Virasoro constraints $C_1=C_2=0$. In the following constraint equations we set $\text{T}=1$, so that
\begin{align}
C_1 &= P_\mu \partial_\sigma X^\mu \,, \\
C_2 &= G^{\mu \nu} P_\mu P_\nu + G_{\mu \nu} \partial_\sigma X^\mu \partial_\sigma X^\nu+2G^{\mu\nu}B_{\nu\kappa}P_\mu\partial_\sigma X^\kappa +G^{\mu\nu}B_{\mu\kappa}B_{\nu\lambda}\partial_\sigma X^\kappa\partial_\sigma X^\lambda\,.
\end{align}
The index $j$ runs over all target-space coordinates except those involved in the light-cone gauge-fixing.
In our case we have $X^j = (Z_1, Z_2, Y_1, Y_2,W_1,W_2,W_3,W_4)^j$.
The zeroth- and first-order Hamiltonians vanish on the branch
connected to the vacuum. Expressing the quadratic massive
Hamiltonian in terms of velocities through the canonical
momentum--velocity relations, we obtain
\begin{equation}
    \begin{aligned}
       \mathcal H_{2,\bullet}&=
    \frac{1}{2}(\dot{Z_1}^2+\dot{Z_2}^2+\dot{Y_1}^2+\dot{Y_2}^2)+\frac{1}{2}(1-\bb_0^2)(\acute{Z_1}^2+\acute{Z_2}^2+\acute{Y_1}^2+\acute{Y_2}^2)~\\
&+\frac{1}{2}(\ga_1^2-\ga_2^2-\ga_3^2)(Z_1^2+Z_2^2+Y_1^2+Y_2^2)-\ga_2\ga_3(Z_1^2-Z_2^2+Y_1^2-Y_2^2)~\\
&+\delta_2(\gamma_1+\gamma_2+\gamma_3)
  (Z_1Y_2'-Z_2'Y_1)
-\delta_1(\gamma_1-\gamma_2+\gamma_3)
  (Z_2Y_1'-Z_1'Y_2)
\\
&-(\delta_0-\delta_q)(\gamma_1+\gamma_2+\gamma_3)
  (Z_1Z_2'+Y_1Y_2')
+\delta_0(\gamma_1-\gamma_2+\gamma_3)
  (Z_1'Z_2+Y_1'Y_2)\,.
    \end{aligned}
\end{equation}
We also find a vanishing cubic interaction, but a non-vanishing quartic Hamiltonian $\mathcal H_4$.

The Lagrangian and Hamiltonian are linked by a Legendre transformation, precisely
\begin{equation}
\mathcal{H}^{\text{g.f.}} = P_j \partial_\tau X^j - \mathcal{L}^{\text{g.f.}}\,, \qquad P_j = \frac{\partial \mathcal{L}^{\text{g.f.}}}{\partial(\partial_\tau X^j)}\,.
\end{equation}
With the conventions specified in the main text, the Hamiltonian
and Lagrangian calculations agree through quartic order in the
massive bosonic sector. When comparing canonical momenta, we
include the shift associated with adding the quadratic total
derivative
$(\gamma_1+\gamma_2)\partial_\tau(Z_1Z_2+Y_1Y_2)$,
which puts the quadratic Lagrangian in the form used in the
main text.

\section{Quartic Lagrangian}
\label{app:L4}
In this appendix we write down the explicit expression for the quartic Lagrangian $\mathcal L_4$ for the massive sector of the theory, in the $a=1/2$ light-cone gauge. It can be decomposed into
\begin{equation}
\mathcal L_4 = \mathcal L_4^{(1)} + \mathcal L_4^{(2)}\,,
\end{equation}
with 
\begin{equation}
\mathcal L_4^{(2)}
=
-\left.\mathcal L_4^{(1)}\right|_{
(Z_1,Z_2,Y_1,Y_2,\delta_1,\delta_2)
\rightarrow
(Y_1,Y_2,Z_1,Z_2,-\delta_1,-\delta_2)
}\,,
\end{equation}
and
\begin{equation}
\begin{aligned}
\mathcal L_4^{(1)}
&=P+A_1\dot Z_1+A_2\dot Z_2+B_1Z_1'+B_2Z_2'
\\
&\qquad +\left(-Q-\frac12J_\tau+\bb_0J_\sigma
                    -\frac12F_\sigma\right)\mathcal K
 +\left(-2R+(1-\bb_0^2)J_\sigma+F_\tau\right)\mathcal H \\
&\qquad -2 \ga_{-+} \ga_{+-} Z_2^2 \mathcal G(Z_1,Z_1) + 2 \ga_{++}^2 Z_1^2 \mathcal G(Z_2,Z_2) + 4 \ga_{++} \ga_{+-} Z_1 Z_2 \mathcal G(Z_1,Z_2) \\
&\qquad +W_{Z_1Z_2}\Omega(Z_1,Z_2)+W_{Z_1Y_2}\Omega(Z_1,Y_2)
   +\frac12W_{Z_1Y_1}\Omega(Z_1,Y_1)+\frac12W_{Z_2Y_2}\Omega(Z_2,Y_2)\,.
\end{aligned}
\end{equation}
We define the functions, where the sums in $\mathcal K$ and $\mathcal H$ run over the four
massive fields, $\mathcal Y\in\{Z_1,Z_2,Y_1,Y_2\}$
\begin{equation}
\begin{aligned}
&\mathcal K = \sum_{\mathcal Y} \left[ \dot{\mathcal Y}^2 + (1-\bb_0^2) \acute{\mathcal Y}^2 \right]\,, \qquad 
\mathcal H = \sum_{\mathcal Y} \left[ \dot{\mathcal Y} \acute{\mathcal Y} - \bb_0 (\acute{\mathcal Y})^2 \right]\,, \\
&\mathcal G(f,g)  = \dot{f} \dot{g} - \bb_0 (\dot{f} g' + f' \dot{g}) + (\bb_0^2-1) f'g'\,, \qquad 
\Omega(f,g) = \dot{f} g' - f' \dot{g}\,,
\end{aligned}
\end{equation}
and the deformation parameters
\begin{equation}
\gamma_{\mu \nu} = \frac{\ga_1 + \mu \ga_2 + \nu \ga_3}{2}\,, \qquad \ga_{++} + \ga_{--} = \ga_{+-} + \ga_{-+}\,,\quad\mu,\nu\in\{+1,-1\}\,.
\end{equation}
For \(a\in\{\tau,\sigma\}\), we also define
\begin{equation}
\begin{aligned}
J_a={}&\gamma_{+-}Z_2\,\partial_aZ_1
+\gamma_{++}Z_1\,\partial_aZ_2\,,\\
F_a={}&-\gamma_{-+}(\bb_0Z_2+\bb_1Y_2)\partial_aZ_1
+\gamma_{++}\bigl[(\bb_0+\bb_q)Z_1+\bb_2Y_1\bigr]\partial_aZ_2\,.
\end{aligned}
\end{equation}

The field polynomials are
\begin{equation}
\begin{aligned}
P&=\frac43\Bigl[
\gamma_{++}^{\,2}\gamma_{--}
(\gamma_{--}-2\gamma_{++})Z_1^4+\gamma_{-+}^{\,2}\gamma_{+-}
(\gamma_{+-}-2\gamma_{-+})Z_2^4
-6\gamma_{-+}\gamma_{++}^{\,2}\gamma_{+-}Z_1^2Z_2^2
\Bigr]\,, \\
Q&=\gamma_{-+}\gamma_{+-}Z_2^2
+\gamma_{++}\gamma_{--}Z_1^2, \qquad
R=\gamma_{-+}^{\,2}(\bb_0Z_2^2+\bb_1Z_2Y_2)
+\gamma_{++}^{\,2}(\bb_0Z_1^2+\bb_2Z_1Y_1)\,.
\end{aligned}
\end{equation}

The coefficients of the terms with one time derivative are
\begin{equation}
\begin{aligned}
A_{1} &= 2 \ga_{+-} Z_2 \left( \ga_{++}(\ga_{--}-2 \ga_{++})  Z_1^2 + \frac{1}{3} \ga_{-+}  (\ga_{+-} -8 \ga_{-+}) Z_2^2 -  \ga_{++} \ga_{--} Y_1^2 -  \ga_{-+} \ga_{+-} Y_2^2 \right)\,, \\
A_{2} &= 2\ga_{++} Z_1 \left(\frac{1}{3} \ga_{++} (7 \ga_{--} - 2 \ga_{++}) Z_1^2 + 3 \ga_{-+} \ga_{+-} Z_2^2 - \ga_{++} \ga_{--} Y_1^2 - \ga_{-+} \ga_{+-} Y_2^2 \right)\,.
\end{aligned}
\end{equation}
The coefficients of the terms with one spatial derivative are
\begin{equation}
\begin{aligned}
B_{1} &= \frac23\bb_0Z_2\Bigl[
3\gamma_{++}\bigl(4\gamma_{++}\gamma_{+-}
-\gamma_{-+}\gamma_{--}\bigr)Z_1^2 +\gamma_{-+}^{\,2}(7\gamma_{+-}-2\gamma_{-+})Z_2^2\\
&\qquad+3\gamma_{-+}\gamma_{++}(2\gamma_{++}+\gamma_{--})Y_1^2
+3\gamma_{-+}^{\,2}(\gamma_{+-}+2\gamma_{-+})Y_2^2
\Bigr]\\
&+\frac23\bb_1\gamma_{-+}Y_2\Bigl[
-3\gamma_{++}\gamma_{--}Z_1^2
+\gamma_{-+}(\gamma_{+-}-2\gamma_{-+})(3Z_2^2-Y_2^2)+3\gamma_{++}(2\gamma_{++}+\gamma_{--})Y_1^2
\Bigr]\,\\
&+8\bb_2\gamma_{+-}\gamma_{++}^{\,2}Z_1Z_2Y_1\, \\
B_{2}&=-\frac23\bb_0\gamma_{++}Z_1\Bigl[
\gamma_{++}(\gamma_{--}-8\gamma_{++})Z_1^2
+3\gamma_{-+}(\gamma_{+-}-2\gamma_{-+})Z_2^2\\
&\qquad+3\gamma_{++}(2\gamma_{++}+\gamma_{--})Y_1^2
+3\gamma_{-+}(\gamma_{+-}+2\gamma_{-+})Y_2^2
\Bigr]\,\\
&+8\bb_1\gamma_{-+}^{\,2}\gamma_{++}Z_1Z_2Y_2\\
&+\frac23\bb_2\gamma_{++}Y_1\Bigl[
3\gamma_{++}(4\gamma_{++}+\gamma_{--})Z_1^2
+\gamma_{++}(\gamma_{--}-2\gamma_{++})Y_1^2+3\gamma_{-+}\gamma_{+-}(Z_2^2+Y_2^2)
\Bigr]\,\\
&+\frac23\bb_q\gamma_{++}Z_1\Bigl[
\gamma_{++}(\gamma_{--}-2\gamma_{++})Z_1^2
+3\gamma_{++}\gamma_{--}Y_1^2+3\gamma_{-+}\gamma_{+-}(Z_2^2+Y_2^2)
\Bigr]\,.
\end{aligned}
\end{equation}
Finally, the coefficients multiplying the antisymmetric derivative
products are
\begin{equation}
\begin{aligned}
W_{Z_1Z_2}={}&2\gamma_{++}\Bigl[
\bigl(\bb_0(\gamma_{++}+\gamma_{--})+\bb_q\gamma_{+-}\bigr)Z_1Z_2
+\bb_1\gamma_{-+}Z_1Y_2+\bb_2\gamma_{+-}Z_2Y_1
\Bigr]\,, \\
W_{Z_1Y_2}={}&2\gamma_{++}\Bigl[
\bb_2\gamma_{+-}Z_1Z_2
+\bigl(\bb_0(\gamma_{++}+\gamma_{--})-\bb_q\gamma_{+-}\bigr)Z_2Y_1
+\bb_1\gamma_{-+}Y_1Y_2
\Bigr]\,, \\
W_{Z_1Y_1}={}&-2\gamma_{-+}\Bigl[
2\bb_0(\gamma_{++}+\gamma_{--})Z_2Y_2
+\bb_1\gamma_{-+}(Y_2^2-Z_2^2)
\Bigr]\,, \\
W_{Z_2Y_2}={}&2\gamma_{++}^{\,2}\Bigl[
\bb_2(Z_1^2-Y_1^2)-2\bb_qZ_1Y_1
\Bigr]\,.
\end{aligned}
\end{equation}

\section{Plots of the amplitude of inelastic processes}
\label{app:plots}

In this appendix we illustrate the dependence of selected inelastic
tree-level coefficients $h_2^2(\vec{p})$ on the elliptic and NSNS deformation
parameters. We set $a=\tfrac12$ and fix the incoming momenta to
$p_1=2$ and $p_2=1$. At each parameter value, the outgoing momenta
$p_3=p_1',p_4=p_2'$ are determined by energy--momentum conservation.
The processes considered are
\begin{equation}
\begin{aligned}
   &z_+(p_1)z_-(p_2)\rightarrow y_-(p_1')y_-(p_2')
 \qquad\text{for the $\bb_0$ deformation}\,,\\
 &z_+(p_1)z_+(p_2)\rightarrow y_-(p_1')y_-(p_2')
 \qquad\text{for the $\bb_1$ and $\bb_2$ deformations}\,.
 \end{aligned}
\end{equation}

\begin{figure}[htbp]
    \centering
    \begin{subfigure}[t]{0.49\linewidth}
        \centering
        \includegraphics[height=6 cm]{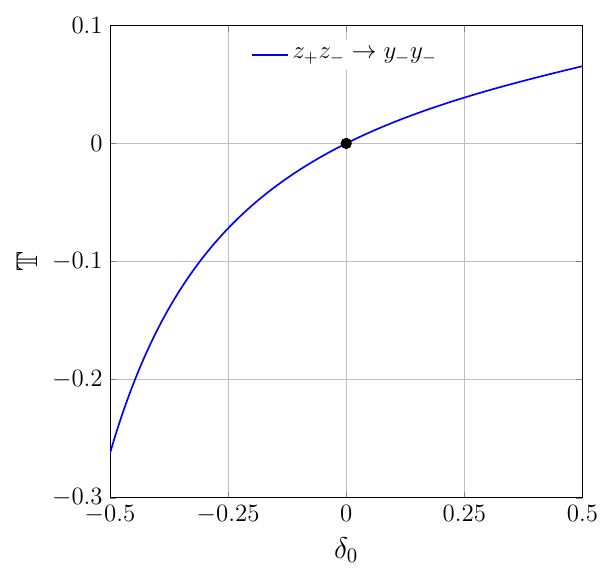}
        \caption{Varying $\bb_0$ at fixed elliptic parameters.}
        \label{fig:Smatrix-delta0-varying}
    \end{subfigure}
    \hfill
    \begin{subfigure}[t]{0.49\linewidth}
        \centering
        \includegraphics[height=6 cm]{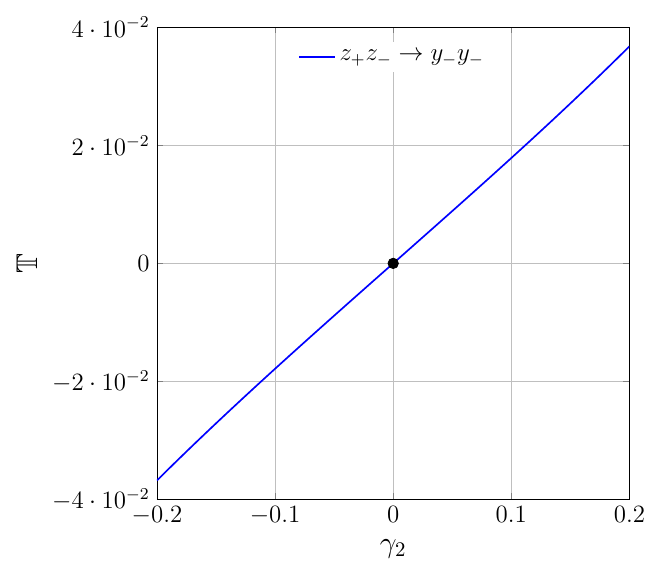}
        \caption{Varying $\gamma_2$ with $\bb_0=0.1$.}
        \label{fig:Smatrix-delta0-fixed}
    \end{subfigure}
    \caption{Inelastic coefficient of $\mathbb{T}$ for
    $z_+z_-\to y_-y_-$ with
    $\bb_1=\bb_2=\bb_q=0$, $a=\tfrac12$, and incoming
    momenta $(p_1,p_2)=(2,1)$.
    Left: $(\gamma_1,\gamma_2,\gamma_3)=(1.1,0.1,0.9)$
    and $\bb_0\in[-0.5,0.5]$. The bound~\eqref{eq:vacuum-stability} is $|\bb_0| < 0.987$.
    Right: $\gamma_1=1.1$, $\gamma_3=0.9$ and $\bb_0=0.1$.
    Black dots mark the zero-flux limit on the left and the
    compatible trigonometric point $\gamma_2=0$ on the right. The dependence on the right plot is not linear.}
    \label{fig:Smatrix-delta0}
\end{figure}

\begin{figure}[htbp]
    \centering
    \begin{subfigure}[t]{0.49\linewidth}
        \centering
        \includegraphics[height=6.2 cm]{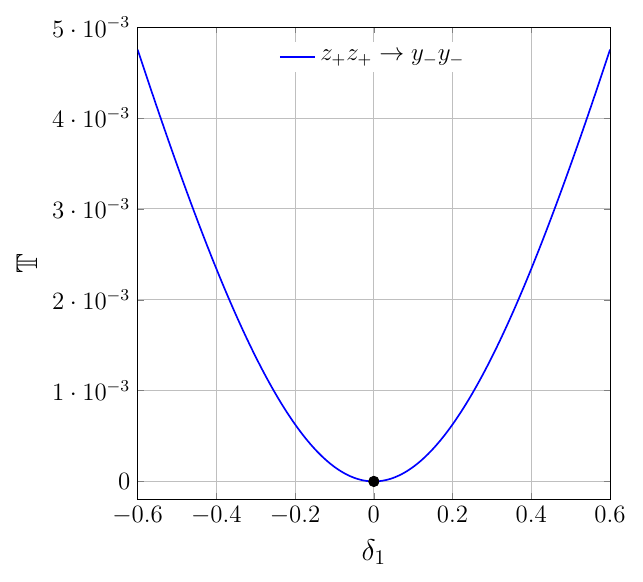}
        \caption{Varying $\bb_1$ at fixed elliptic parameters.}
        \label{fig:Smatrix-delta1-varying}
    \end{subfigure}
    \hfill
    \begin{subfigure}[t]{0.49\linewidth}
        \centering
        \includegraphics[height=6 cm]{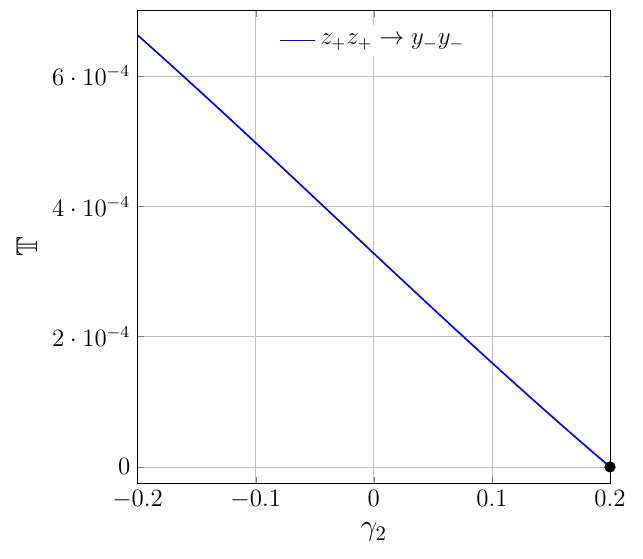}
        \caption{Varying $\gamma_2$ with $\bb_1=0.1$.}
        \label{fig:Smatrix-delta1-fixed}
    \end{subfigure}
    \caption{Inelastic coefficient of $\mathbb{T}$ for
    $z_+z_+\to y_-y_-$ with
    $\bb_0=\bb_2=\bb_q=0$, $a=\tfrac12$, and incoming
    momenta $(p_1,p_2)=(2,1)$.
    Left: $(\gamma_1,\gamma_2,\gamma_3)=(1.1,0.1,0.9)$
    and $\bb_1\in[-0.6,0.6]$.  The bound~\eqref{eq:vacuum-stability} is $|\bb_1| < 0.639$. 
    Right: $\gamma_1=1.1$, $\gamma_3=0.9$ and $\bb_1=0.1$.
    Black dots mark the zero $B$-field limit on the left and the
    compatible space-like trigonometric point
    $\gamma_2=\gamma_1-\gamma_3=0.2$ on the right.}
    \label{fig:Smatrix-delta1}
\end{figure}

\begin{figure}[htbp]
    \centering
    \begin{subfigure}[t]{0.49\linewidth}
        \centering
        \includegraphics[height=6 cm]{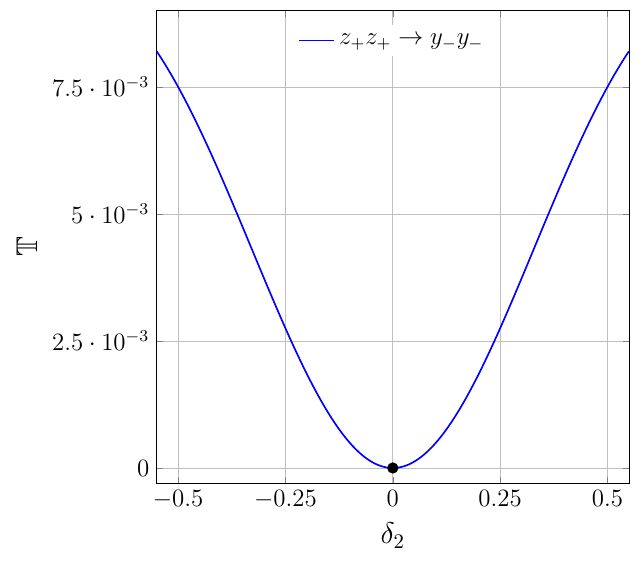}
        \caption{Varying $\bb_2$ at fixed elliptic parameters.}
        \label{fig:Smatrix-delta2-varying}
    \end{subfigure}
    \hfill
    \begin{subfigure}[t]{0.49\linewidth}
        \centering
        \includegraphics[height=6 cm]{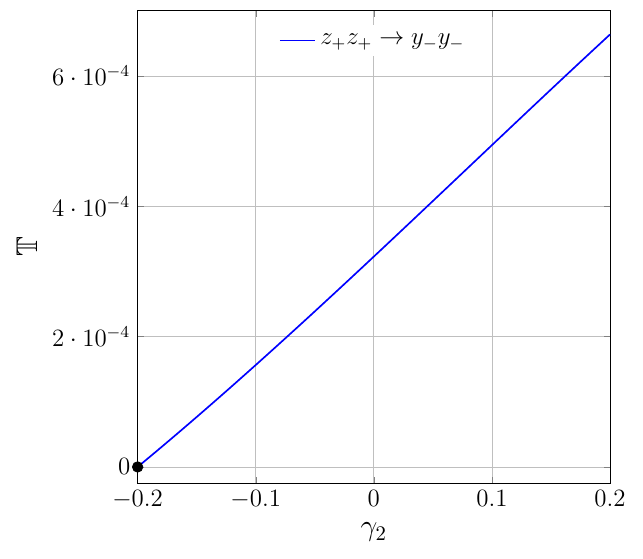}
        \caption{Varying $\gamma_2$ with $\bb_2=0.1$.}
        \label{fig:Smatrix-delta2-fixed}
    \end{subfigure}
    \caption{Inelastic coefficient of $\mathbb{T}$ for
    $z_+z_+\to y_-y_-$ with
    $\bb_0=\bb_1=\bb_q=0$, $a=\tfrac12$, and incoming
    momenta $(p_1,p_2)=(2,1)$.
    Left: $(\gamma_1,\gamma_2,\gamma_3)=(1.1,0.1,0.9)$
    and $\bb_2\in[-0.55,0.55]$.  The bound~\eqref{eq:vacuum-stability} is $|\bb_2| < 0.578$. 
    Right: $\gamma_1=1.1$, $\gamma_3=0.9$ and $\bb_2=0.1$.
    Black dots mark the zero $B$-field limit on the left and the
    compatible space-like trigonometric point
    $\gamma_2=\gamma_3-\gamma_1=-0.2$ on the right.}
    \label{fig:Smatrix-delta2}
\end{figure}

\newpage

\bibliographystyle{JHEP}
\bibliography{references}

\end{document}